\documentclass[twocolumn,tighten]{aastex63}
\usepackage{graphicx}
\usepackage{gensymb}
\usepackage{amsmath}
\def\kms{$\rm km~s^{-1}$}

\begin{document}

\title{Far UV and Optical Emissions from Three Very Large Supernova Remnants \\  Located at Unusually High Galactic Latitudes}

\author[0000-0003-3829-2056]{Robert A.\ Fesen}
\affiliation{6127 Wilder Lab, Department of Physics and Astronomy, Dartmouth College, Hanover, NH 03755 USA}

\author{Marcel Drechsler}
\affiliation{Sternwarte B{\"a}renstein, Feldstraße 17, 09471 B{\"a}renstein, Germany}

\author[00000-0002-4471-9960]{Kathryn E.\ Weil}
\affiliation{Department of Physics and Astronomy, Purdue University, 525 Northwestern Avenue, West Lafayette, IN 47907 USA}

\author{Xavier Strottner}
\affiliation{Montfraze, 01370 Saint Etienne Du Bois France}

\author[0000-0002-7868-1622]{John C.\ Raymond}
\affiliation{Harvard-Smithsonian Center for Astrophysics, 60 Garden St., Cambridge, MA 02138, USA}

\author{Justin Rupert}
\affil{MDM Observatory, Kitt Peak National Observatory, 950 N. Cherry Ave., Tucson, AZ 85719, USA}

\author[0000-0002-0763-3885]{Dan Milisavljevic}
\affiliation{Department of Physics and Astronomy, Purdue University, 525 Northwestern Avenue, West Lafayette, IN 47907 USA}

\author[0000-0001-8073-8731]{Bhagya M.\ Subrayan}
\affiliation{Department of Physics and Astronomy, Purdue University, 525 Northwestern Avenue, West Lafayette, IN 47907 USA}

\author{Dennis di Cicco}
\affil{MDW Sky Survey, New Mexico Skies Observatory, Mayhill, NM 88339, USA}

\author{Sean Walker}
\affil{MDW Sky Survey, New Mexico Skies Observatory, Mayhill, NM 88339, USA}

\author{David Mittelman}
\affil{MDW Sky Survey, New Mexico Skies Observatory, Mayhill, NM 88339, USA}

\author{Mathew Ludgate}
\affil{Ross Creek Observatory, Dunedin 9010, New Zealand}

%*******************************
\begin{abstract}

Galactic supernova remnants (SNRs) with angular dimensions greater than a few
degrees are relatively rare, as are remnants located more than ten degrees off
the Galactic plane. Here we report a UV and optical
investigation of two previously suspected SNRs more than ten degrees
in both angular diameter and Galactic latitude. One is a proposed remnant
discovered in 2008 through 1420 MHz polarization maps near Galactic coordinates
$l$ = 353\degr, $b$ = $-$34\degr.  GALEX far UV (FUV) and H$\alpha$ emission
mosaics show the object's radio emission coincident with a 
$11\degr \times 14\degr$ shell of UV filaments which surrounds a 
diffuse H$\alpha$ emission ring. Another proposed high latitude SNR is the
$20\degr \times 26\degr$ Antlia nebula (G275.5+18.4) discovered in 2002 through
low-resolution all-sky H$\alpha$ and ROSAT soft X-ray emissions. GALEX
FUV and H$\alpha$ mosaics along with optical spectra indicate the presence of
shocks throughout the Antlia nebula with estimated shock velocities of 70 to
over 100 km s$^{-1}$. We also present evidence that it has collided with the NE rim of the Gum Nebula. We find both of these large nebulae
are bona fide SNRs with ages less than 10$^{5}$ yr despite their
unusually large angular dimensions.  We also present FUV and optical images along with optical spectra of a new high-latitude SNR (G249.7+24.7) some $4.5\degr$ in
diameter which has also been independently discovered in 
X-rays and radio (Becker at al. 2021). We find this remnant's
distance to be $\leq$400 pc based on the detection of red and 
blue Na I absorption features in the spectra of two background stars.

\end{abstract}
\bigskip
\keywords{SN: individual objects: ISM: supernova remnants } 

%*******************************

\section{Introduction}

In spite of an increasing variety of supernovae (SNe) sub-types, the majority of SNe are generally believed to release $0.5 - 2 \times 10^{51}$ erg, although superluminous SNe may release substantially more \citep{Howell2017}.  Only a small percentage of this comes out as visible light, with most of a SN's energy initially carried away
in the form of kinetic energy. It is this enormous point deposition of energy that has a significant impact on the structure and energy content of a galaxy's interstellar medium through an expanding remnant that can last largely intact up to $\simeq$10$^{5}$ yr  \citep{Cox1974,McKee1977,Shull1989,Blondin1998}.

Currently, there are some 300 confirmed Galactic supernova remnants (SNRs) catalogued  with dozens of other suspected SNRs and more added every few years \citep{Safi2012,Green2019}. Most Galactic SNRs are less than a degree in angular size, more than 1 kpc distant and well evolved, with typical estimated ages between 10$^{4}$ and 10$^{5}$ yr. 

%%%%%%%%%%%%%%%%%%%%%%%%%%%%%%%%%%%%%%%%%%%%%%%%%%%%%%%%%%%%%%%%
%%% Table of Galex SNRs %%%
\begin{deluxetable*}{lcccc}[ht]
\tablecolumns{5}
\tablecaption{Locations and Dimensions of SNRs  }
\tablewidth{0pt}
\tablehead{ 
\colhead{SNR ID} & \colhead{Galactic Coordinates} &  \colhead{Approximate Center (J2000)} & \colhead{Angular Size}  & \colhead{Estimated Distance} }
\startdata
G249+24 & $l$ = 249.7 $b$ = $+24.7$ & RA = 09$^{\rm h}$33$^{\rm m}$  Dec = $-17\degr$00$'$ & $4.5\degr $  &   $<$390 pc   \\
Antlia  & $l$ = 275.5 $b$ = $+18.4$ & RA = 10$^{\rm h}$38$^{\rm m}$  Dec = $-37\degr$20$'$ & $20\degr \times 26\degr$    &   $\sim$250 pc   \\
G354-33 & $l$ = 354.0 $b$ = $-33.5$ & RA = 20$^{\rm h}$16$^{\rm m}$  Dec = $-45\degr$50$'$ & $11\degr \times 14\degr$    &   $\sim$500 pc   \\
\enddata
\end{deluxetable*}
%%%%%%%%%%%%%%%%%%%%%%%%%%%%%%%%%%%%%%%%%%%%%%%%%%%%%%%%%%%%%%%%

Out-sized attention relative to their population percentage has been directed to the handful of the Milky Way's young remnants with ages less than 5000 yr. These include the Crab Nebula, Cassiopeia A, the remnants of Tycho and Kepler, the Vela remnant, and Puppis A. This increased attention is due to their high expansion velocities, metal-rich ejecta, bright emission across the entire electromagnetic spectrum, and clearer connections to the various core-collapse and thermonuclear SN sub-types. 

The majority of Galactic SNRs are first detected through radio observations due to their characteristic synchrotron nonthermal radio emission associated with shocked gas
\citep{Downes1971,Chevalier1977,Green1984,Green2004}. Such nonthermal emission leads to a power law flux density, S, with S $\varpropto$  $\nu^{-\alpha}$ where $\alpha$ is the emission spectral index with typical  SNR values between -0.3 and -0.7. Rarer pulsar wind-driven SNRs exhibit a much flatter radio spectrum, with spectral indices around zero.
 
While multi-frequency radio surveys have been historically the dominant tool for finding SNRs,
X-ray studies have also led to the discovery of several additional SNRs. These include RX J1713.7-3946 (G347.3-0.5; \citealt{Pfeff1996}) and the young ``Vela Junior'' SNR (G266.2-1.2; \citealt{Aschenbach1998}) coincident with the much larger Vela SNR. Recent examples of X-ray confirmed or discovered remnants include several new SNRs in the LMC \citep{Maggi2014} and X-ray confirmation of the suspected  Galactic remnant G53.4+0.0 \citep{Driessen2018}.

Despite the fact that less than 50\% of Galactic SNRs exhibit any appreciable associated optical emission, a remnant's optical emission can be useful in confirming the presence of high-velocity shocks and in defining a remnant's overall size and
morphology. Although discoveries of new SNRs in the optical is relatively rare, there have been several recently reported discoveries \citep{Stupar2007,Boumis2009,Stupar2012,Fesen2015,Neustadt2017,Stupar2018,How2018,Fesen2019} along with many proposed SNR candidates \citep{Stupar2008,Stupar2011,Boumis2009,Ali2012,Sabin2013}.

The main criterion for optical SNR identification is a line ratio of I([\ion{S}{2}])/I(H$\alpha$) $\geq 0.4$ which has proven useful in identifying the shocked emission of SNRs in both the Milky Way and nearby local group galaxies \citep{Blair1981,Dopita1984,Fesen1985,Leonidaki2013,Long2017}. Recently, the near infrared [\ion{Fe}{2}] line emissions at 1.27 and 1.64 $\mu$m have also been used to detect dust obscured SNRs in nearby galaxies (see review by \citealt{Long2017}).

One wavelength regime that has not yet been fully exploited to search for new Galactic SNRs is the ultraviolet (UV). 
Here we present results of an initial study of SNRs located far away from the Galactic plane using wide field-of-view (FOV) UV images assembled from the Galaxy Evolution Explorer (GALEX) All-Sky survey \citep{Bianchi2009}. We began this research by investigating two unusually high latitude suspected supernova remnants including the exceptionally large Antlia remnant \citep{McCull2002} (see Table 1).
During this work, we also found a new apparent SNR.
However, after our initial submission, W. Becker kindly sent us a
preprint of their independent discovery of this new remnant based
on SRG/eRosita X-ray and Parks All-Sky Survey observations.
Since their paper has now been published \citep{Becker2021}, we have cited their
results in reference to our independent findings. UV and optical
images plus some follow-up optical spectra are described in
Section 2, with results presented in Section 3. We discuss the
general properties of these SNRs in Section 4, with our
conclusions and discussion of helpful follow-up observations
given in Section 5.

%During this work, we also found a new apparent SNR.  UV and optical images plus some follow-up optical spectra are described in $\S$2, with results presented in $\S$3. We discuss the general properties of these SNRs in $\S$4, with our conclusions and discussion of helpful follow-up observations given in $\S$5.

%%%%%%%%%%%%%%%%%%%%%%%%%%%%%%%%%%%%%%%%%%%%%%%%%%%%%%%%%%%%%%%%%%%%%
%%% Figure 1: Galex map  
%%%%%%%
\begin{figure*}[ht]
\centerline{\includegraphics[angle=0,width=10.0cm]{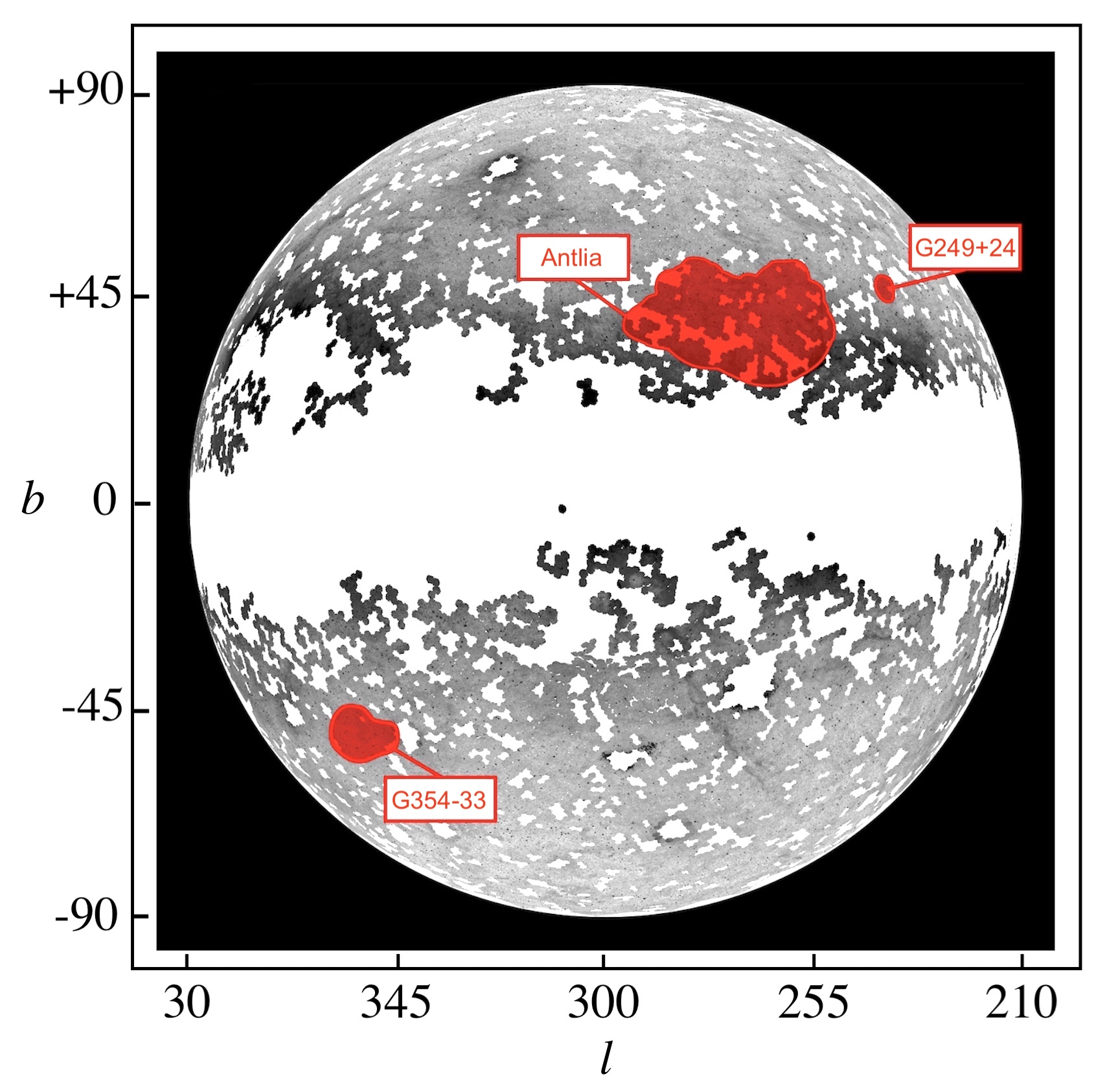}}
\caption{GALEX FUV intensity map of half of the sky in Galactic coordinates, showing locations of two apparent SNRs, G249+24 and G354-33, and the exceptionally large Antlia SNR. White areas denote regions with no GALEX imaging data. \label{globe}  }
\end{figure*}
%%%%%%%%%%%%%%%%%%%%%%%%%%%%%%%%%%%%%%%%%%%%%%%%%%%%%%%%%%%%%%%%%%%%%%

%%%%%%%%%%%%%%%%%%%%%%%%%%%%%%%%%%%%%%%%%%%%%%%%%%%%%%%%%%%%%%%%%
%%% Table of Optical Data %%%
\begin{deluxetable*}{llcccccc}[ht]
\tablecolumns{8}
\tablecaption{Optical Data on SNRs }
\tablehead{\colhead{SNR} &  \colhead{Survey or ~~~~~~~~~~~ } & \colhead{Camera or} &
\multicolumn{3}{c}{\underline{~~~~~~~~~~~~~~~~~~~~~~~~Direct Images~~~~~~~~~~~~~~~~~~~~~~~~}} & \multicolumn{2}{c}{\underline{~~~~~~~~~~Spectra~~~~~~~~~~}  } \\
\colhead{ID} & Observatory  & Telescope & \colhead{Filters} & \colhead{FOV}  & Pixel Scale &
\colhead{Wavelength} &  \colhead{FWHM}  } 
\startdata
G249+24  & MDW H$\alpha$ Survey & 130 mm lens   &  H$\alpha$ & 
           $3.4\degr \times 3.4\degr$ & 3.17$''$  & \nodata & \nodata \\
         & MDM Observatory & 2.4 m  & H$\alpha$ + [\ion{N}{2}]   &
          $18' \times 18'$ & 0.55$''$  & $\lambda$4000 - $\lambda$6900 & 3.5 \AA \\
          & MDM Observatory & 2.4 m  & \nodata & \nodata & \nodata & $\lambda$5735 - $\lambda$6065 & 0.55 \AA  \\
Antlia   & SHASSA Survey & f/1.6 52 mm lens & H$\alpha$, $\lambda$6440, $\lambda$6770 &
         $13\degr \times 13\degr$ & 47.6$''$  & \nodata & \nodata \\
         & MDW H$\alpha$ Survey &  130 mm lens   &  H$\alpha$ & 
           $3.4\degr \times 3.4\degr$ & 3.17$''$  & \nodata & \nodata \\
         & Ross Creek Obs. & f/2.0 200 mm lens & 
           H$\alpha$ + [\ion{N}{2}], [\ion{O}{3}], [\ion{S}{2}] & 
          $10.3\degr \times 6.9\degr$ & 3.87$''$ & \nodata & \nodata \\
         & SALT Observatory  & 10 m & \nodata & \nodata & \nodata & $\lambda$4050 - $\lambda$7120 & 5.0 \AA  \\  
G354-33  & SHASSA Survey & f/1.6 52 mm lens & H$\alpha$, $\lambda$6440, $\lambda$6770 &
         $13\degr \times 13\degr$ & 47.6$''$  & \nodata & \nodata \\
         & CHILESCOPE   & 0.5 m &  H$\alpha$ + [\ion{N}{2}], [\ion{O}{3}]  &  
          $1.1\degr \times 1.1\degr$ & 0.96$''$  & \nodata   & \nodata \\
\enddata
\label{Table_2}
\end{deluxetable*}
%%%%%%%%%%%%%%%%%%%%%%%%%%%%%%%%%%%%%%%%%%%%%%%%%%%%%%%%%%%%%%%%%%%%

\section{Data and Observations}
\subsection{GALEX UV Images}

The GALEX satellite was a NASA science mission led by the California Institute of Technology who operated it from July 2003 through February 2012.  The main instrument was a 50 cm diameter modified Ritchey-Chrétien telescope, a dichroic beam splitter and astigmatism corrector, and two microchannel plate detectors to simultaneously cover two wavelength bands with a 1.25 degree field of view with a resolution of $1.5''$ pixel$^{-1}$.  Direct images were obtained in two broad bandpasses: a far-UV (FUV) channel sensitive to light in the 1344 to 1786 \AA\/ range, and a near-UV (NUV) channel covering 1771 to 2831 \AA \ \citep{Morrissey2007}. Resulting images are circular in shape with an image FWHM resolution of $\sim4.2''$ and $\sim5.3''$ in the FUV and NUV bands, respectively.

Being mainly a mission to study the UV properties of galaxies in the local universe, GALEX survey images largely avoided the complex and external galaxy poor Galactic plane. However, even its All-Sky Survey program focused away from the plane of the Milky Way did not cover all possible areas, leading to numerous gaps (see Fig.~\ref{globe}). Nonetheless, some 26,000 square degrees were imaged to a depth of  m$_{\rm AB}$ = 20.5 \citep{Bianchi2009}. Using the on-line GALEX images, large FUV and NUV image mosaics were examined of several regions typically more than 10 degrees off the Galactic plane. 

%Using GALEX mosaic images, we initiated follow-up investigations on the Antlia remnant and G249.2+24.8, including both wide FOV optical imaging and spectra.

\subsection{Optical images}

Several different sources of narrow passband optical images were used in our investigation of the three SNR candidates.
Table 2 lists details of these optical images.

Wide-field, low-resolution H$\alpha$ images of regions around the suspected SNRs G249+24 and Antlia
obtained as part of the MDW Hydrogen-Alpha Survey\footnote{https://www.mdwskysurvey.org} were used to provide general optical size and emission features.
This survey uses a 130 mm f/2 camera at the New Mexico Skies Observatory, with a FLI ProLine 16803 CCD and 3 nm filter centered on H$\alpha$.  This telescope-camera system produced field-of-view (FOV) of approximately $3.5\degr \times3.5\degr$ with a pixel size of $3\farcs17$. Each field was observed 12 times each with an exposure of 1200 s. 

We have also made large mosaics from Southern H$\alpha$ Sky Survey Atlas (SHASSA; \citealt{Gaustad2001}). This wide-angle robotic survey covered $\delta$ = $+15\degr$ to $-90\degr$
with $13\degr$ square images with an angular resolution of $\simeq0.8'$. Narrow passband continuum filters on blue and red sides of H$\alpha$ centered at 6440 and 6770 \AA \ allow for stellar and background subtraction. The survey had a native sensitive level of $1.2 \times 10^{-17}$
erg cm$^{-2}$ s$^{-1}$ arcsec$^{-2}$ which could be significantly lowered with smoothing.

The western region of the Antlia remnant where it abuts the northeastern limb of the Gum Nebula was imaged using  H$\alpha$ + [\ion{N}{2}],
[\ion{O}{3}] and [\ion{S}{2}] filters using a 200 mm f/2 lens and a $9576 \times 6388$ pixel ZWO ASI6200mm Pro
CCD. This system provided a FOV of $10.3\degr \times 6.9\degr$  with a $3.87''$ pixel$^{-1}$ image scale. Total exposure times for the three filters listed above were 6.9 hr, 13.3 hr, and 6.8 hr respectively.

Higher resolution images of the the northwestern limb of the suspected G354-33 remnant were obtained with a 0.5 m Newtonian telescope at the CHILESCOPE Observatory located 25 km south of the Gemini South telescope in
the Chilean Andes. A $4096 \times 4096$ FLI PROLINE 16803 CCD camera was used that provided a 
FOV of $1.1\degr \times 1.1\degr$ with a 
a pixel scale of 0\farcs963 pixel$^{-1}$. Several 10 minutes exposures were taken with H$\alpha$ + [\ion{N}{2}] 
and [\ion{O}{3}] filters.

Images of optical filaments detected in MDW images associated with GALEX FUV filaments in G249+24 were obtained with a 2.4 m telescope at the MDM Observatory at Kitt Peak, Arizona using 
using the Ohio State Multi-Object Spectrograph (OSMOS; \citealt{Martini2011}) in direct imaging mode. With a $4096 \times 4096$ CCD, this telescope/camera system yielded a clear FOV of $18' \times 18'$. With on-chip $2 \times 2$ 
or $4 \times 4$ pixel binning, spatial resolution was $0.55''$ and $1.10''$, respectively. Three exposures of 600 to 1200 s each using an H$\alpha$ + [\ion{N}{2}] filter were taken at four positions where long slit OSMOS spectra were also obtained (see below).

%Finally, a few wide-field, narrow passband H$\alpha$, [\ion{O}{3}], and [\ion{S}{2}]
%images were obtained using a 200mm f/2 Nikon lens attached to a $9576 \times 6388$ pixel
%ZWO ASI6200mm camera. This system produced 15.5\degr diagonal FOV images with
%scale of $3\farc87$ pixel$^{-1}$.

%%%%%%%%%%%%%%%%%%%%%%%%%%%%%%%%%%%%%%%%%%%%%%%%%%%%%%%%%%%%%%%%%%%
%%% Figure 2: G1 FUV 
%%%%%%%
\begin{figure*}[htp]
\centerline{\includegraphics[angle=0,width=19.0cm]{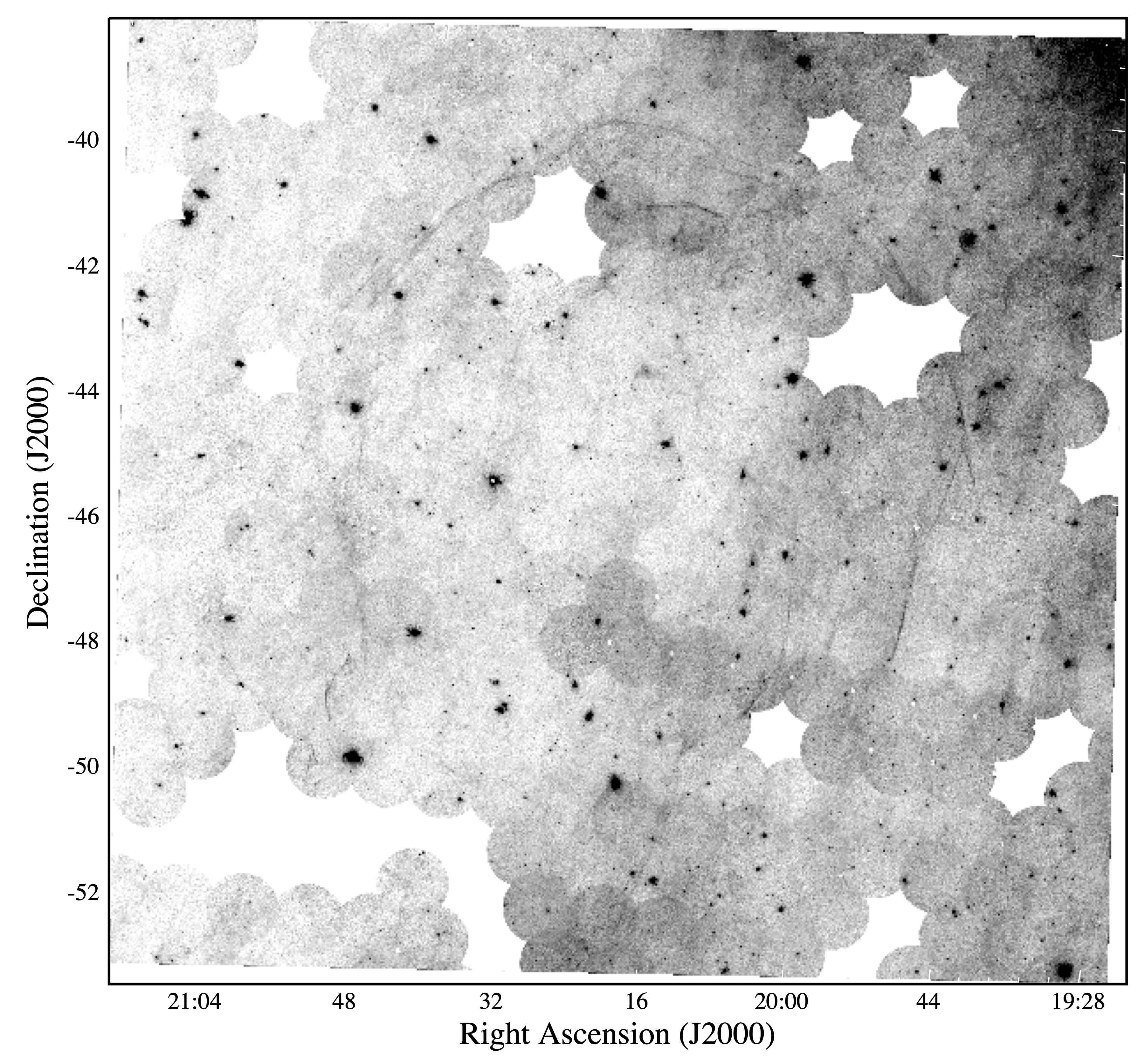}}
\caption{GALEX FUV intensity map of G354-33 showing a roughly spherical shell of UV emission filaments. Approximate shell center: 20$^{\rm h}$ 16.0$^{\rm m}$ 36$^{\rm s}$ $-45$\degr 40$'$.  Note: Individual circular GALEX 
images are $1.2\degr$ in diameter. \label{G1_FUV}
}
\end{figure*}
%%%%%%%%%%%%%%%%%%%%%%%%%%%%%%%%%%%%%%%%%%%%%%%%%%%%%%%%%%%%%%%%%%%%%%

%%%%%%%%%%%%%%%%%%%%%%%%%%%%%%%%%%%%%%%%%%%%%%%%%%%%%%%%%%%%%%%%%%
%%% Figure 3: G1 FUV  and Halpha
%%%%%%%
\begin{figure*}
\begin{center}
\includegraphics[angle=0,width=0.49\textwidth]{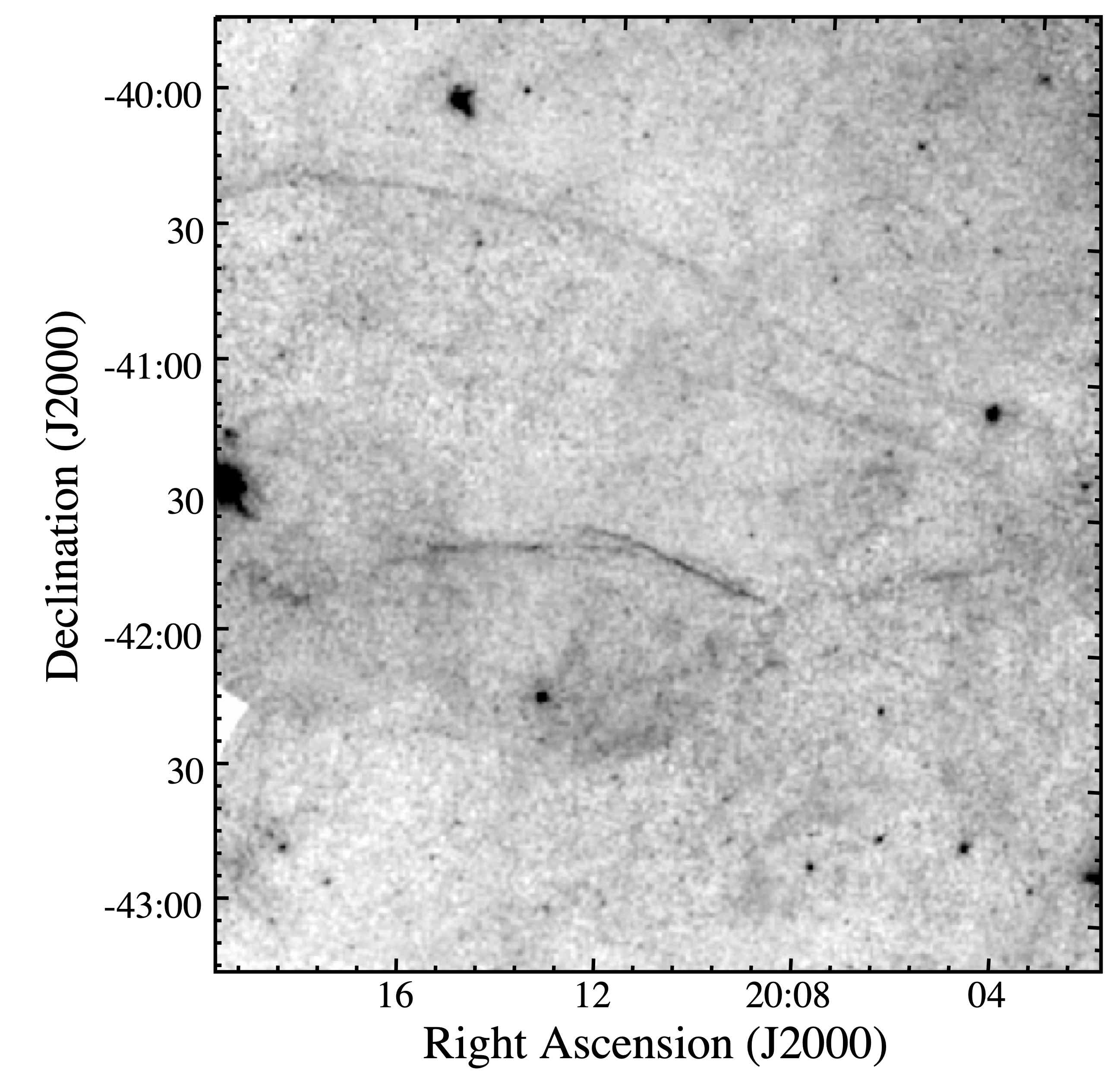}
\includegraphics[angle=0,width=0.49\textwidth]{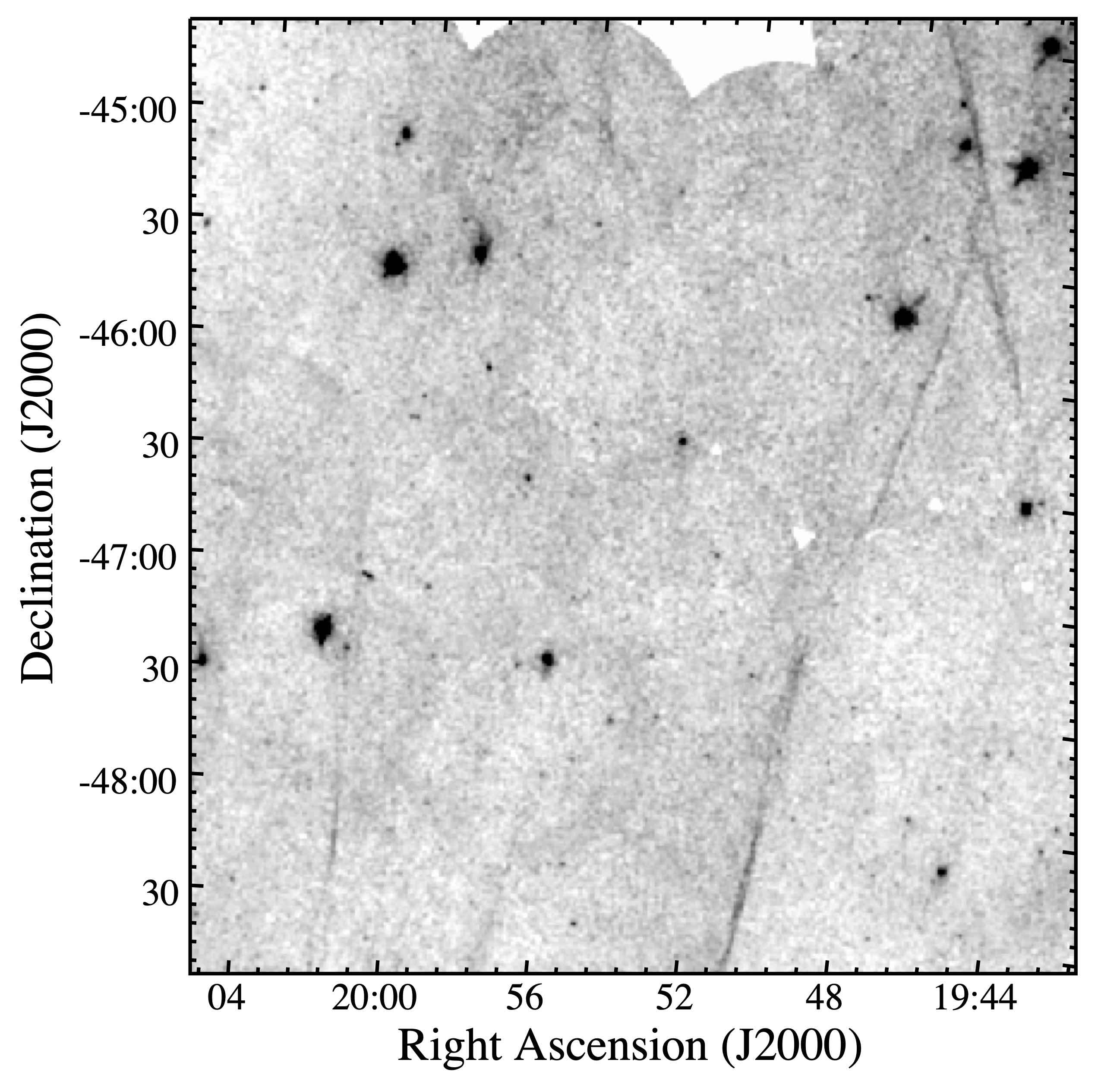}
\includegraphics[angle=0,width=0.49\textwidth]{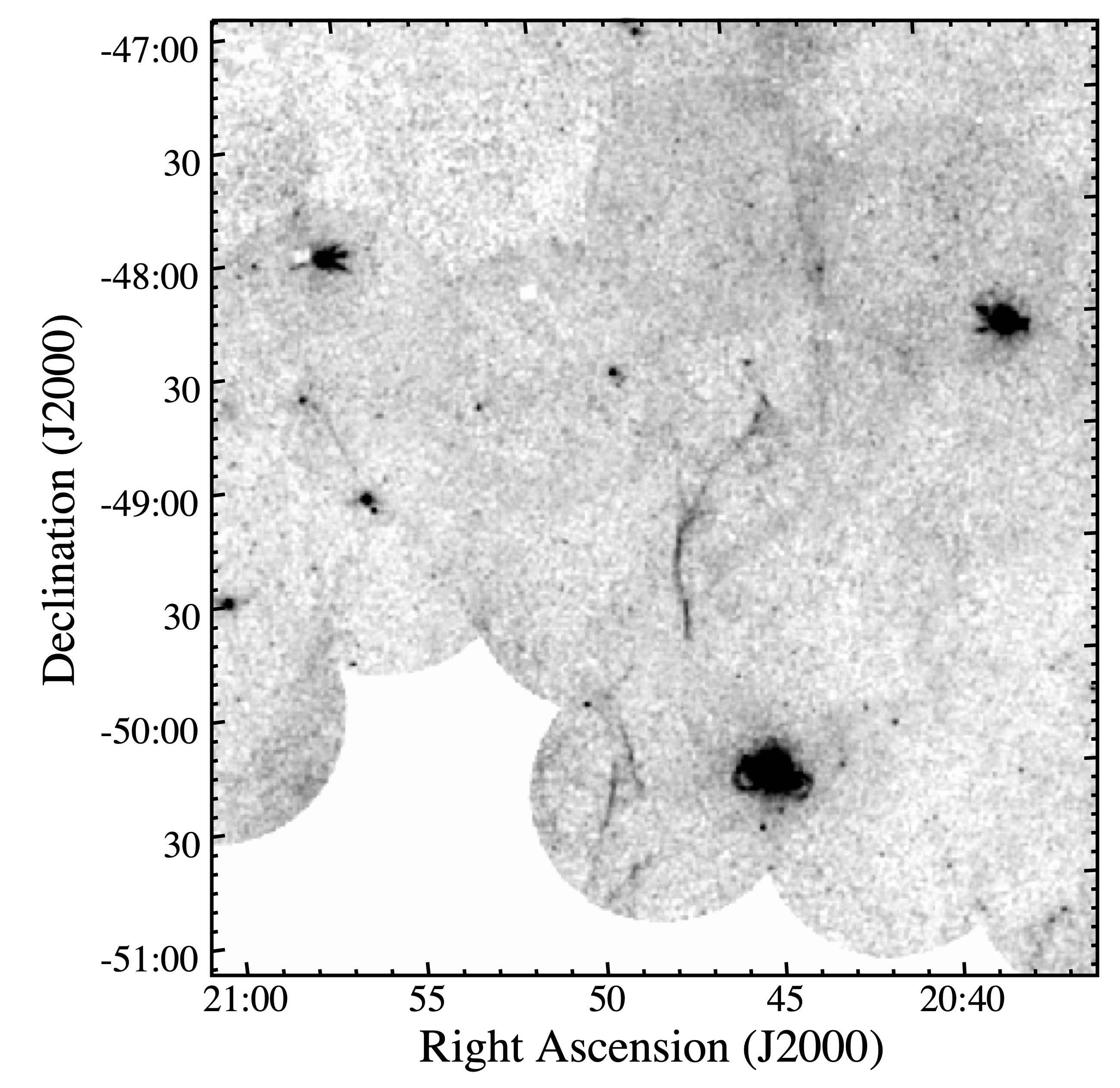}
\includegraphics[angle=0,width=0.49\textwidth]{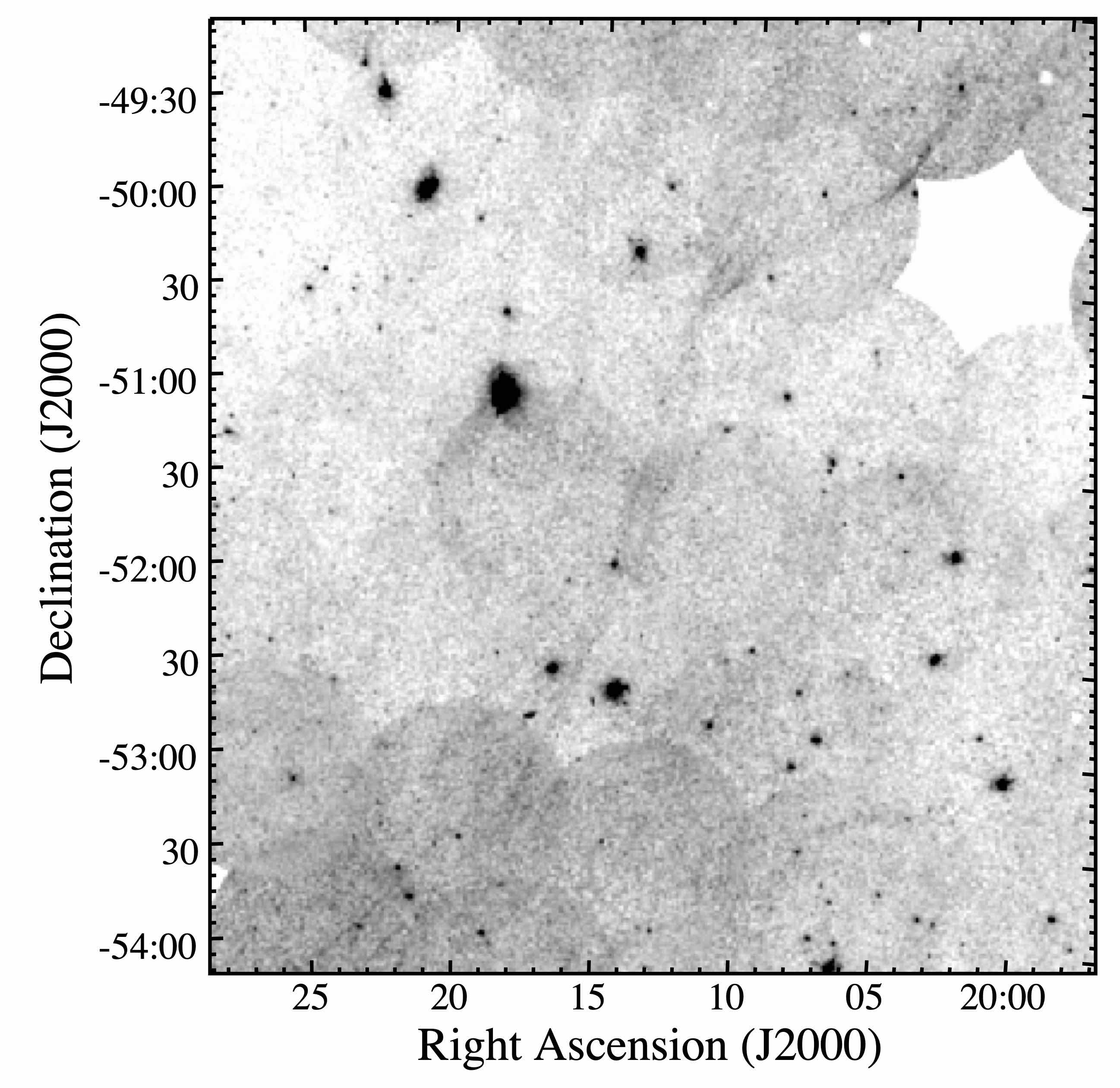}
\caption{Blow-ups of the GALEX FUV mosaic of G354-33 along the remnant's northern
(upper left), western (upper right), southeastern (lower left) and southwestern (lower right) limbs. 
\label{G1_blow-ups}
}
\end{center}
\end{figure*}
%%%%%%%%%%%%%%%%%%%%%%%%%%%%%%%%%%%%%%%%%%%%%%%%%%%%%%%%%%%%%%%%%%

\subsection{Optical Spectra}

Low-dispersion optical spectra of filaments in the northern suspected SNR, G249+24, were obtained with the MDM 2.4m Hiltner telescope using Ohio State Multi-Object Spectrograph (OSMOS; \citealt{Martini2011}). Employing a blue VPH grism (R = 1600) and a 1.2 arcsec wide slit,  exposures of $2 \times 900$ s were taken covering 4000--6900 \AA \ with a spectral resolution of 1.2 \AA \ pixel$^{-1}$ and a FWHM = 3.5 \AA.  Spectra were extracted from regions clean of appreciable emission along each of the $15'$ long slits. 

As part of a program looking for high-velocity interstellar \ion{Na}{1} absorption features in background stars in the direction of Galactic SNRs, moderate-dispersion spectra of over a dozen early-type stars
with projected locations toward G249+24 and with estimated Gaia distances
between 300 and 2000 pc were also observed on with the MDM 2.4m Hiltner 
The data were taken using a Boller \& Chivens
Spectrograph (CCDS) which employs a Loral 1200 $\times$ 800 CCD detector. Using a 1800 grooves mm$^{-1}$ grating blazed at 4700 \AA \ yielded a wavelength coverage of 330 \AA \ with a spectral scale of 0.275 \AA \ per pixel. Exposure times varied from $1-2 \times 400$ s to $3 \times 1200$ s depending on star brightness and sky conditions.

A $1.3''$ wide slit was used which resulted in a measured FWHM of 2.9 pixels, corresponding to a spectral resolution of 0.80 \AA \ providing an effective R $\simeq$7400 at the \ion{Na}{1} $\lambda\lambda$5890,5896 lines. 
Although this spectral resolution is low
compared to conventional interstellar absorption studies where R values typically equal or exceed 30~000, this spectrograph provided sufficient velocity resolution at the \ion{Na}{1}  lines for detecting $\geq$ 40 km s$^{-1}$ interstellar absorption features seen in SNRs \citep{Jenkins1976J,Danks1995,Welsh2003,Sallmen2004,Fesen2018,Ritchely2020}.
Wavelength calibration was done using Hg and Ne comparison lamps and  have  been  corrected  to local standard of rest (LSR) values. Deblended absorption feature velocities are believed accurate to $\pm$4 km s$^{-1}$.

For southern hemisphere Antlia nebula, spectra were taken of suspected SNR filaments using the Robert Stobie Spectrograph (RSS) on the 10~m SALT telescope in South Africa. Using a 900 lines per mm grating and a 1.5$''$ wide slit, spectra were obtained covering 4050 to 7120 \AA \ region with a  FWHM resolution of 5 \AA \  and a dispersion of 1.0 Å pix$^{-1}$. Exposures ranged from $2 \times 300$s to $2 \times 600$s.

 MDM spectra were reduced using using Pyraf software and OSMOS reduction pipelines\footnote{https://github.com/jrthorstensen/thorosmos} in Astropy. and PYRAF\footnote{PYRAF is a product of the Space Telescope Science Institute, which is operated by AURA for NASA.}. L.A. Cosmic \citep{vanDokkum2001} was used to remove cosmic rays and calibrated using a HgNe or Ar lamp and spectroscopic standard stars \citep{Oke1974,Massey1990}. SALT spectra were reduced using specific SALT reduction software.

%%%%%%%%%%%%%%%%%%% RESULTS  %%%%%%%%%%%%%%%%%%%%%%%%%%%%%%%%%%%%% 
\section{Results}

A few large and unusually high Galactic latitude nebulae have been recently proposed as previously unrecognized SNRs. These objects have angular diameters well in excess of the largest confirmed SNRs, such as the 8\degr \ diameter Vela SNR. If these objects were to be found to be true SNRs, they would expand the notion of how large an identifiable Galactic remnant can be and how far off the Galactic plane should SNR researchers be looking.

The proposed new SNRs include
the huge $\sim$24\degr\ diameter Antlia nebula (G275.5+18.4; \citealt{McCull2002}) and the $\sim$10\degr\ diameter radio remnant G354-33  \citep{Testori2008}. Due to limited data on it, Antlia is listed only as a possible SNR in the most recent catalogue of Galactic supernova remnants \citep{Green2019}, whereas no mention is made of G354-33 in either this 2019 SNR catalogue or in online SNR lists\footnote{http://www.mrao.cam.ac.uk/surveys/snrs/snrs.info.html 
http://snrcat.physics.umanitoba.ca/index.php?}.
Below we discuss GALEX far UV (FUV) and optical properties of these two suspected SNRs, plus a third object, G249.2+24.4, which appears to be a new
SNR based both on our FUV and optical data, and X-ray and
radio data presented by \citet{Becker2021}. The presentation
order follows that of our research work.

%which we believe is a new SNR. The presentation order follows that of our research work.

%------------------------------------------------------------------

%%%%%%%%%%%%%%%%%%%%%%%%%%%%%%%%%%%%%%%%%%%%%%%%%%%%%%%%%%%%%%%%%%%%
%%% Figure 4: G1 Halpha
\begin{figure*}
\begin{center}
\centerline{\includegraphics[width=18.0cm]{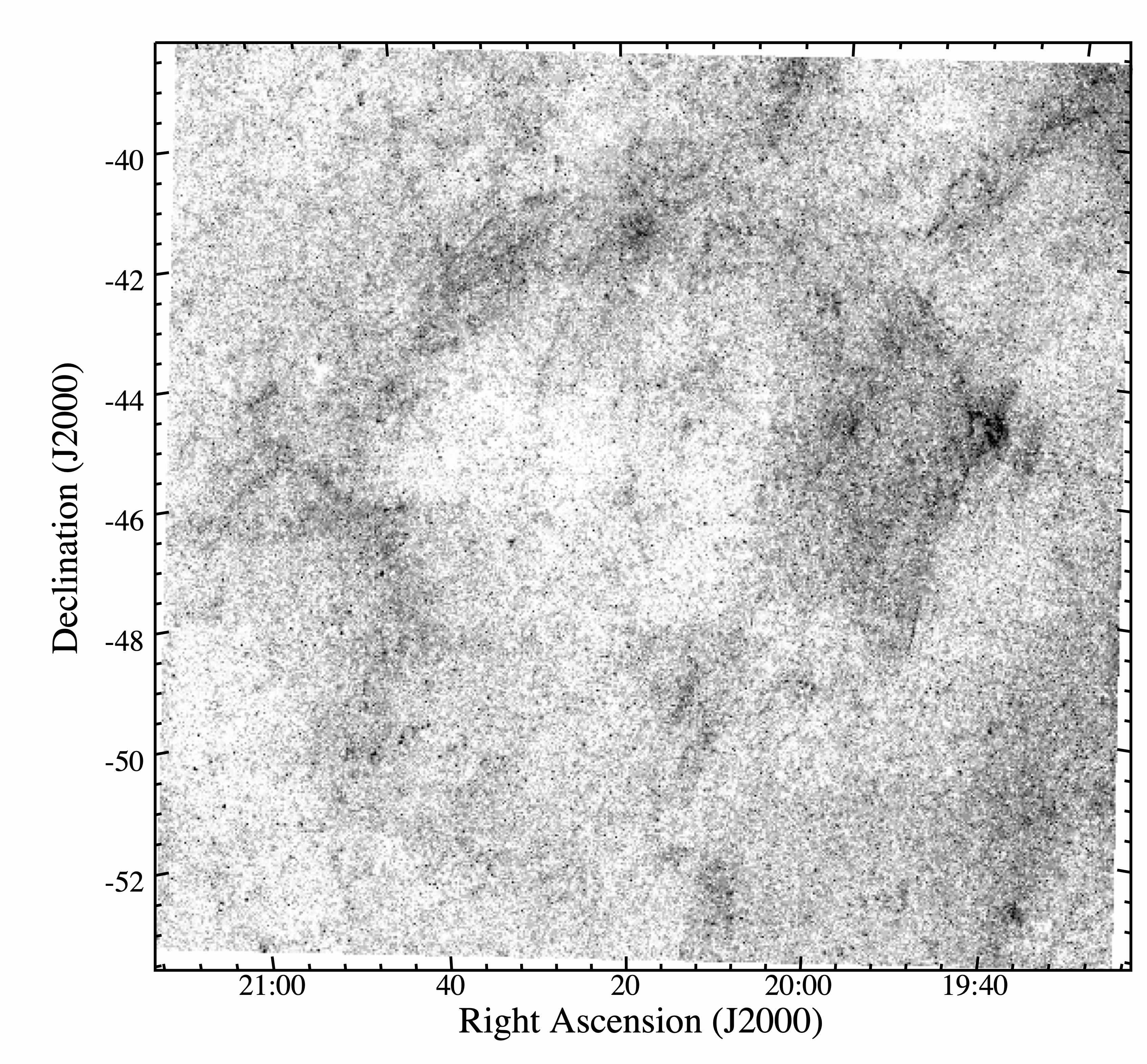} }
\caption{Continuum subtracted SHASSA H$\alpha$ image of G354-33.
\label{G1_Ha}
}
\end{center}
\end{figure*}
%%%%%%%%%%%%%%%%%%%%%%%%%%%%%%%%%%%%%%%%%%%%%%%%%%%%%%%%%%%%%%%%%%

%------------------------------------------------------------------
\subsection{The Suspected Radio SNR G354.0-33.5}

Using a 1420 MHz linear polarization survey of the southern hemisphere sky, \citet{Testori2008} identified a large $\sim$10\degr\ depolarized shell at $\alpha$(J2000) = 20$^{\rm h}$ 25$^{\rm m}$, $\delta$(J2000) =  $-47\degr$,  roughly corresponding to Galactic coordinates $l$ = 353\degr \ $b = -34\degr$. This shell could also be seen in a 1.4 GHz polarization all-sky survey map \citep{Reich2009}.
Due to the object's spherical morphology and radio properties, \citet{Testori2008} suggested a SNR identification with a size suggesting a distance between 300 and 500 pc, a physical diameter of 17.4 pc $\times$ d$_{\rm 100 pc}$
and a z distance of 57.4 pc  $\times$d$_{\rm 100 pc}$. \citet{Testori2008} noted that the object could be seen in 408 MHz data  \citep{Haslam1982} and weakly in the SHASSA H$\alpha$ images \citep{Gaustad2001}.  This object also shows up in gradients of linearly polarized synchrotron radio emission \citep{Iacob2014}.
More recently, \citet{Bracco2020} briefly commented on the presence of some GALEX FUV filaments coincident with this radio shell. 

However, to our knowledge, there is no work showing this suspected SNR's FUV emissions or any associated optical emissions. Below we present large mosaics of GALEX FUV images along with a SHASSA H$\alpha$ image and compare these to the low resolution 1420 MHz radio images.
%%%%%%%%%%%%%%%%%%%%%%%%%%%%%%%%%%%%%%%%%%%%%%%%%%%%%%%%%%%%%%%%%%
%%% Figure 5: G1 FUV vs Halpha
%%%%%%%
\begin{figure*}
\begin{center}
%\centerline{\includegraphics[angle=0,width=13.0cm]{FUV_vs_Ha_west_v2.jpg}}
%\centerline{\includegraphics[angle=0,width=13.0cm]{FUV_vs_Ha_NE_v3.jpg}}
\includegraphics[angle=0,width=0.49\textwidth]{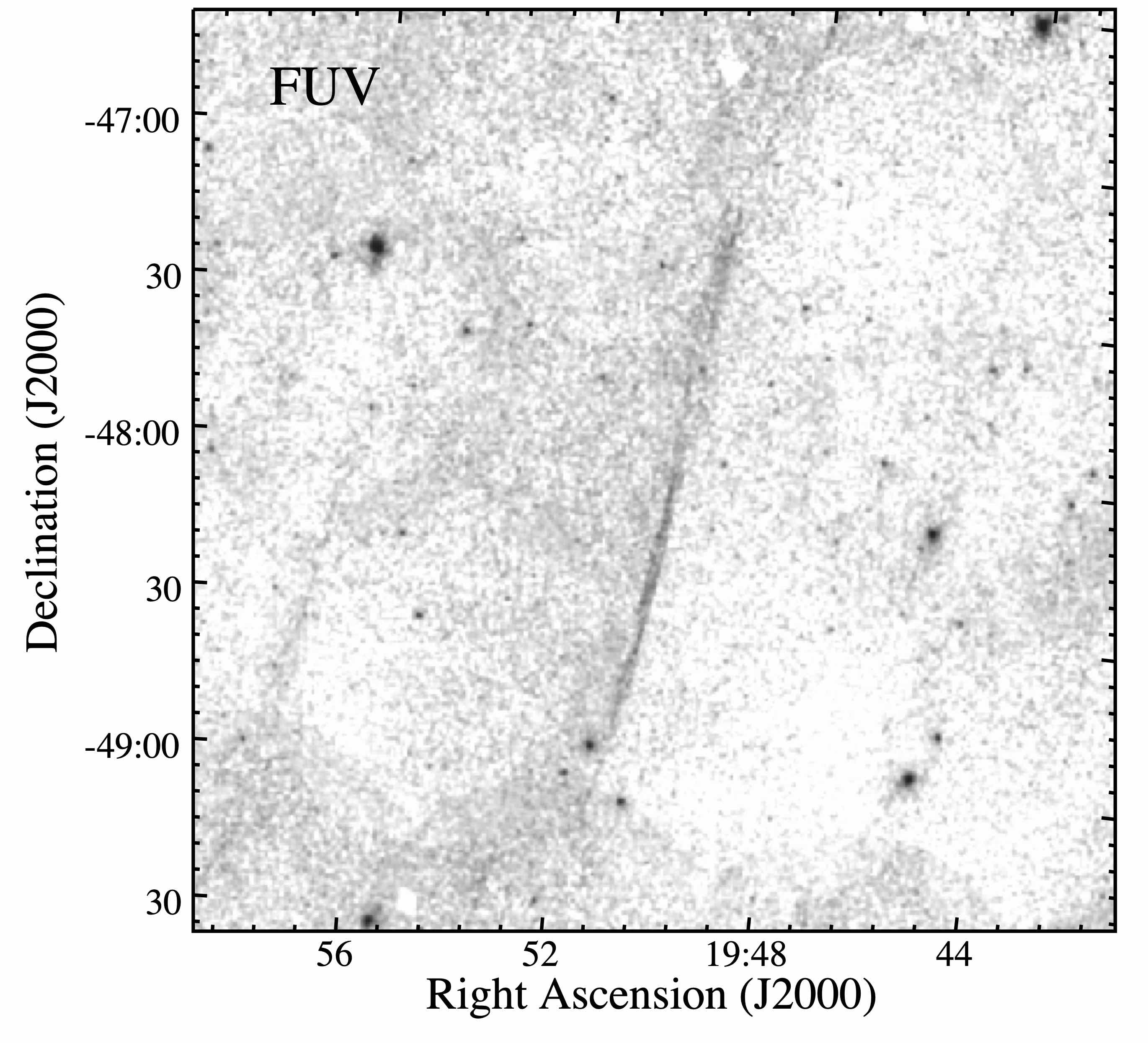}
\includegraphics[angle=0,width=0.49\textwidth]{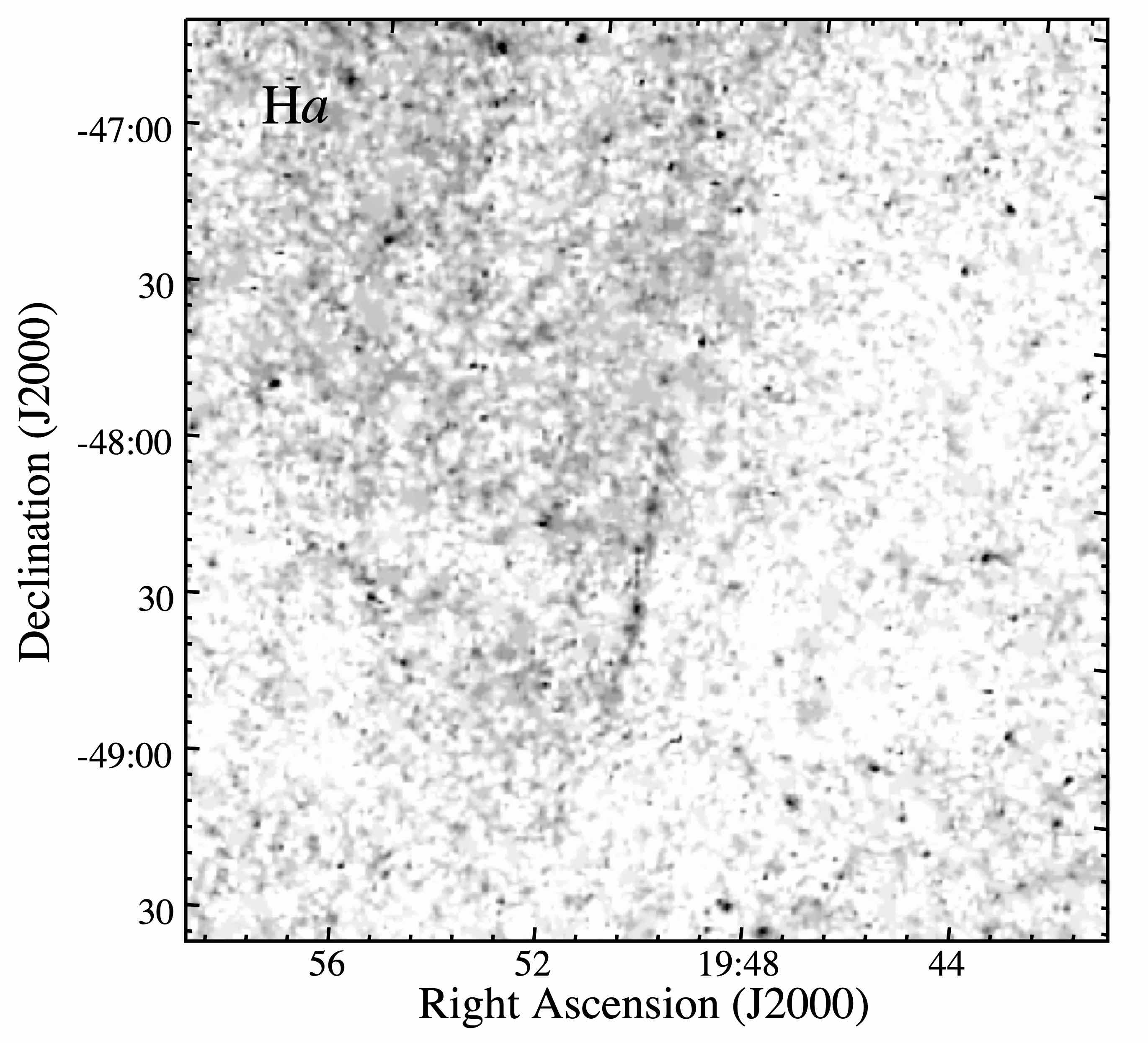}
\includegraphics[angle=0,width=0.49\textwidth]{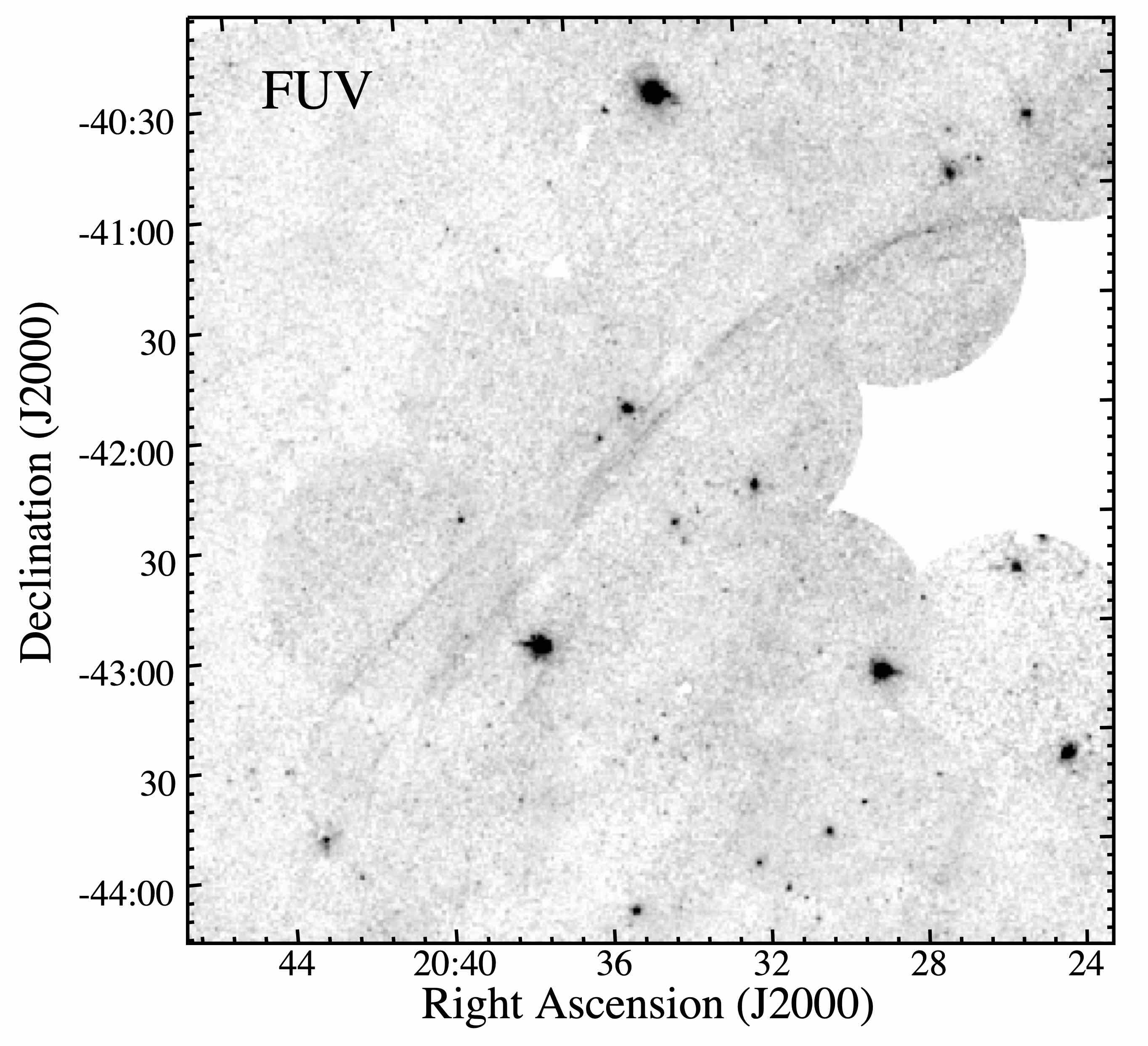}
\includegraphics[angle=0,width=0.49\textwidth]{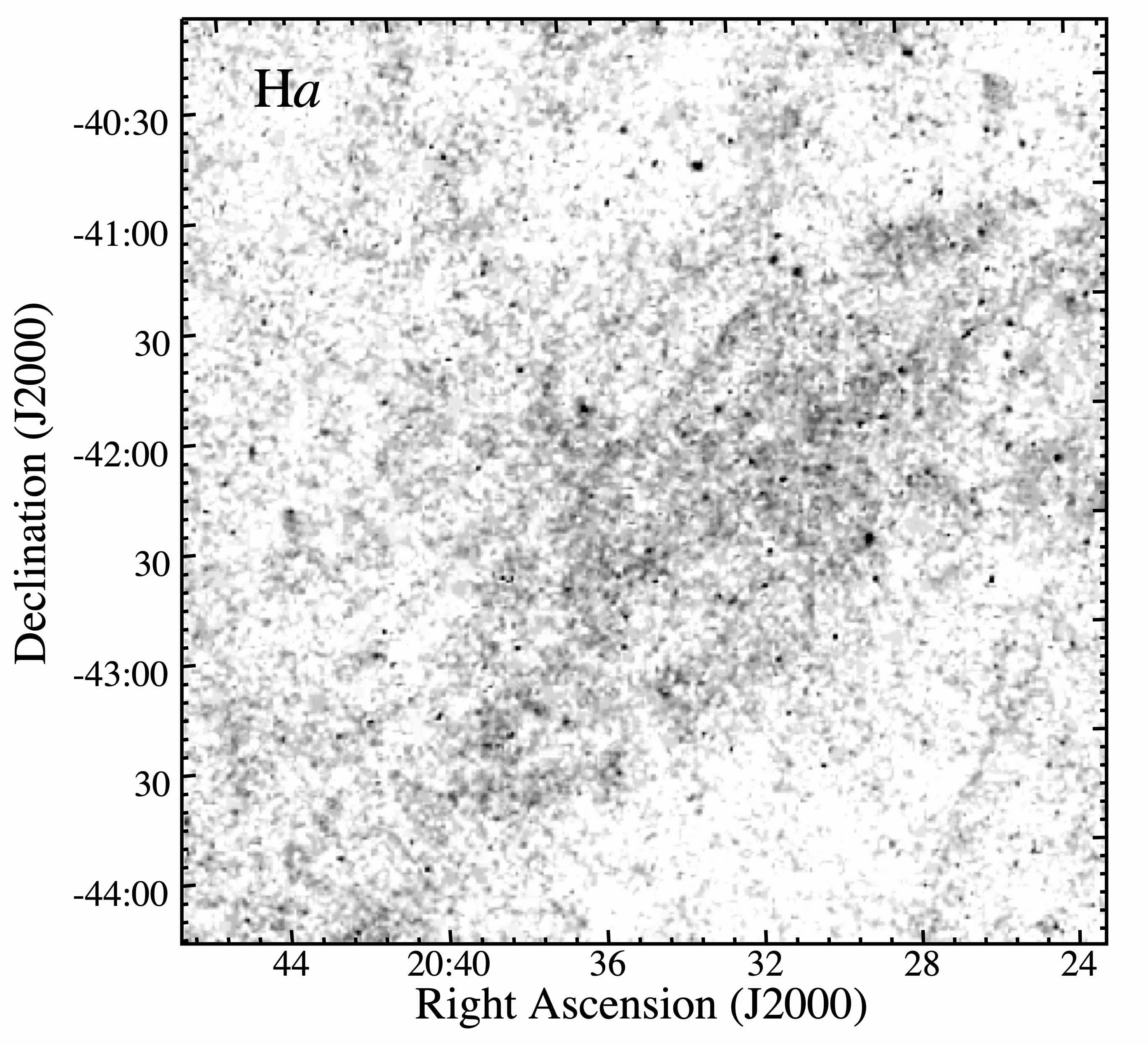}
\caption{Matched GALEX FUV and continuum subtracted SHASSA H$\alpha$ images for a bright filament along G354-33's western limb (upper panels) and northeastern limb (lower panels). 
\label{G1_FUV_vs_Ha}
}
\end{center}
\end{figure*}
%%%%%%%%%%%%%%%%%%%%%%%%%%%%%%%%%%%%%%%%%%%%%%%%%%%%%%%%%%%%%%%%%%%
\subsubsection{Far UV and H$\alpha$ Images}

Figure~\ref{G1_FUV} shows a wide-field mosaic of GALEX FUV images of G354-34 smoothed by a nine point Gaussian. The image mosaic
reveals a large, coherent set of sharp UV emitting filaments arranged in a fairly circular and complete shell-like structure which is 
positionally coincident with the 1420 MHz radio shell.\footnote{Because of distortions due to the large size of this emission structure, coordinates shown are only accurate to a few arc minutes.} 

Multiple and overlapping filaments are common along the shell's northern and western limbs.
Because these filaments appear brightest in the GALEX FUV images compared to NUV images, we will concentrate mainly on the shell's far UV emission structure.

Although many of the FUV filaments are unresolved at GALEX's 4.6$''$ FWHM resolution, some filaments appear partially resolved in places, especially along its northernmost edge. It is not clear if this appearance is due to closely spaced multiple filaments or resolution of a single emission filament.

The strong shock-like morphology of these filaments and their arrangement in a nearly continuous shell lends support to  \citet{Testori2008}'s suggestion for it to be a previously unrecognized Galactic supernova remnant, albeit with unusually large angular dimensions and distance from the Galactic plane.

%%%%%%%%%%%%%%%%%%%%%%%%%%%%%%%%%%%%%%%%%%%%%%%%%%%%%%%%%%%%%%%%%%%%
%%% Figure 6: G1 FUV  and Halpha
\begin{figure*}
\begin{center}
\includegraphics[width=15.0cm]{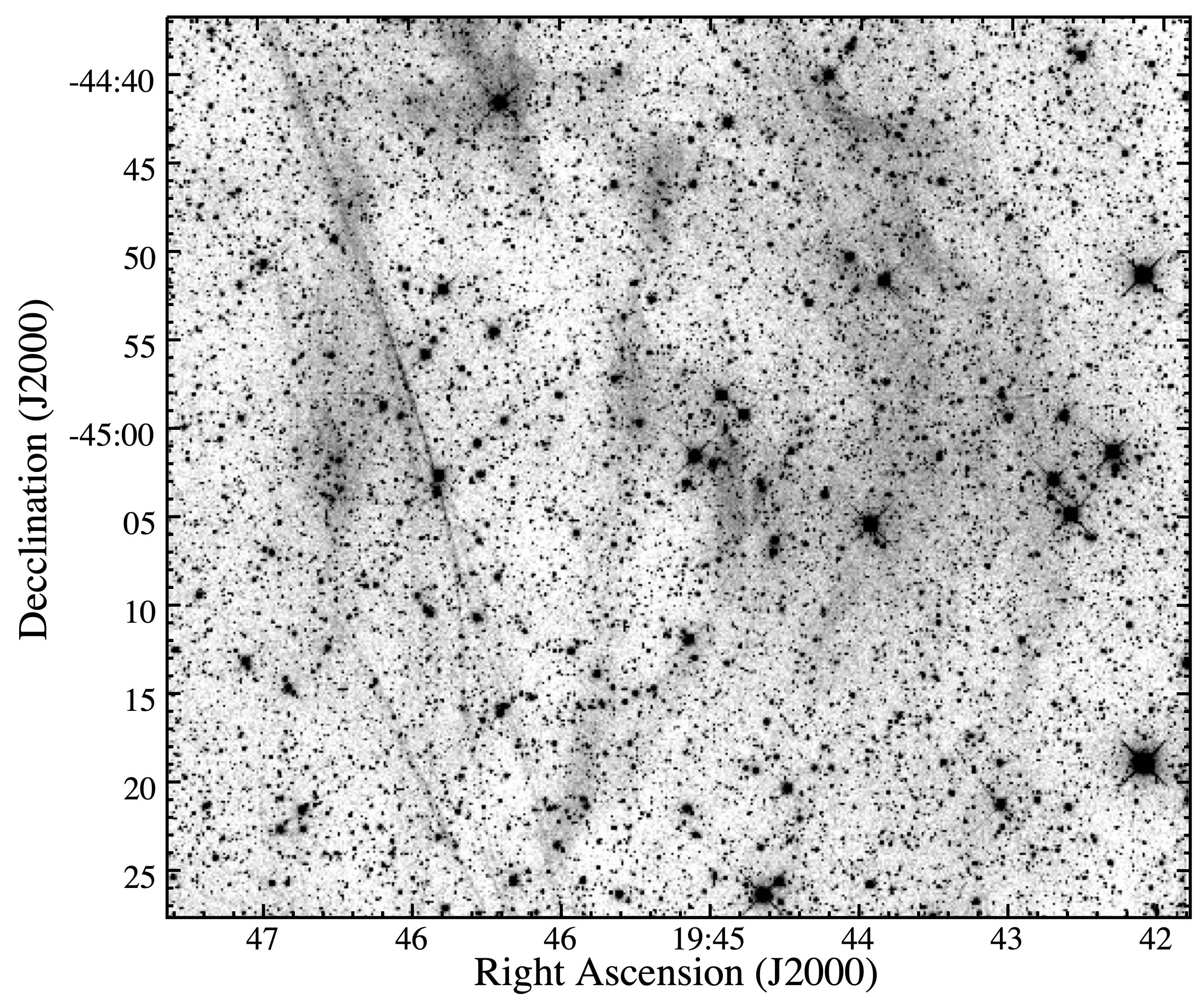} 
\includegraphics[angle=0,width=17.0cm]{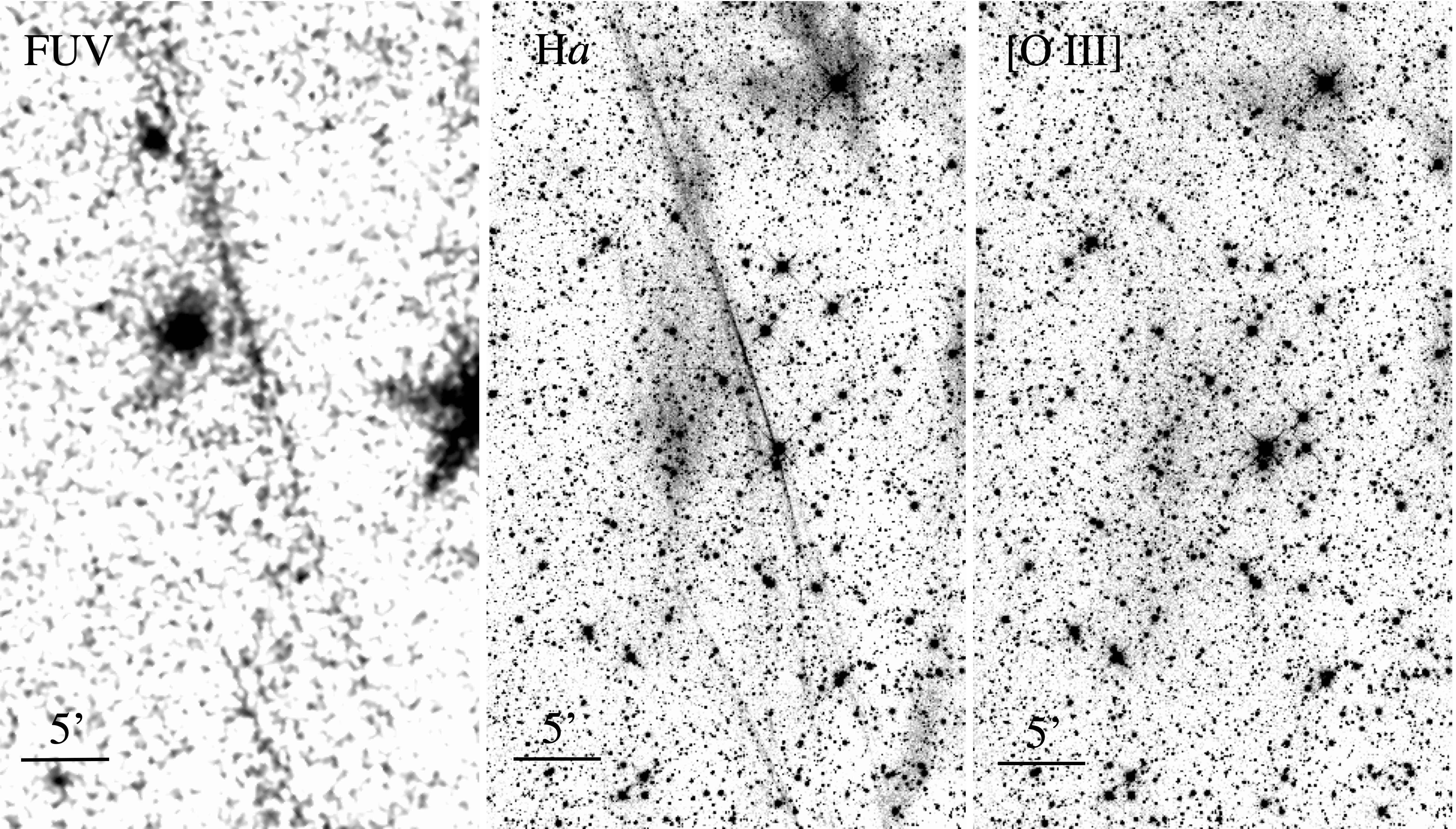}
\caption{{\it{Top:}} CHILESCOPE H$\alpha$ + [\ion{N}{2}] image of G354-33's northwest limb revealing emission structure details lost in the low-resolution SHASSA image shown in Fig.\ 4. {\it{Bottom:}} Matched GALEX FUV and Chilescope H$\alpha$ and [\ion{O}{3}] images for the thin filaments seen in this NW region. The thin filament is clearly seen in the FUV and H$\alpha$ images but not in [\ion{O}{3}].
\label{G1_Chile}
}
\end{center}
\end{figure*}
%%%%%%%%%%%%%%%%%%%%%%%%%%%%%%%%%%%%%%%%%%%%%%%%%%%%%%%%%%%%%%%%%%

Based on its FUV filaments, we estimate somewhat larger angular dimensions than the 10\degr \ cited by \citet{Testori2008}. Instead, we find dimensions of $\simeq$11\degr \ East-West and $\simeq$14\degr \ North-South, with a distinct indentation along its south-western limb. Its southern extent is poorly constrained in our GALEX FUV mosaic but FUV filaments extend south to at least a Declination of $-53.5\degr$. 
While largely coincident in location with the 1420 MHZ radio emission map of \citet{Testori2008}, the  $\sim$11\degr \ East-West diameter seen in the FUV images is centered around 20$^{\rm h}$ 16$^{\rm m}$, -45\degr\ 50$'$ ($l$ = 354.0\degr,  $b$ = $-33.5$\degr). Hence, we will use the shorten name G354-33 based on the FUV images (see Table 1) which is slightly different from the G353-34 name cited by \citet {Testori2008}.

Due to its large size, many of the filamentary details visible in the full resolution FUV mosaic are lost in Figure~\ref{G1_FUV}. Consequently, we show in Figure~\ref{G1_blow-ups} blow-ups of four regions, along G354-33's northern, western, southeastern, and southwestern rims to give a better sense of the object's fine-scale UV emission structure and morphology.
Note that some individual filaments are two to three degrees in length, equivalent to the diameters of some of the largest confirmed Galactic SNRs.

A continuum subtracted SHASSA H$\alpha$ image mosaic of the G354-33 remnant is shown in
Figure~\ref{G1_Ha}. This image
reveals a thick diffuse shell of emission located within the boundaries of the FUV filaments. The H$\alpha$ emission structure is far more diffuse and broader than that seen in the GALEX FUV image with very few obvious thin filaments. We estimate an H$\alpha$ flux of around $1 - 2 \times 10^{-17}$ erg
cm$^{-2}$ s$^{-1}$ arcsec$^{-2}$ for most of the diffuse emission and about 3 times times this for the area of brighter emission along the nebula's northwestern limb.

While its H$\alpha$ emission's location matches that seen in the FUV image, some weak H$\alpha$ emission is seen to extend farther to the east and south than is readily visible in the GALEX FUV mosaic image. In general, however, the FUV filaments appear to mark the outer edges of the diffuse H$\alpha$ emission shell. 

This is shown in Figure~\ref{G1_FUV_vs_Ha} where we compare FUV vs.\ H$\alpha$ images for two northeastern sections along its limb. The upper images show the presence of weak H$\alpha$ emission coincident with the bright main section of the FUV filament, with diffuse H$\alpha$ emission extending toward the east bordered by the long FUV filament. The lower panels show a section along the nebula's northeastern limb where again the UV filaments are seen to mark the extent of the diffuse H$\alpha$ emission. The SHASSA image also gives a hint of filamentary structure that resembles
that seen in the FUV image.

Higher resolution optical images give a better sense of the remnant's true emission structure.
Figure~\ref{G1_Chile} presents a higher resolution CHILISCOPE 
H$\alpha$ + [\ion{N}{2}] image of a small portion of G354-33's
northwestern limb where Figure 4 revealed a particularly bright
emission patch. The upper panel of 
this figure
shows this emission patch to be diffuse and situated mainly west of a $\sim45'$ long, thin H$\alpha$ filament which likely marks the western extent of the remnant in this region.

The three lower panels of Figure~\ref{G1_Chile}
compare the emission appearance of this H$\alpha$ filament in the
GALEX FUV image and in an equally deep [\ion{O}{3}] image. The similarity of the filament as seen in the FUV and H$\alpha$ images suggests G354-33's optical structure
would consist of numerous thin, unresolved filaments like that seen in the FUV mosaic image shown in Figure~\ref{G1_FUV}.

Based on the  filament's absence of 
[\ion{O}{3}] emission, its thin morphology and its brightness in the UV, it is likely to represent a non-radiative or a `Balmer dominated' emission filament. Given the sharpness of the other G354-33 filament's in the FUV, they are also probably Balmer dominated emission filaments marking
the location of the object's shock front and outer boundary.

%%%%%%%%%%%%%%%%%%%%%%%%%%%%%%%%%%%%%%%%%%%%%%%%%%%%%%%%%%%%%%%%%%%
%%% Figure 7: G1  Villa Elisa radio image. 
%%%%%%%
\begin{figure*}
\begin{center}
\includegraphics[angle=0,width=17.50cm]{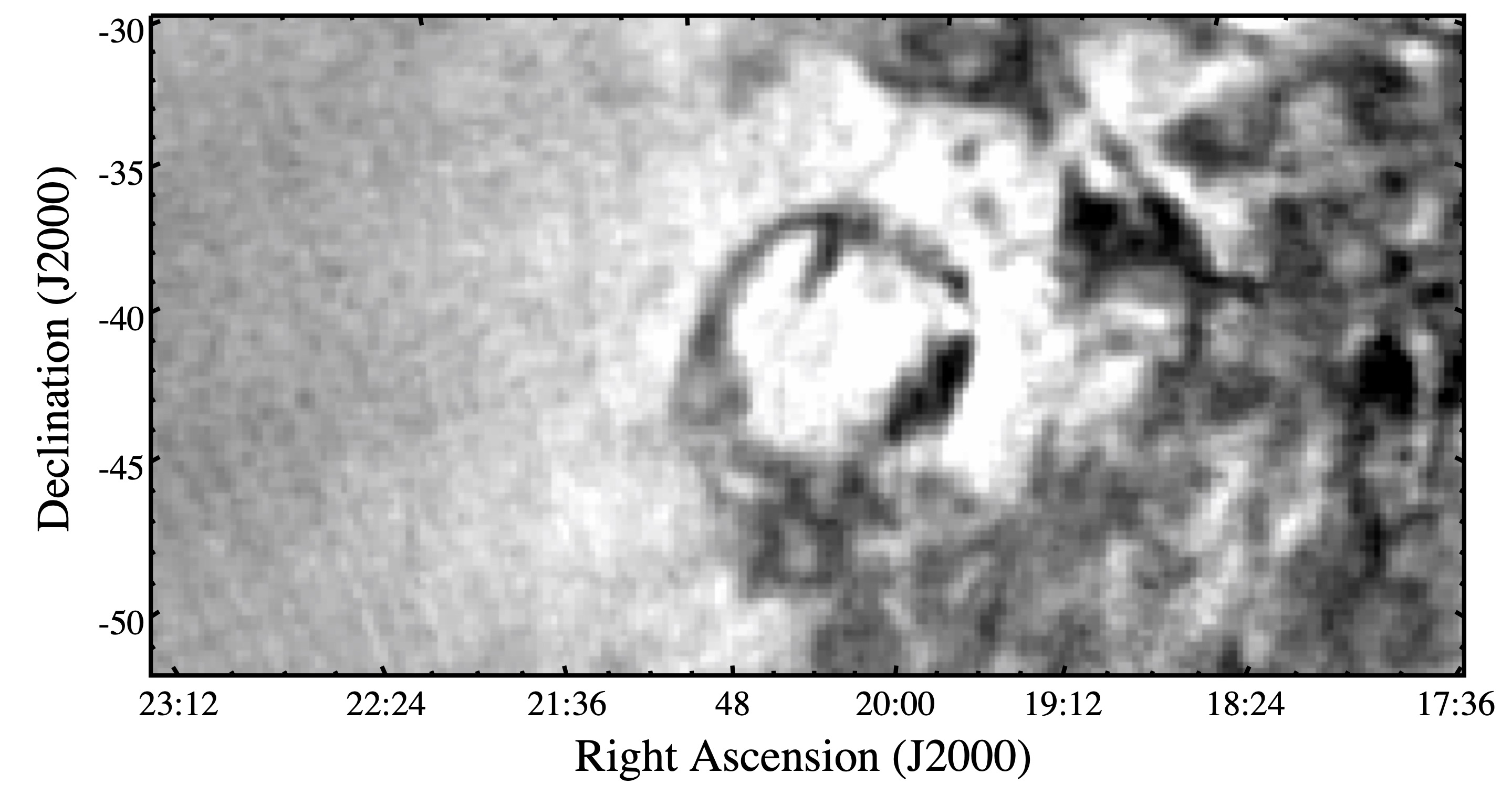}
\includegraphics[width=0.49\textwidth]{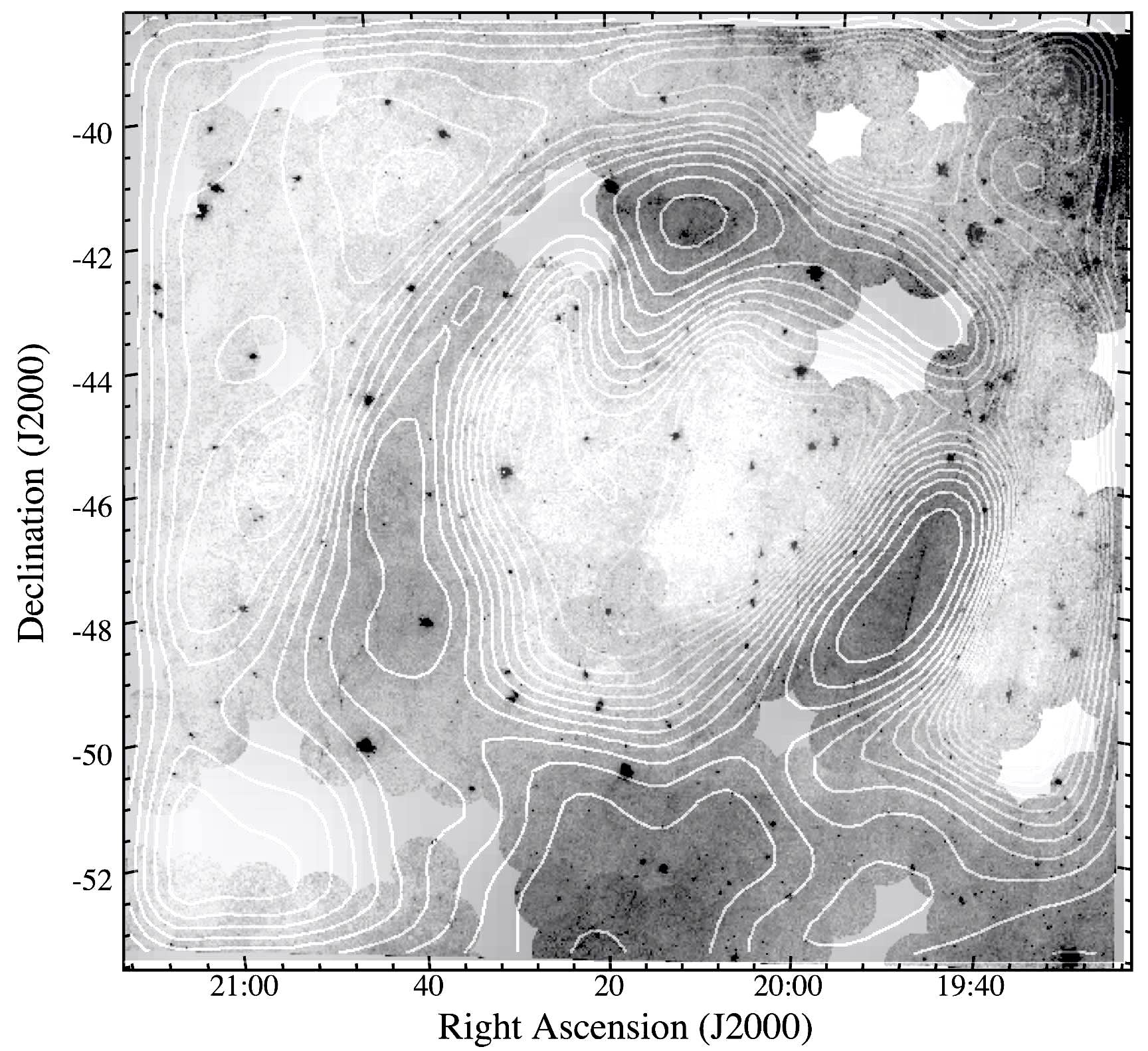}
\includegraphics[width=0.49\textwidth]{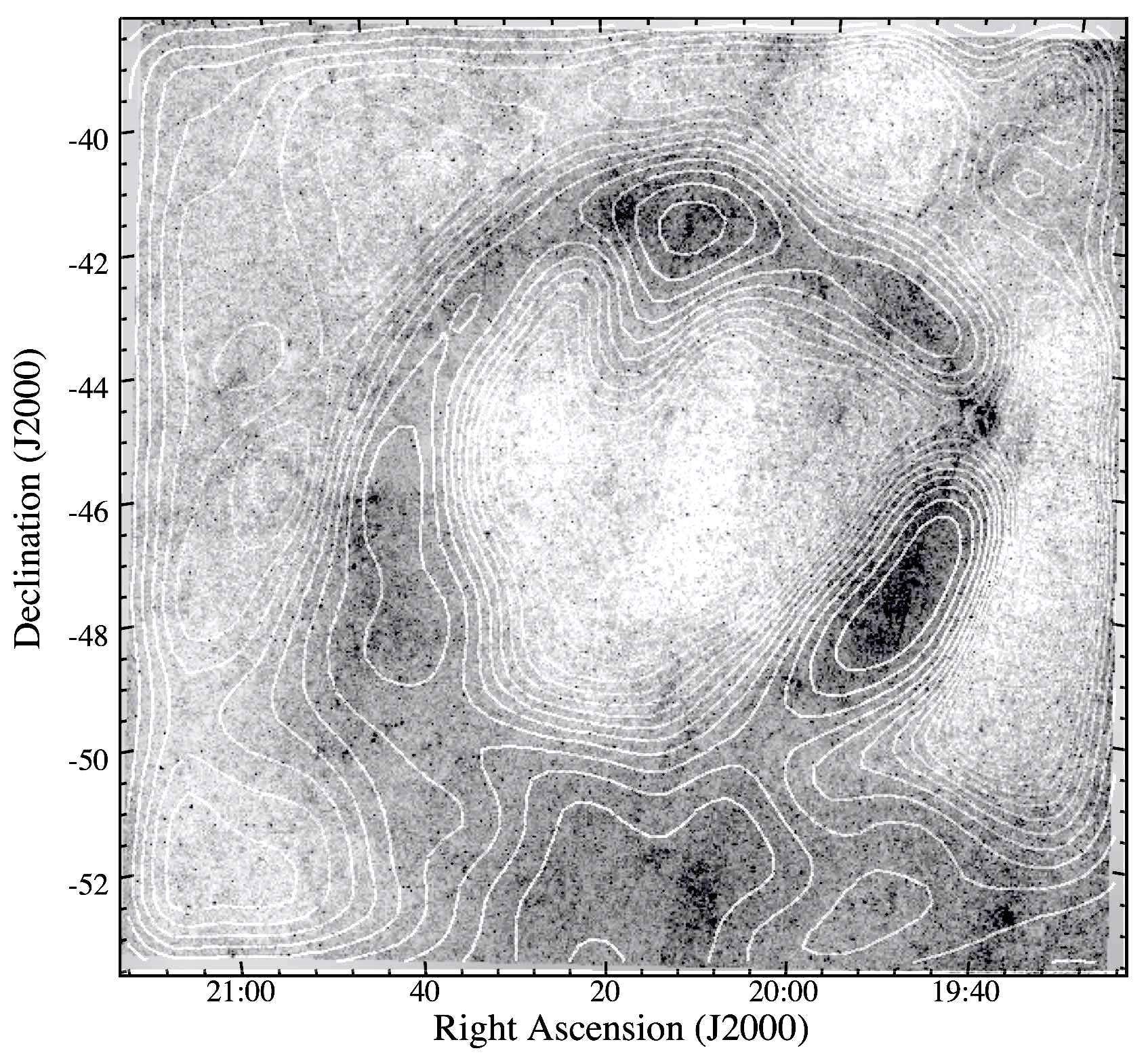}
\caption{$Upper~Panel$ 1420 MHz Villa Elisa Stokes Q polarization intensity image covering a 60\degr $\times$ 25\degr region centered on G354-33. $Lower~ Panels:$ Contour overlays of the
Villa Elisa radio emission with the GALEX FUV (left) and SHASSA H$\alpha$ (right) images 
of the G354-33 region showing excellent size and positional agreement of the 1420 MHz radio emission ring  with the filamentary shell of far UV and H$\alpha$ emission filaments.
\label{Villa}
}
\end{center}
\end{figure*}
%%%%%%%%%%%%%%%%%%%%%%%%%%%%%%%%%%%%%%%%%%%%%%%%%%%%%%%%%%%%%%%%%%

\subsubsection{Radio and X-rays}

As already noted, \citet{Testori2008} used 1420 MHz Villa Elisa radio data \citep{Testori2001} to identify the radio emission ring,
G354-33, as a possible SNR. The upper panel of Figure~\ref{Villa} shows a $60\degr \times 25\degr$ wide section of the Villa Elisa 1420 MHz polarization southern sky survey data. The emission ring appears distinct in shape and separate from other emission features along the southern edge of the Galactic plane (i.e., the right-hand side of the image).

The images in the lower panels of Figure~\ref{Villa} show the  GALEX FUV image mosaic (left panel) and the SHASSA H$\alpha$ image (right panel) overlayed with the 1420 MHz Villa Elisa radio image contours. Overall, there is good agreement in both size and location for the FUV  and H$\alpha$ emission shells and radio ring seen in these images. This includes the greater extent toward the southern limb area. We note that the brightest portion of the radio emission lies on the object's western limb, the side facing the Galactic plane.

Examination of ROSAT on-line data showed no obvious associated X-ray emission with this nebula. This is not surprising given its location 33.5\degr \ below the Galactic plane. Moreover, in view of the lack of any bright FUV emission nebulae along its limbs, this suggests that it lies in a region with few or none  large-scale interstellar clouds which might generate significant X-ray flux.

%%%%%%%%%%%%%%%%%%%%%%%%%%%%%%%%%%%%%%%%%%%%%%%%%%%%%%%%%%%%%%%%%%
%%% Figure 8: G2 FUV
%%%%%%%
\begin{figure*}[htp]
\centerline{\includegraphics[angle=0,width=16.0cm]{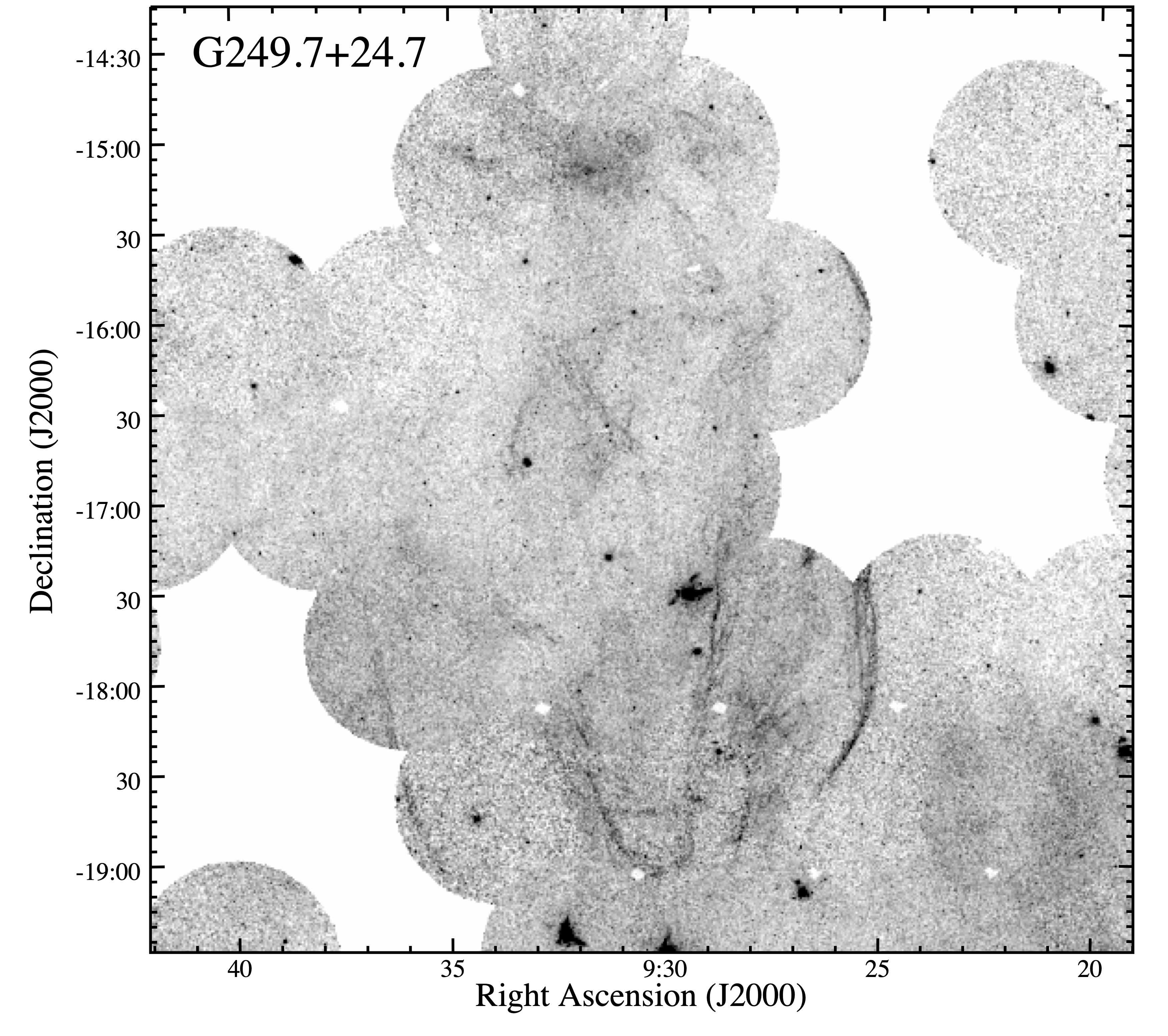}}
\caption{FUV intensity map of G249+24 showing an elliptical shaped shell of UV emission filaments. 
\label{G2_FUV}
}
\end{figure*}
%%%%%%%%%%%%%%%%%%%%%%%%%%%%%%%%%%%%%%%%%%%%%%%%%%%%%%%%%%%%%%%%%%

%------------------------------------------------------------------
%------------------------------------------------------------------
\subsection{SNR G249.7+24.7}

Using GALEX FUV mosaics, we discovered a region containing several long, thin, shock-like morphology filaments aligned mostly N-S and arranged in a roughly elliptical shape  $\sim2.5 \times 4.0\degr$ in size.
These filaments appear centered at $\alpha$(J2000) = 9$^{\rm h}$ 30.5$^{\rm m}$ $\delta$(J2000) = $-16$\degr 40$'$. However, as discussed below, we believe the object to be larger and centered
$\alpha$(J2000) = 9$^{\rm h}$ 33$^{\rm m}$ $\delta$(J2000) = $-17$\degr 00$'$
which corresponds to Galactic coordinates $l$ = 249.7\degr,  $b$ = +24.7\degr. We call this object G249+24 for short. Below, we present images of this object's UV and H$\alpha$ emission structure along with optical spectra of four of its filamentary regions.

\subsubsection{Far UV Mosaic Image}

Figure~\ref{G2_FUV} presents a $5.2\degr \times 5.2\degr$ mosaic of GALEX FUV images centered on G249+24's. The image shows several UV filaments along with a few smaller diffuse emission patches. Although GALEX imaging of this region is fairly incomplete, enough imaging exists to indicate G249+24's minimum size and extent\footnote{The bright feature at
9$^{\rm h}$30$^{\rm m}$, $-17$\degr 30$'$ is HD 82093, an Ap(EuSrCr) star; V = 7.08.}.

The nebula's brightest UV filaments are concentrated in its southern region where there is a bright,  long gently curved filament
along its southwestern edge. This feature itself consists of several separate, closely aligned filaments. Although a smaller but a similarly bright filament caught on the western edge of a more northern GALEX image might give the impression that this filament's full length and extent is partially missing in this mosaic, this is not the case based on H$\alpha$ images (see below).

\subsubsection{H$\alpha$ Emission}

Whereas the object's FUV emission filaments are readily apparent in the GALEX image mosaic shown in Figure~\ref{G2_FUV}, its optical emissions are relatively faint. Although some of its brighter filaments are weakly visible on broad red passband  Digital Sky Survey images, they are so faint and scattered over such a large area that it is not surprising that this nebula had not attracted prior attention.

\begin{figure*}[htp]
\begin{center}
\includegraphics[angle=0,width=8.5cm]{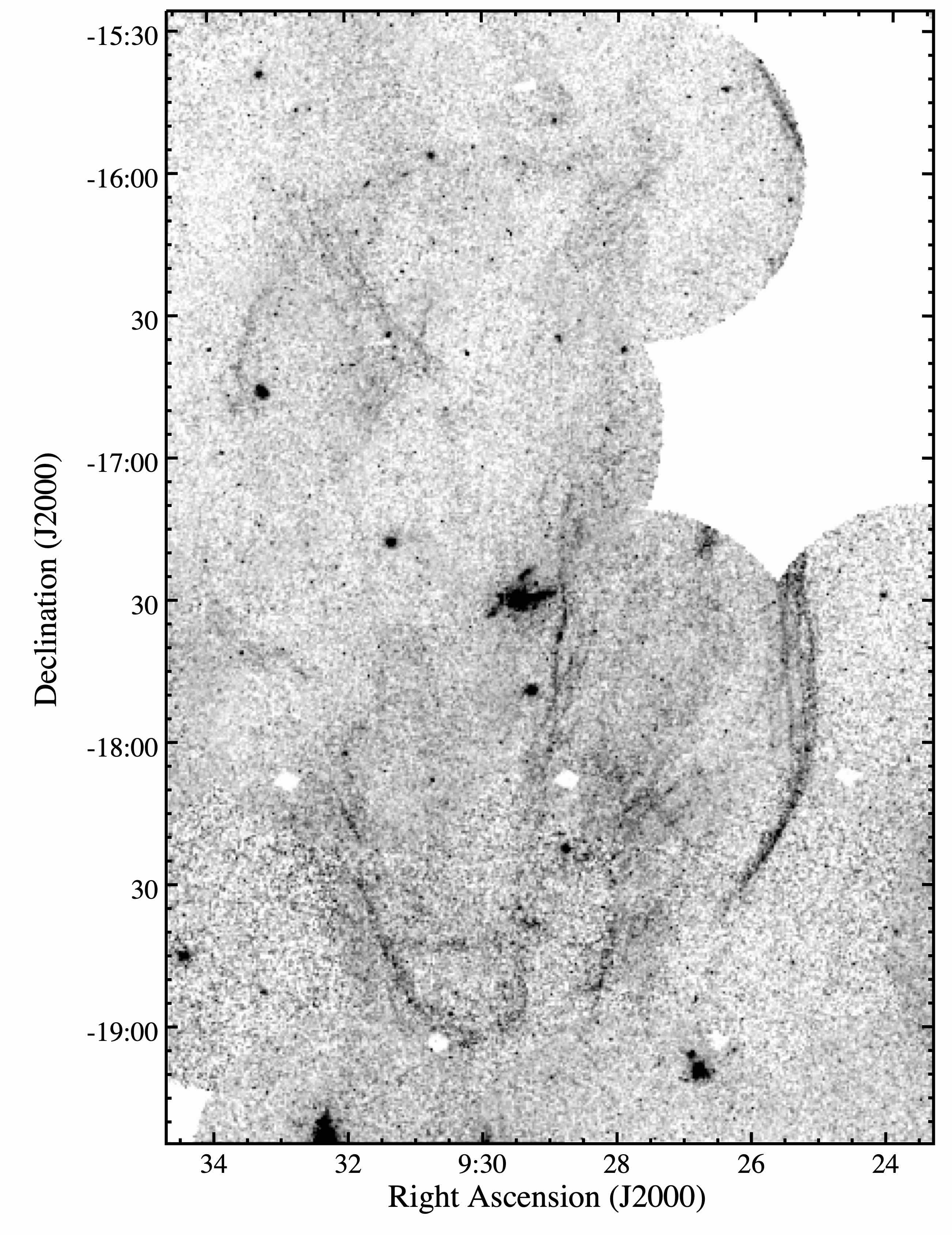} 
\includegraphics[angle=0,width=8.5cm]{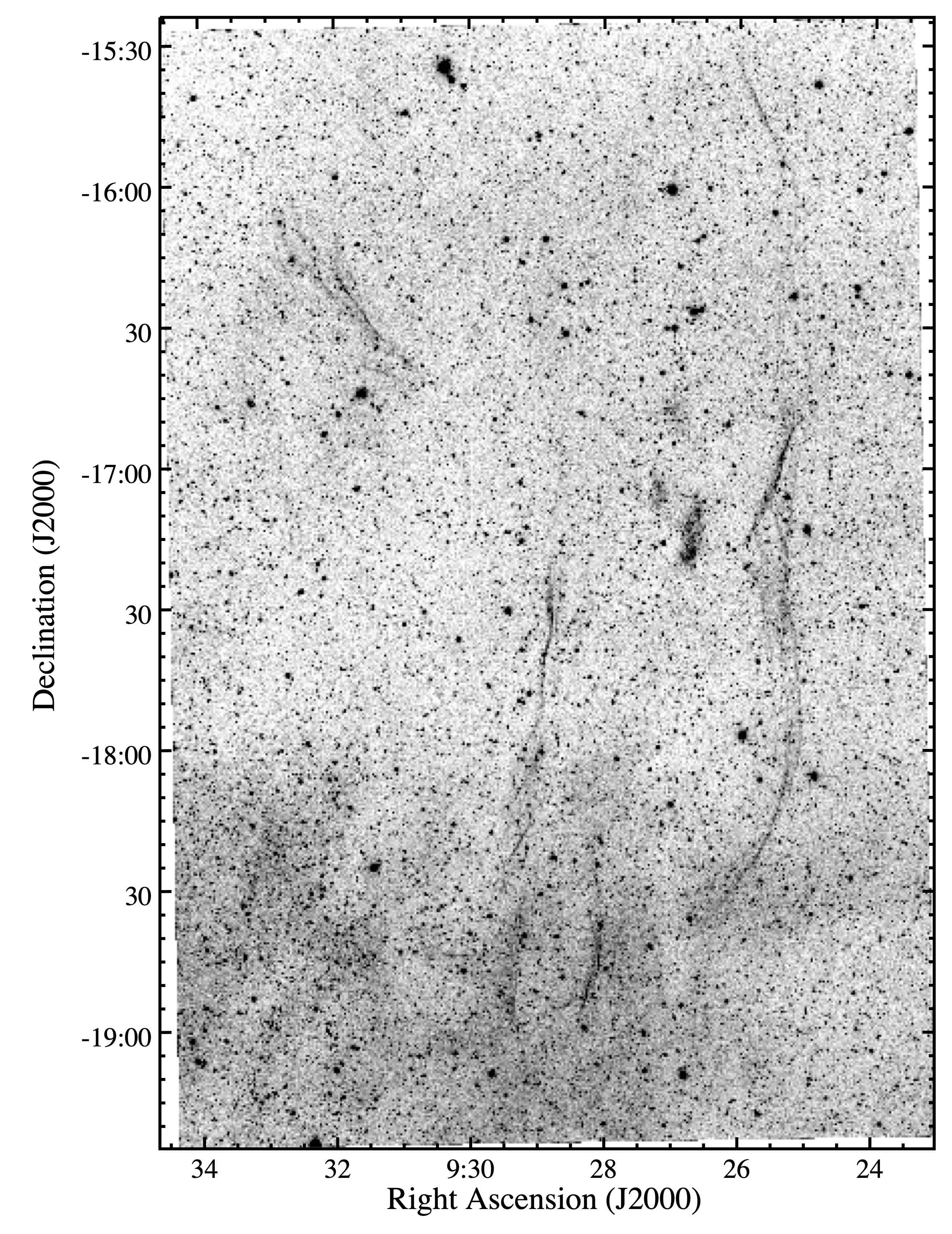} 
\caption{G249+24's FUV emissions (left) compared to its H$\alpha$ emission as seen in 
MDW's All-Sky H$\alpha$  Survey images (right).  \label{G2_FUV_n_Ha}
}
\end{center}
\end{figure*}

The H$\alpha$ images of the MDW All-Sky survey reveal more fully G249+24's optical emissions. In Figure~\ref{G2_FUV_n_Ha}, we show a side-by-side comparison of G249+24's central emission structure in GALEX FUV and the MDW H$\alpha$ images. In general, there is a good agreement between the nebula's UV and H$\alpha$ emissions. The bright FUV and curved southwestern filament shows up strongly in H$\alpha$ as does a shorter filament a degree to its east. 
Correlated UV -- H$\alpha$ emission is also seen along the upper left-hand portion of Figure 8, where diffuse and filament type of emission is seen in both.
(Note: Faint diffuse emission seen along the lower portion of the H$\alpha$ image extends several degrees to both the east and west of the  region and hence does not seem to be connected to G249+24's other optical features (see also Figure~\ref{G2_slits} below).

However, little in the way of H$\alpha$ emission can be seen along the object's southeast limb, where in contrast one finds considerable FUV emission. A similar difference between FUV and H$\alpha$ fluxes is seen for the south-central area around 
9$^{\rm h}$ 28$^{\rm m}$, $-18$\degr 15$'$. 
Consequently, it is clear that this nebula's overall structure is more readily visible in the FUV than in even fairly deep H$\alpha$ images.

Higher resolution MDM H$\alpha$ images of selected regions of the nebula were also obtained and presented in Figure~\ref{G2_MDM_images}. Although these images were taken mainly in preparation for follow-up spectral observations, they showed a surprisingly amount filamentary detail to G249+24's optical emission than initially indicated in Figure~\ref{G2_FUV_n_Ha}.
Numerous long and delicate filaments are visible in several on these images, indicating a rich and beautiful optical shock emission structure largely hidden due to its relative faintness.

%%%%%%%%%%%%%%%%%%%%%%%%%%%%%%%%%%%%%%%%%%%%%%%%%%%%%%%%%%%%%%%%%%%%
%%% Figure 9: G2    FUV  and Halpha
%%%%%%%
%%%%%%%%%%%%%%%%%%%%%%%%%%%%%%%%%%%%%%%%%%%%%%%%%%%%%%%%%%%%%%%%%%

%%%%%%%%%%%%%%%%%%%%%%%%%%%%%%%%%%%%%%%%%%%%%%%%%%%%%%%%%%%%%%%%%%
%%% Figure 10  G2: Slit Positions
%%%%%%%
\begin{figure*}
\begin{center}
\includegraphics[angle=0,width=12.71cm]{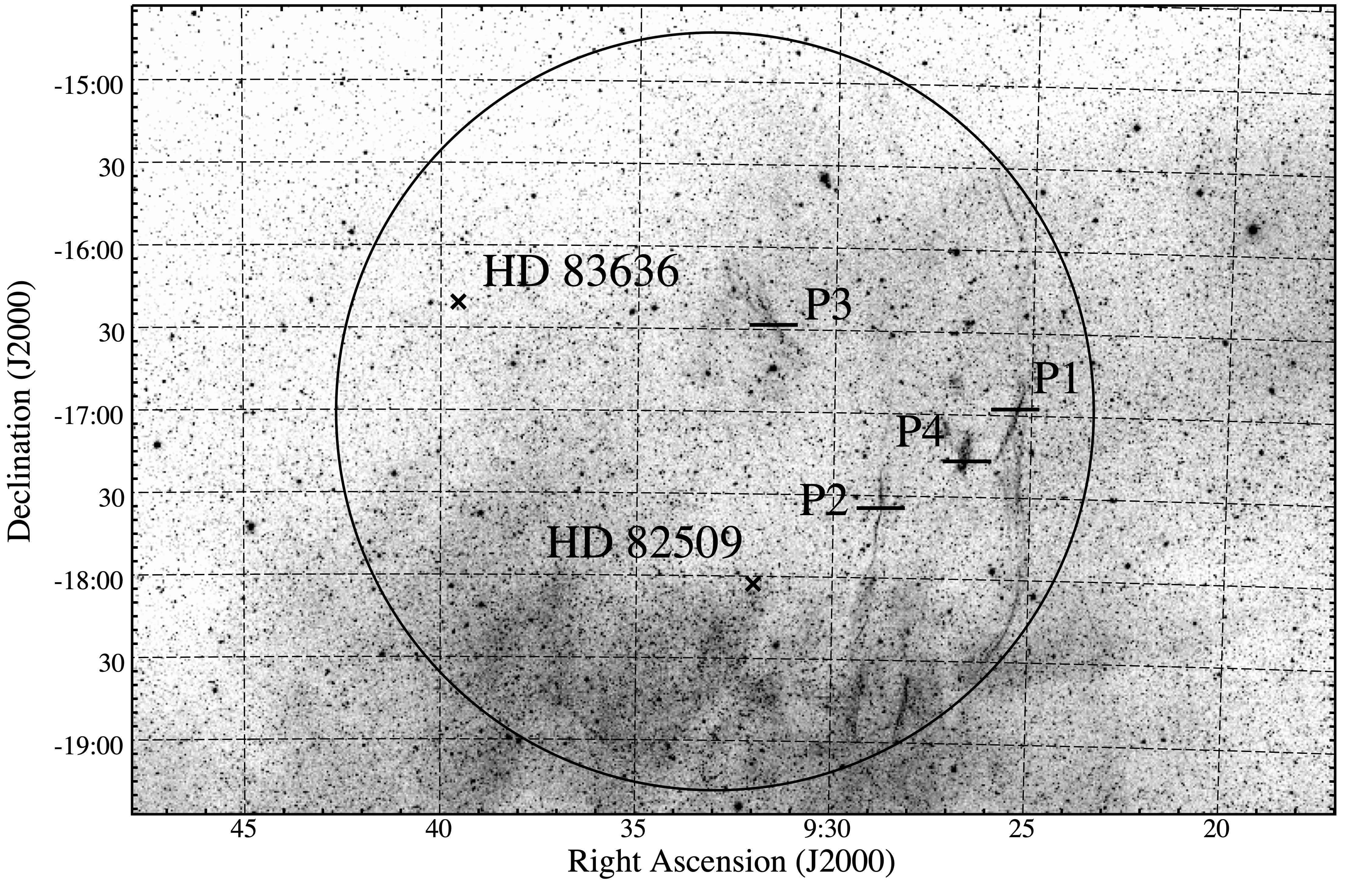}
\caption{MDW H$\alpha$ image of G249.7+24.7 showing
where low-dispersion optical spectra were taken. Also shown are the locations early type stars HD 82509 and HD 83636 which were found to exhibit high-velocity \ion{Na}{1} absorption features. The black circle (dia.\  = 4.5$\degr$) marks the approximate size of the remnant.
\label{G2_slits}
}
\end{center}
\end{figure*}
%%%%%%%%%%%%%%%%%%%%%%%%%%%%%%%%%%%%%%%%%%%%%%%%%%%%%%%%%%%%%%%%%%

%%%%%%%%%%%%%%%%%%%%%%%%%%%%%%%%%%%%%%%%%%%%%%%%%%%%%%%%%%%%%%%%%%
%%% Figure 11 G2: MDM images
%%%%%%%
\begin{figure*}
\begin{center}
\includegraphics[angle=0,width=8.4cm]{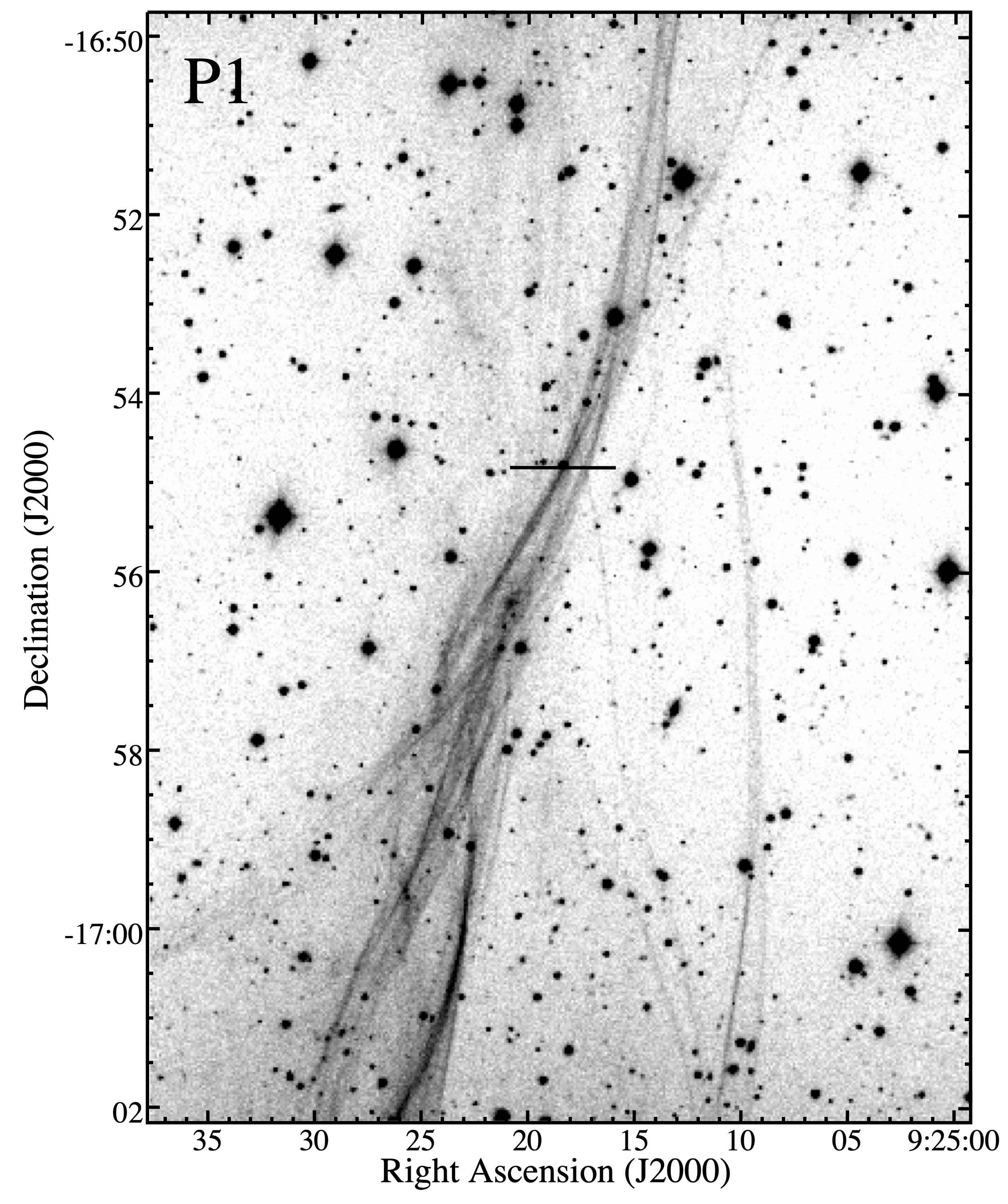} 
\includegraphics[angle=0,width=8.5cm]{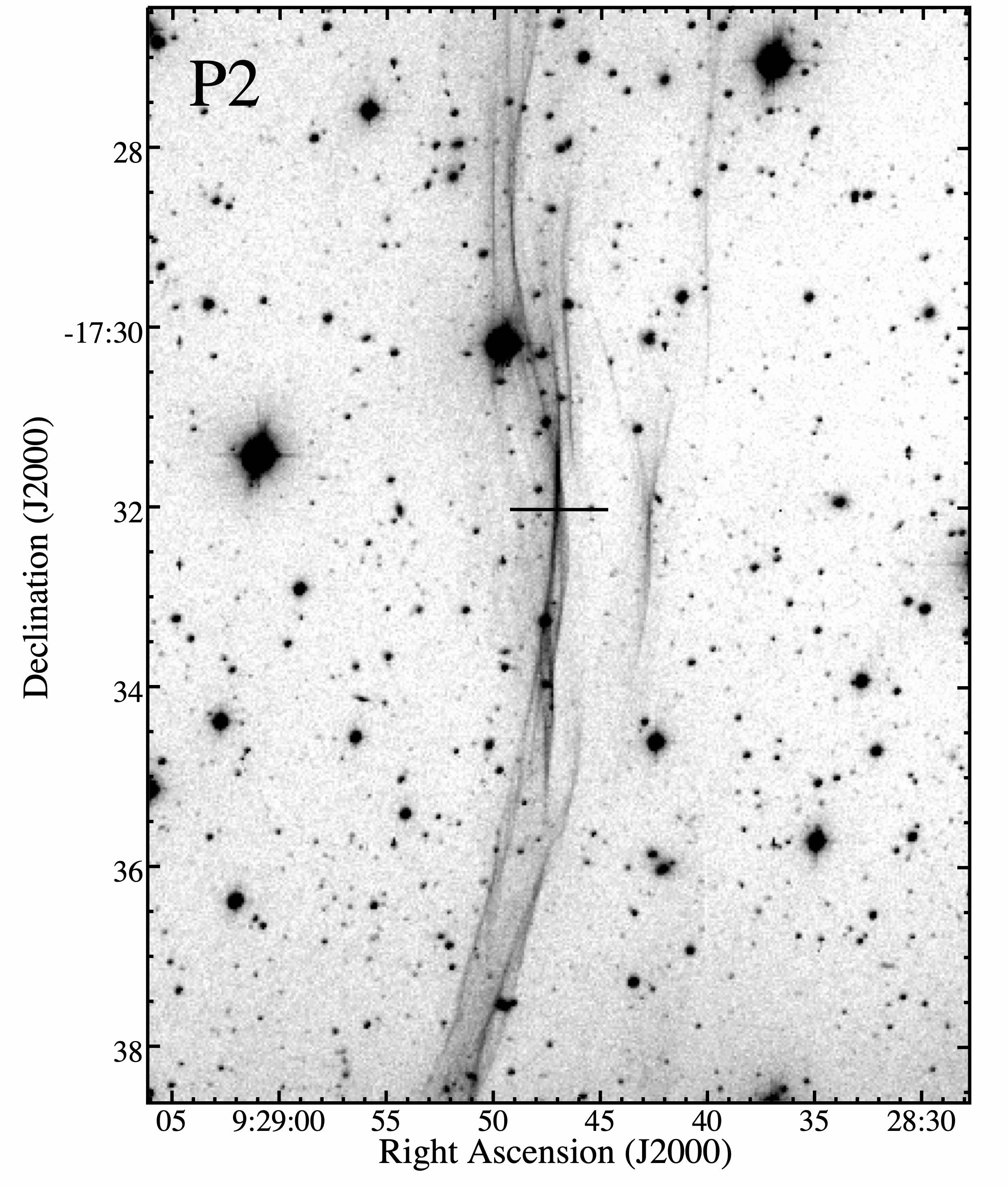} 
\includegraphics[angle=0,width=8.5cm]{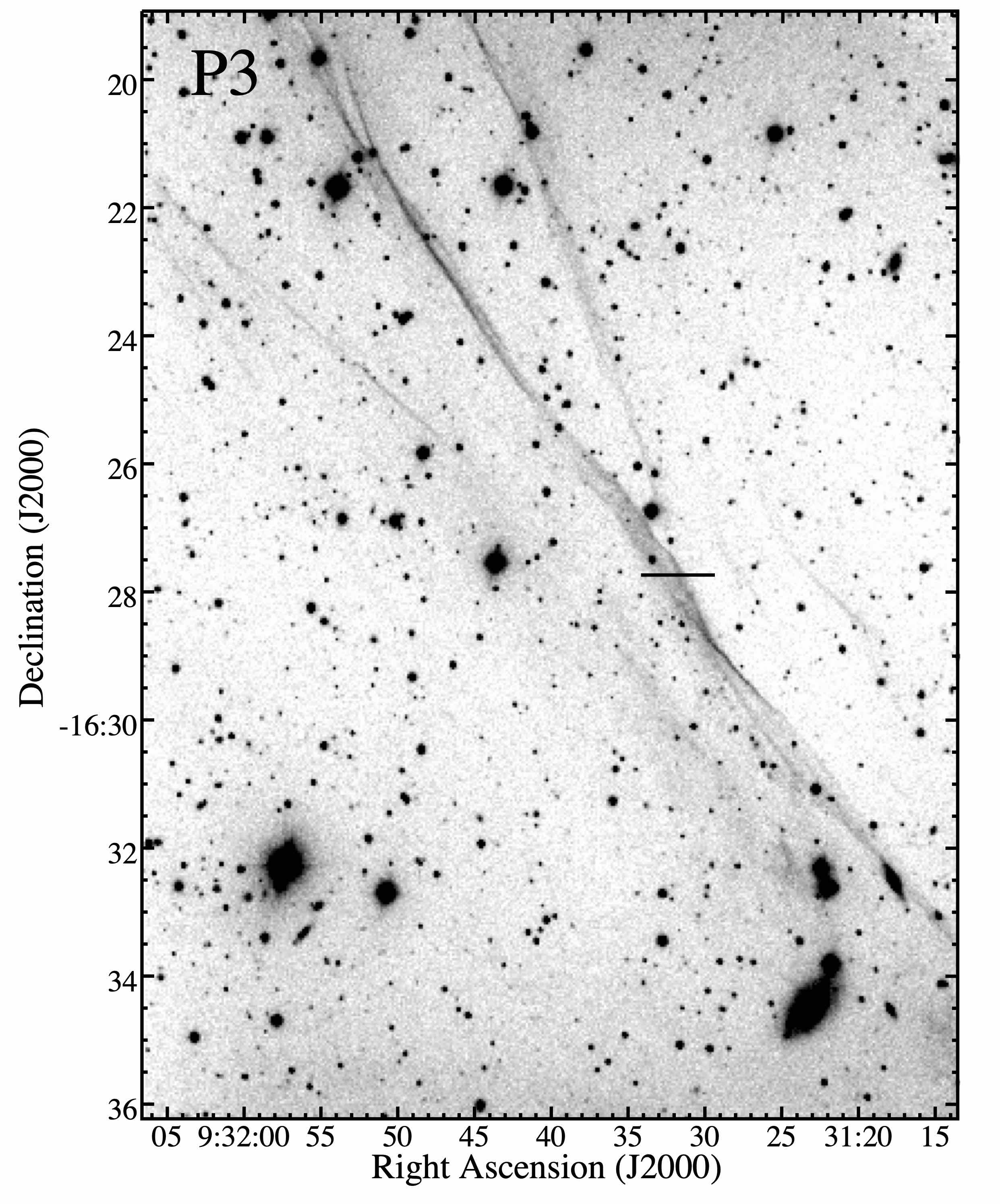} 
\includegraphics[angle=0,width=8.3cm]{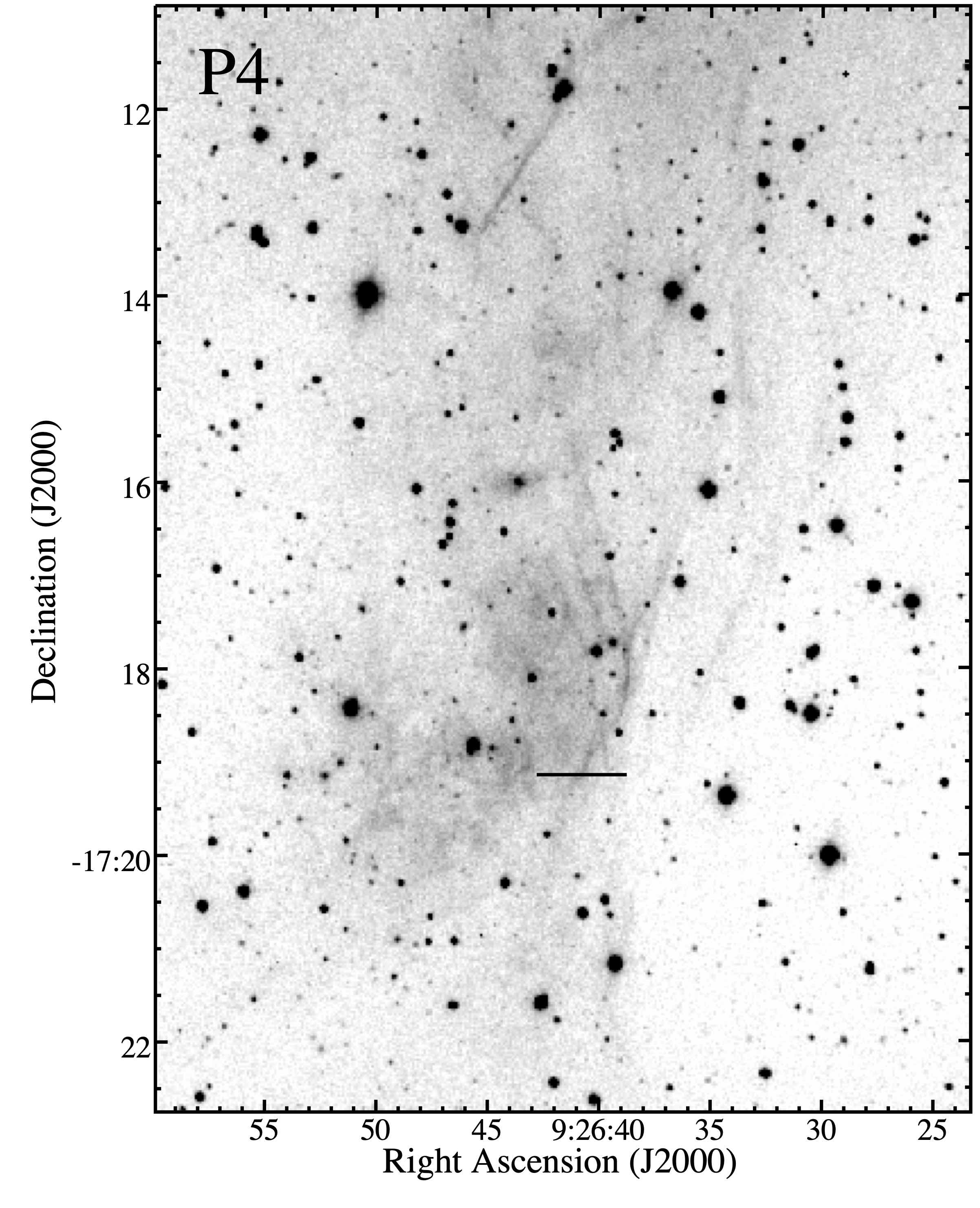} 
\caption{MDM H$\alpha$ + [\ion{N}{2}] images of filaments showing locations of slit Positions 1 and 2.  Slit lengths as shown are $1'$ in length.
Note the numerous long, fine filaments in these regions which are common characteristics of SNR shocks.
\label{G2_MDM_images}
}
\end{center}
\end{figure*}
%%%%%%%%%%%%%%%%%%%%%%%%%%%%%%%%%%%%%%%%%%%%%%%%%%%%%%%%%%%%%%%%%%

%%%%%%%%%%%%%%%%%%%%%%%%%%%%%%%%%%%%%%%%%%%%%%%%%%%%%%%%%%%%%%%%%%%
%%% Figure 12:  G2: SPECTRA
%%%%%%%
\begin{figure*}[ht]
\begin{center}
\includegraphics[angle=0,width=15.0cm]{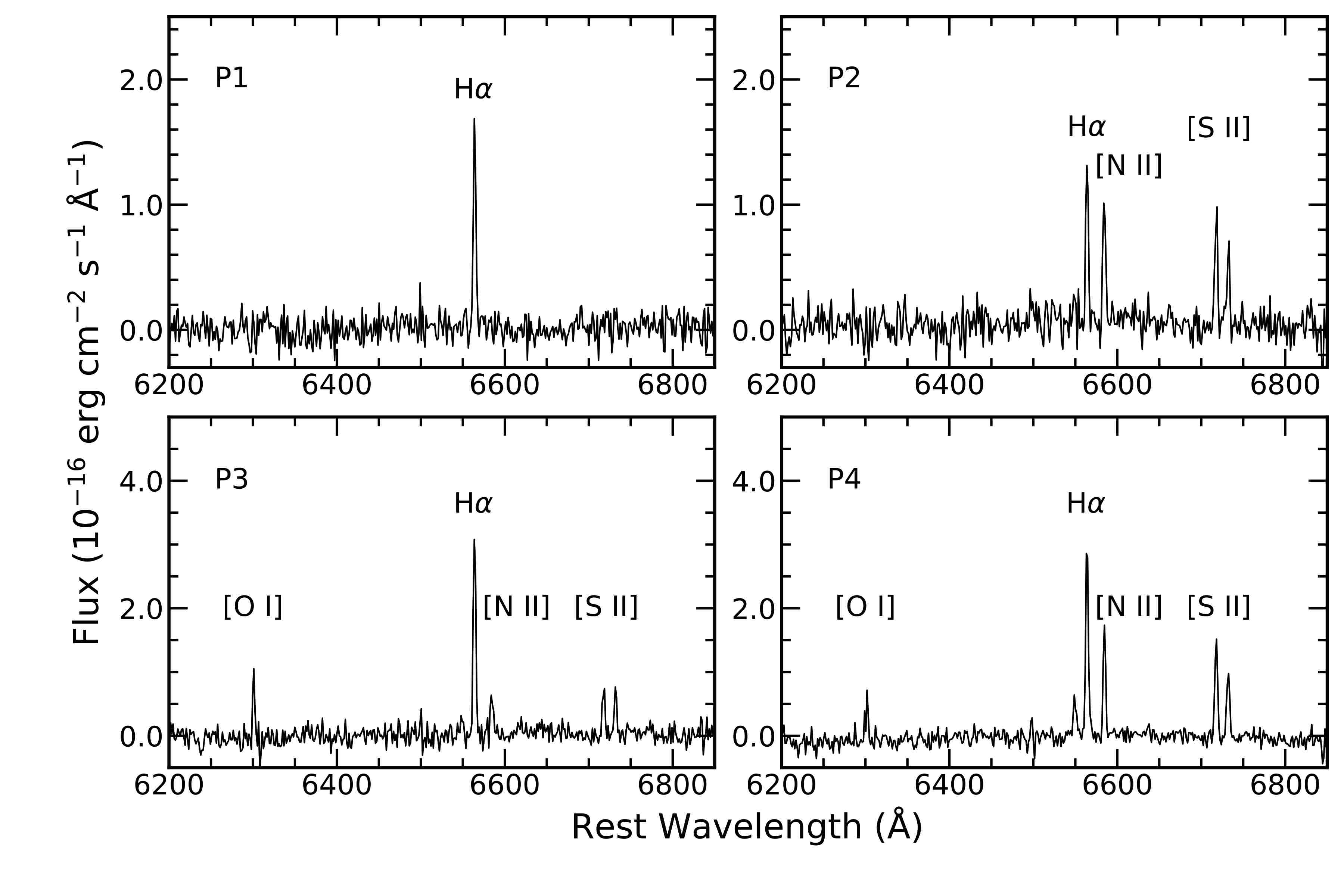}
\caption{Optical spectra of filamentary emission at positions in G249+24.
\label{G2_spectra}
}
\end{center}
\end{figure*}
%%%%%%%%%%%%%%%%%%%%%%%%%%%%%%%%%%%%%%%%%%%%%%%%%%%%%%%%%%%%%%%%%%%

%%%%%%%%%%%%%%%%%%%%%%%%%%%%%%%%%%%%%%%%%%%%%%%%%%%%%%%%%%%%%%%%%%%
%%% Figure 13: Position 1 in two pieces
%%%%%%%
\begin{figure*}
\centerline{\includegraphics[angle=0,width=9.0cm]{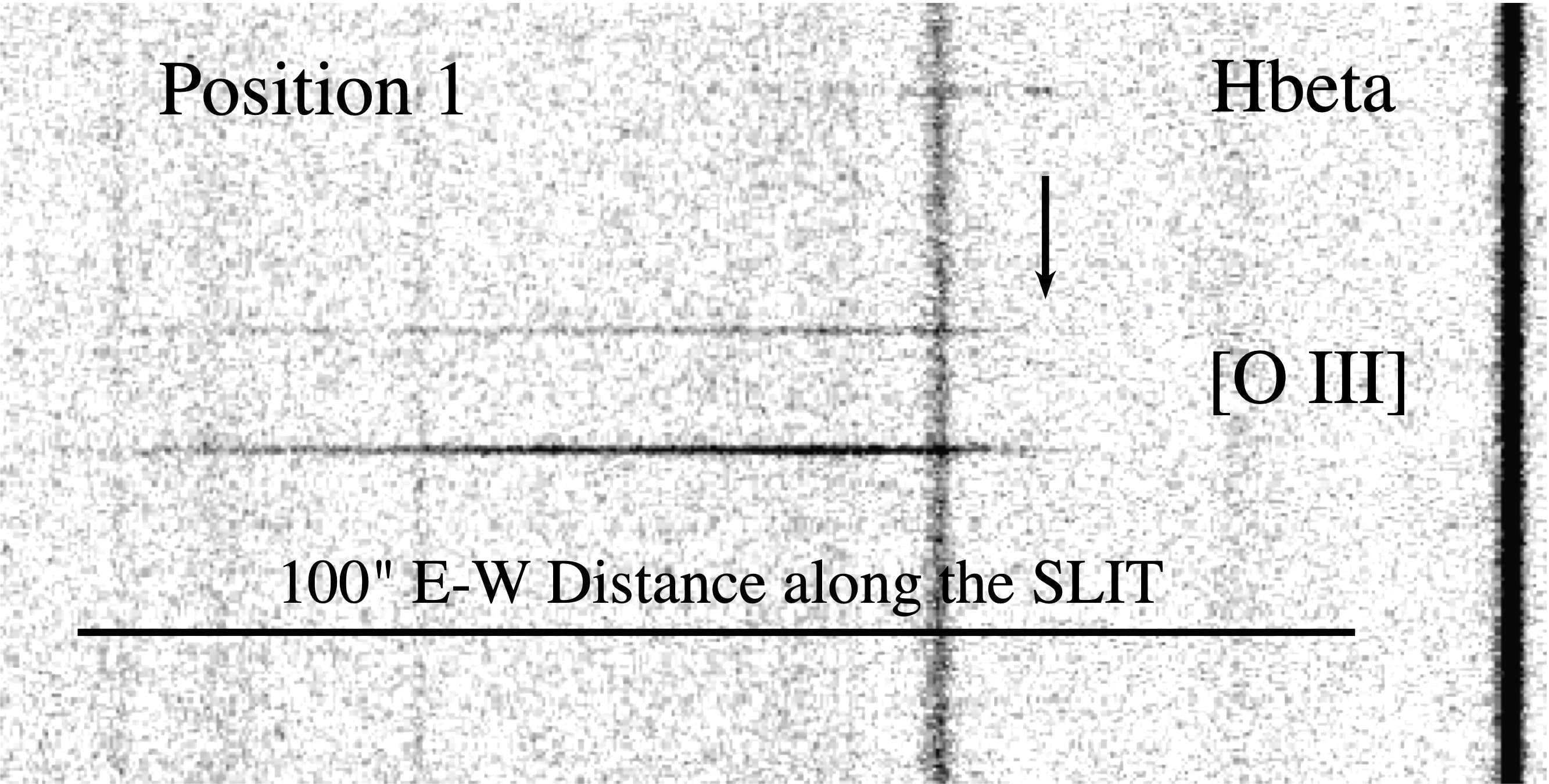}}
\centerline{\includegraphics[angle=0,width=9.0cm]{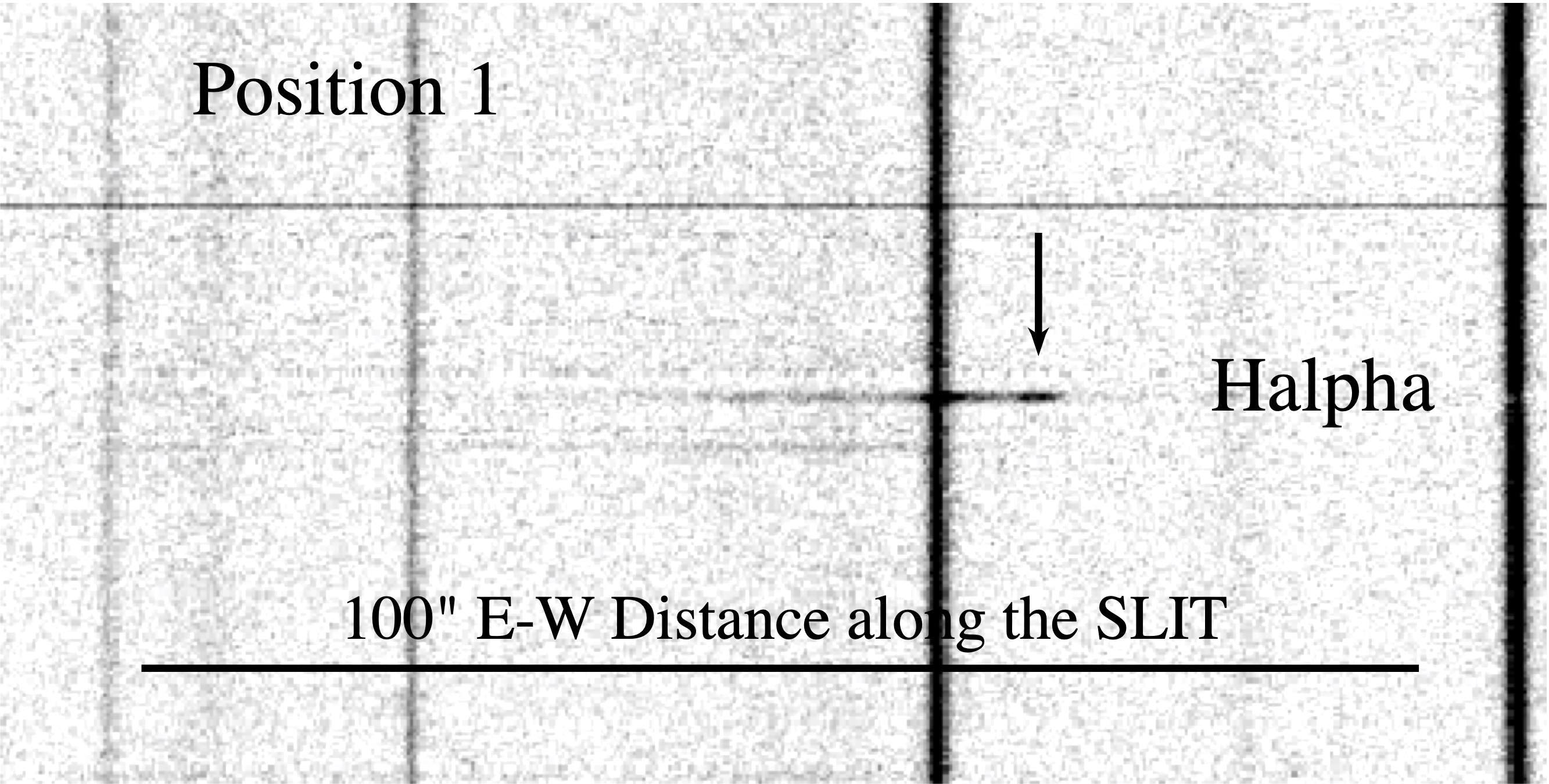}}
\caption{Sections of the long slit spectra for Position 1. Both 
[\ion{O}{3}] $\lambda$4959 and $\lambda$5007 lines are visible in the upper panel.
\label{G2_pos1}
}
\end{figure*}
%%%%%%%%%%%%%%%%%%%%%%%%%%%%%%%%%%%%%%%%%%%%%%%%%%%%%%%%%%%%%%%%%%%%

\subsubsection{Low-Dispersion Optical Spectra}

Low-dispersion, exploratory slit spectra were taken at four positions (P1 through P4) in G249+24 as shown in Figures~\ref{G2_slits} and \ref{G2_MDM_images} in order to investigate the emission nature G249+24's optical filaments.
Since only the spectra taken at P1 exhibited any [\ion{O}{3}] emission, the four spectral plots shown in  Figure~\ref{G2_spectra} cover only the wavelength region 6200 to 6900 \AA.

The resulting spectra reveal clear evidence for its filaments being shock heated ISM like that commonly seen in SNRs.
For example, spectra obtained at P2, P3 and P4 exhibited [\ion{S}{2}]/H$\alpha$ ratios of 1.26$\pm0.20$, 
0.45$\pm0.04$, and 0.91$\pm0.15$, respectively, 
thus well above the standard criteria ratio of 0.40 indicative of shock emission. Added support for shocks is the detection of [\ion{O}{1}] $\lambda$6300 emission in the spectra of both P3 and P4. The presence of [\ion{O}{1}] is a secondary
indicator for shock emission commonly observed in evolved SNRs \citep{Fesen1985,Kop2020}. 

The [\ion{S}{2}]  
$\lambda$6716/$\lambda$6731 line ratio can be used to estimate the electron
density in the S$^{+}$ recombination zone and is nearly independent of electron
temperature \citep{Osterbrock2006}. The $\lambda$6716/$\lambda$6731 ratio was found to be near the low density limit of 1.43 indicating n$_{\rm e}$ $<$100 cm$^{-3}$; $1.48\pm0.04$ for P2, and $1.42\pm0.03$ for P4.
A bit surprisingly, this ratio for P3 was observed to be $\simeq$1.15 indicating an n$_{\rm e}$ density  much higher than the other locations of around 500 cm$^{-1}$. Interestingly, the spectrum seen at P3 also showed much weaker [\ion{N}{2}] $\lambda$6583 line emission compared to that seen at P2 and P4.

The spectrum taken at P1 deserves special attention due to significant line emission variations with respect to distance along the E-W aligned slit. This is shown in Figure~\ref{G2_pos1}, where line emission can be seen to start at the location of the mainly N-S aligned filament extending many arcseconds to the east (left).
The emission line ratio of H$\alpha$/[\ion{O}{3}] can be seen to vary considerably down stream from the shock front.

Figure~\ref{G2_pos1}  also shows the 2D image of the background subtracted P1 spectrum for the areas around H$\alpha$, and [\ion{O}{3}] 
$\lambda$4959, $\lambda$5007 and H$\beta$. This figure shows a clear separation of the filament's Balmer dominated emission followed by strong [\ion{O}{3}] emission resulting in large   [\ion{O}{3}] $\lambda$5007/H$\alpha$ ratios. The arrows indicate the section where the P1 spectrum shown in Figure~\ref{G2_MDM_images} was taken.

Behind the leading edge of the shock front marked by start of line emissions, H$\alpha$ emission becomes strong with little or no emission in other lines. The intensity of H$\alpha$ then briefly drops in strength within a distance of a few arcseconds, then increases for a short distance followed by a long, gradual decline. 
The [\ion{O}{3}] emission is barely detectable at the shock front where the H$\alpha$ starts but substantially increases some 10-20 arcseconds behind the shock front, to where it dominates the downstream spectrum. This is consistent with the picture of an extended post-shock cooling zone. The presence of such strong [\ion{O}{3}] emission behind the shock front suggests shock velocities at least 100 km s$^{-1}$ are present in certain filaments. In contrast, spectra at P2, P3, and P4 showed no appreciable  [\ion{O}{3}] emission indicating much lower velocity shocks, less than about 70 km s$^{-1}$.

%%%%%%%%%%%%%%%%%%%%%%%%%%%%%%%%%%%%%%%%%%%%%%%%%%%%%%%%%%%%%%%%%%%
%%% Figure 14 --  CCDS spectra plots
%%%%%%%
\begin{figure*}[ht]
\begin{center}
\includegraphics[angle=0,width=15.cm]{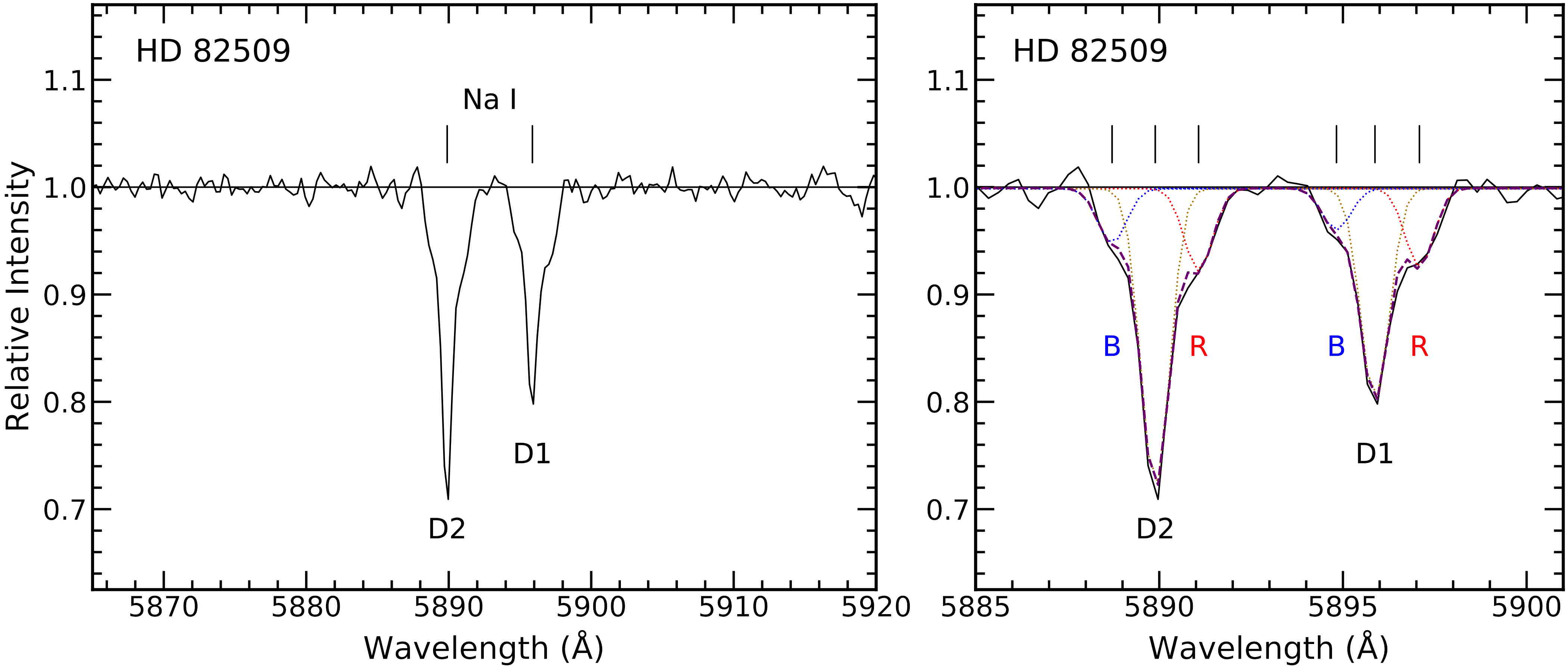} 
\includegraphics[angle=0,width=15.0cm]{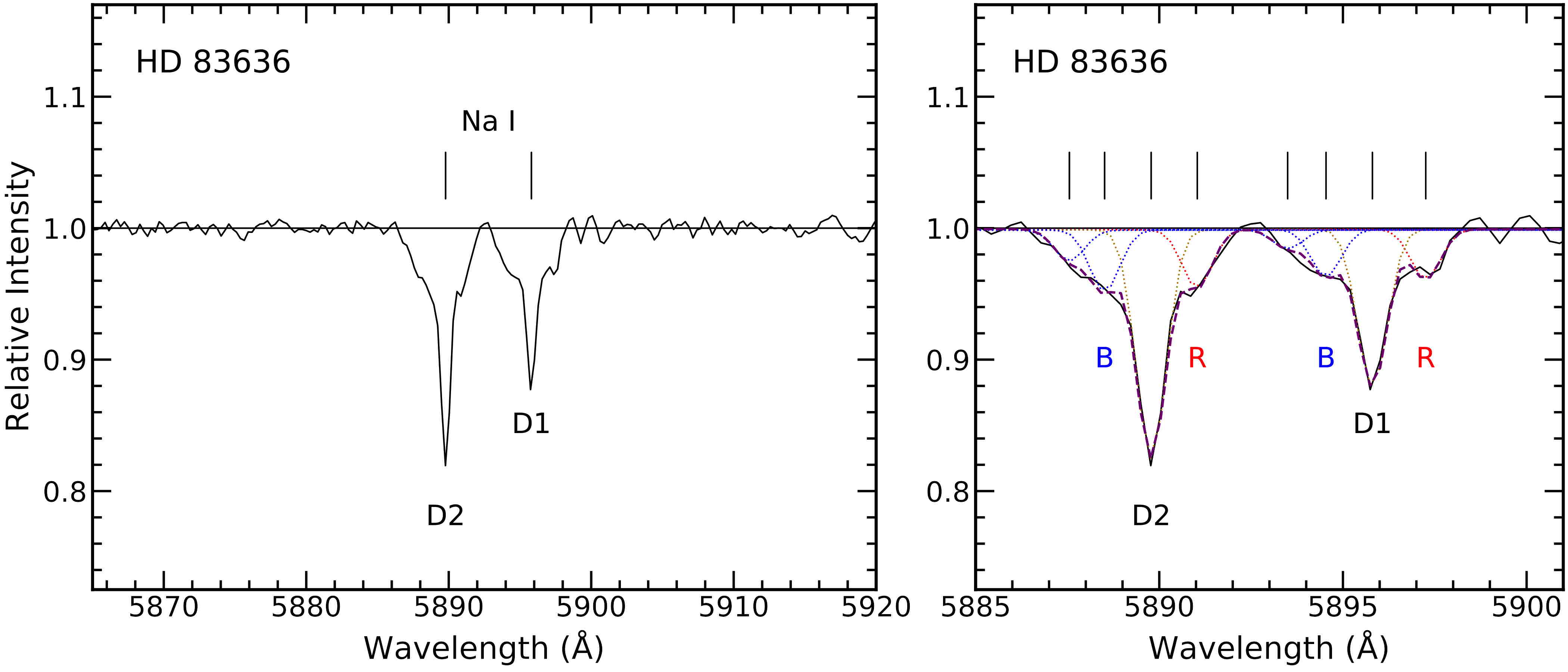}
\caption{Moderate dispersion CCDS spectra of HD~82509 (top) and HD 83636 (bottom) showing deblending of the \ion{Na}{1} lines into high-velocity blue and redshifted absorption components associated with G249+24. See Table 3.
\label{CCDS_spectra}
}
\end{center}
\end{figure*}
%%%%%%%%%%%%%%%%%%%%%%%%%%%%%%%%%%%%%%%%%%%%%%%%%%%%%%%%%%%%%%%%%%%%

Similar 2D emission structures have been observed in the high latitude Galactic halo SNRs G70.0-21.5 \citep{Raymond2020} and G107.0+9.0 \citep{Fesen2020}.  They can be interpreted as emission from a $\sim100$ km s$^{-1}$ shock in partially neutral gas.  Neutral hydrogen atoms swept up by the shock are quickly ionized, but before that happens, some of them are excited to produce H$\alpha$.  This emission region is very thin.  It is followed by a thicker ionization zone where \ion{O}{2} is ionized to \ion{O}{3}.  As the gas cools, the [\ion{O}{3}] fades and H$\alpha$ brightens.  In the case of G107.0+9.0, the 10$"$ gap between H$\alpha$ and [\ion{O}{3}] and the $\sim$ 10$"$ wide diffuse patch of [\ion{O}{3}]  emission are well matched  by models of 100 km s$^{-1}$ shocks in a gas of density 0.1 cm$^{-3}$ at a distance of 1 kpc \citep{Fesen2020}.  The wider patch of  [\ion{O}{3}] diffuse emission at Position 1 of G249+24 suggests a longer cooling length, as would occur in a slightly faster shock, perhaps $\sim$120 km s$^{-1}$. These structures are spatially resolved thanks to the low density of the Galactic halo and the relatively small distances to these SNRs.

The flux level of H$\beta$ was either undetected or too faint at P2, P3, and P4 to accurately measure an H$\beta$/H$\alpha$ ratio which could be used to estimate extinction.  Only the spectrum P1 showed a weak but measurable level of H$\beta$ flux. There the measured H$\alpha$/H$\beta$ ratio of $2.85\pm 0.20$ suggests a very low amount of extinction.  Adopting a theoretical H$\alpha$/H$\beta$ ratio of 2.87 for 10$^{4}$ K we estimate a foreground  $E(B-V) < 0.07$. However, in view of the weakness of the H$\beta$ detection, this estimate comes with a large uncertainty. Nonetheless, such a low interstellar extinction is not unexpected due to the remnant's high Galactic latitude of $24.4\degr$.

If this low extinction estimate is correct, then the lack of [\ion{O}{3}] emission at P2, P3, and P4 suggests a fairly low shock velocity, around 70 km s$^{-1}$ or less, in contrast to that indicated at P1. Consequently, it appears that a variety of shock speeds, from $<$70 to $\geq$120 km s$^{-1}$ appear to be present throughout the nebula's structure. This conclusion is supported by \ion{Na}{1} absorptions described below.

%%%%%%%%%%%%%%%%%%%%%%%%%%%%%%%%%
% Table 3 %%%
\begin{deluxetable*}{lcccccccccc}[ht]
\tablecolumns{10}
\tablecaption{Measured Interstellar Na I Velocity Components and Equivalent Widths}
\tablewidth{0pt}
\tablehead{
\colhead{Star} & \colhead{Distance} & \colhead{\ion{Na}{1}}  & \colhead{$\mathrm{V}_0$}      & \colhead{$\mathrm{EW}_0$}     &
                                   \colhead{$\mathrm{V_{R1}}$}   & \colhead{$\mathrm{EW_{R1}}$}  &
                                   \colhead{$\mathrm{V_{B1}}$}   & \colhead{$\mathrm{EW_{B1}}$}  &
                                   \colhead{$\mathrm{V_{B2}}$}   & \colhead{$\mathrm{EW_{B2}}$} \\
\colhead{ID} &  \colhead{(pc)}       &\colhead{Line}       & \colhead{(km s$^{-1}$)}         & \colhead{(\AA)} & 
                                   \colhead{(km s$^{-1}$)}         & \colhead{(\AA)}                 &   
                                   \colhead{(km s$^{-1}$)}         & \colhead{(\AA)}                 &  
                                   \colhead{(km s$^{-1}$)}         & \colhead{(\AA)}   }                          
\startdata
HD 82509 & $825\pm25$ & $\lambda$5889.95  & $-3$   & 0.28  & +57 & 0.08  & $-63$ & 0.05 & \nodata  & \nodata  \\
          &           & $\lambda$5895.92  & $-2$   & 0.20  & +59 & 0.08  & $-56$ & 0.04 & \nodata  & \nodata  \\
HD 83636 & $386\pm4$  & $\lambda$5889.95  & $-9$   & 0.17  & +55 & 0.05 & $-73$ & 0.05 & $-122$ & 0.02  \\
          &           & $\lambda$5895.92  & $-6$   & 0.12  & +70 & 0.04 & $-70$ & 0.04 & $-123$ & 0.01  \\
\enddata
\label{CCDS_table}
\end{deluxetable*}
%%%%%%%%%%%%%%%%%%%%%%%%%%%%%%%%

\bigskip

\subsubsection{High-Velocity Na I Absorption Lines}

Moderate-dispersion optical spectra (R $\simeq$ 7200) of several stars
with projected positions within the estimated radio boundary of
G249+24 (see \citet{Becker2021} and Section 3.2.5 below) were
obtained in a search for high-velocity Na I interstellar absorption components, which could yield both a direct measurement of its expansion velocity plus maximum remnant distance information. Of the stars examined, two showed evidence for the presence of high-velocity ($>$ 40 km s$^{-1}$) \ion{Na}{1} absorption features.  

The two stars were: HD~82509 (B = 10.29, V = 10.4; A2 IV) with an estimated Gaia Early Release DR3 (EDR3) distance estimate of $825\pm25$ pc, and  HD~83636 (B = 9.74, V = 9.57: A2 IV/V) with an EDR3 distance estimate of $386\pm4$ pc. Figure~\ref{G2_slits} shows the locations of these stars relative to the object's FUV and optical emissions.

Reduced  and  normalized  spectra  of both stars for the wavelength region around the \ion{Na}{1} D lines at $\lambda$5889.95  and $\lambda$5895.92 indicating the presence of high-velocity \ion{Na}{1} absorptions are shown in  Figure~\ref{CCDS_spectra}.  Using IRAF deblending software and allowing initially chosen absorption component wavelengths to vary but keeping the Gaussian line profiles and FWHMs all the same, we found red and blue absorption features in the observed \ion{Na}{1}  profiles in both stars with LSR velocities around +55  to $-73$ km s$^{-1}$. In addition, to achieve a good fit to the Na absorption in the  spectrum of HD~83636, an additional blue-shifted components with a radial velocity of around $-123$ km s$^{-1}$ was needed (see Table 3). 
We note that the lower velocity components are consistent with the observed spectra seen for slit positions P2, P3, and P4 where no [\ion{O}{3}] was seen indicating shock velocities below $\simeq$ 70 km s$^{-1}$, while the presence of the higher velocity 123 km s$^{-1}$ component is consistent with the strong [\ion{O}{3}] seen at slit position P1. 

The detection of these high-velocity \ion{Na}{1} absorptions in these two stars can be used to estimate a maximum distance to G249+24.  HD~82509's Gaia EDR3 parallax derived distance sets a distance limit of less than 825 pc, while HD~83636 sets a maximum distance of less than 390 pc. Assuming an angular diameter of 4.5\degr  (see below), this shorter maximum 390 pc distance implies a maximum physical size for G249+24 of $\sim$30 pc, similar to that estimated for the Cygnus Loop SNR \citep{Fesen2018}. \\

\subsubsection{Associated Radio and X-Ray Emissions}

Examination of several on-line radio maps including the NRAO VLA Sky Survey (NVSS;  \citealt{Condon1998}) did not show obvious radio emission in the G249+24's region. However, the 1420 MHz Bonn survey \citep{Reich1982} did show some emission along G249+24's FUV and optical filaments near $\alpha$(J2000) = 9$^{\rm h}$ 27.5$^{\rm m}$, $\delta$(J2000) = $-16$\degr 10$'$.

%%%%%%%%%%%%%%%%%%%%%%%%%%%%%%%%%%%%%%%%%%%%%%%%%%%%%%%%%%%%%%%%%%%%
%%% Figure 15: G2 radio emission
%%%%%%%
\begin{figure*}[htp]
\begin{center}
\includegraphics[width=0.55\textwidth]{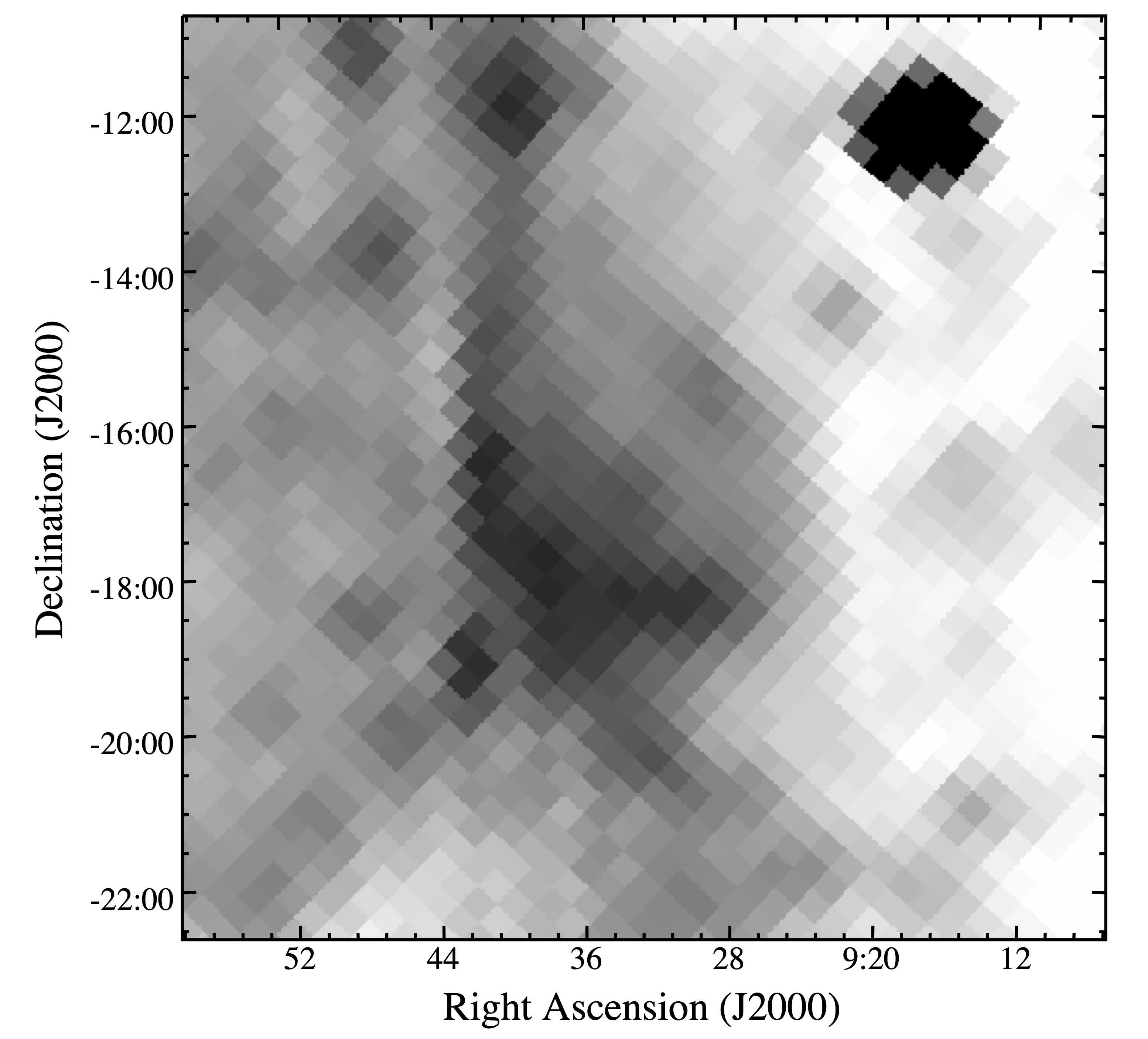}
\caption{A 408 MHz intensity image of the G249+24 region from the \citet{Haslam1982} All-Sky survey.
\label{G2_radio}
}
\end{center}
\end{figure*}
%%%%%%%%%%%%%%%%%%%%%%%%%%%%%%%%%%%%%%%%%%%%%%%%%%%%%%%%%%%%%%%%%%%

The 408 MHz all-sky survey \citep{Haslam1981,Haslam1982} also showed enhanced emission in a roughly spherical shape approximately 4.5\degr \ in diameter centered  at 9$^{\rm h}$ 32.7$^{\rm m}$, $-16$\degr 45$'$, a position that is nearly the same indicated by FUV and optical images (see Fig.~\ref{G2_radio}).
A seemingly unrelated broad, vertical emission feature borders the emission shell 
on its eastern side and complicates the assessment of this emission. Assuming 
this emission shell is associated with the nebula,
it would suggest a far more spherical SNR than indicated by either the GALEX FUV 
or MDW H$\alpha$ images. 
Taking the object's largest N-S angular dimensions based of its 
FUV filaments (see Figs.\ 8 and 9) and requiring that both stars, HD~82509 and HD~83636, 
must lie inside the remnant's shell, we estimate G249+24 is $\simeq4.5\degr$ in diameter assuming a circular morphology. This size estimate is in
excellent agreement with that found by \citet{Becker2021}
using eRosita X-ray images and archival Parks All-Sky Survey radio data.

\citet{Becker2021} report the detection of diffuse X-ray emission filling almost the entire remnant with a gas temperture of order kT = 0.1 keV, along with radio data indicating a spectral index of $-0.69$ fully consistent with its SNR nature.

\subsection{The Exceptionally Large Suspected Antlia SNR}

\citet{McCull2002} reported a discovery of a 
very large $\sim$24\degr \ diameter H$\alpha$ emission shell with interior ROSAT 0.25 keV X-ray emission located in the southern hemisphere and at a high Galactic latitude of +19\degr. They proposed it to be a previously unrecognized SNR and named it Antlia for the constellation it is found in. \citet{McCull2002} suggested it was an extremely old remnant with an estimated age of $\sim$1 Myr  and located relatively nearby with an uncertain distance of between 60 pc and 340 pc. 

The Antlia remnant does appears in the catalogue of high energy SNRs \citep{Safi2012} but not in the 
\citet{Green2019} catalogue of confirmed Galactic SNRs,
which cited the need for further observations to confirm its nature and parameters. Except for a 2007 far UV study finding weak coincident \ion{C}{3} $\lambda$977 and \ion{C}{4} $\lambda\lambda$1548,1551 line emissions from the object's interior \citep{Shinn2007}, and a brief AAS abstract by 
\citet{Orchard2015} who reported Wisconsin H-alpha Mapper data indicating 
[\ion{S}{2}] emission from the Antlia Nebula consistent with a SNR interpretation,
there has been no detailed follow-up investigation of this exceptionally large suspected SNR. This situation plus its location at an unusually high Galactic latitude led us to include it in our high Galactic latitude GALEX FUV investigation of SNRs.

\subsubsection{GALEX FUV and MDW H$\alpha$ Images}

\begin{figure*}[htp]
\begin{center}
\centerline{\includegraphics[angle=0,width=17.12cm]{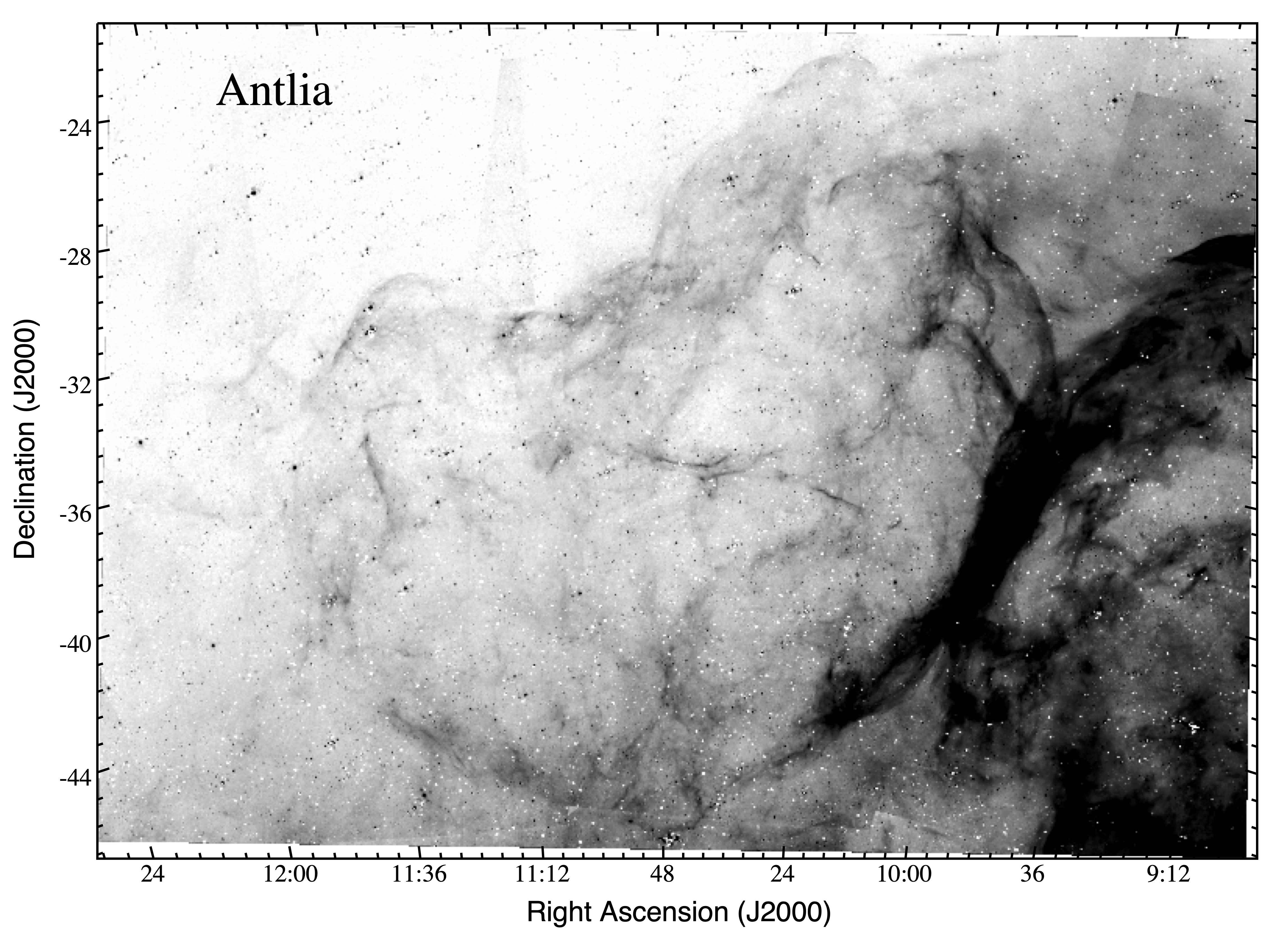} }
\centerline{\includegraphics[angle=0,width=17.12cm]{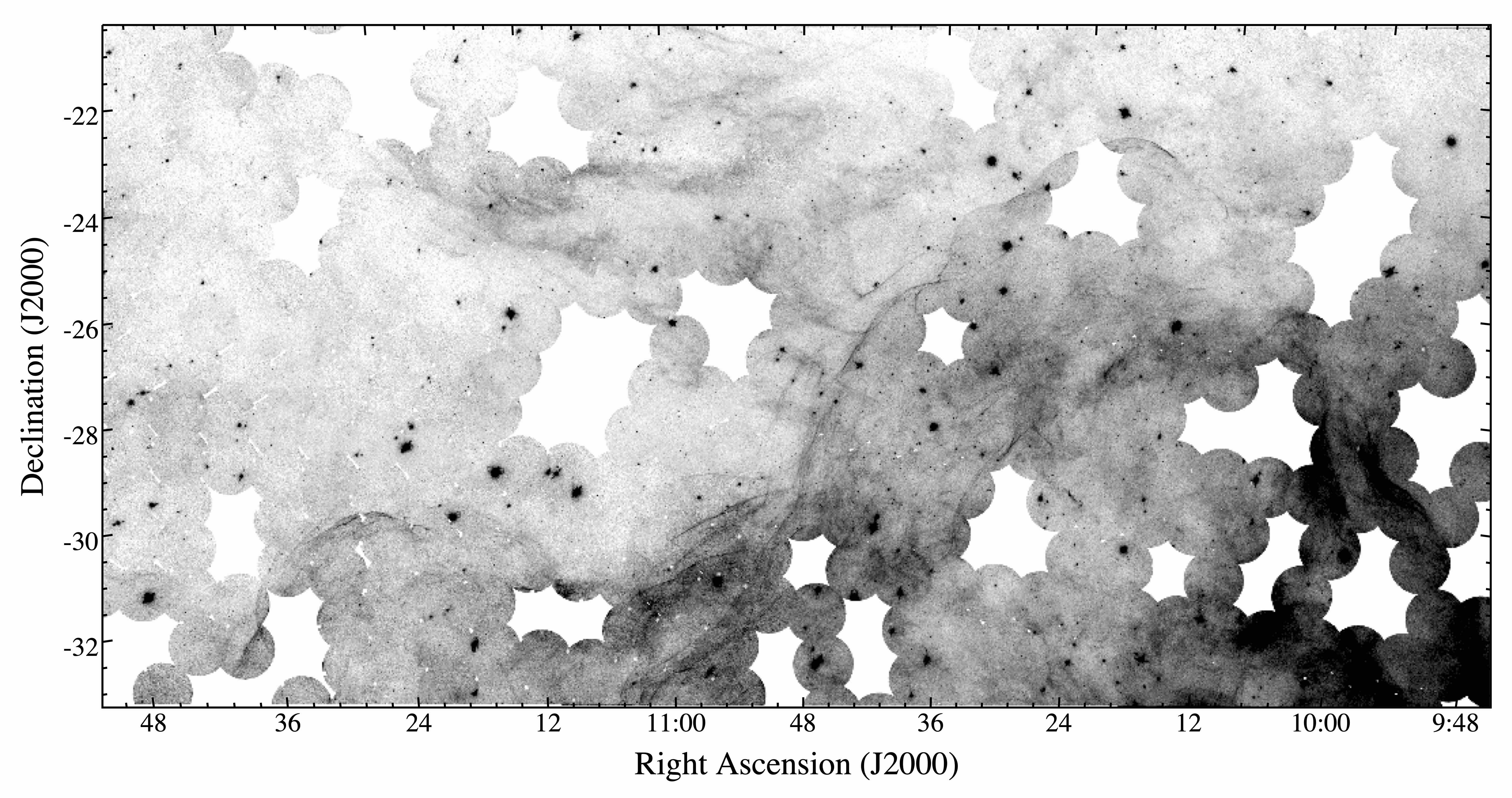}}
\caption{$Top:$ Continuum subtracted SHASSA H$\alpha$ image mosaic of the Antlia remnant. $Bottom:$ GALEX FUV mosaic showing a nearly unbroken line of UV emission filaments
along Antlia's northern and eastern boundary.  \label{G3_SHASSA}
}
\end{center}
\end{figure*}

In the top panel of Figure~\ref{G3_SHASSA}, we present a mosaic of SHASSA H$\alpha$ images of  the Antlia remnant at a higher resolution than the VTSS H$\alpha$ image \citep{Fink2003} which led \citet{McCull2002} to  discover it. It shows a $\sim 20\degr \times 26\degr$ H$\alpha$ emission shell with a well determined boundary roughly centered at 
$\alpha$(J2000) = 10$^{\rm h}$ 38$^{\rm m}$, $\delta$(J2000) = $-37\degr$ 18$'$ corresponding to Galactic coordinates $l = 275.5\degr$ $b = +18.4\degr$. (Note: These Galactic coordinates differ slightly from those given by \citealt{McCull2002}.)

A mosaic of FUV GALEX images covering just the remnant's northern limb is shown in the lower panel of Figure 13. This image reveals a long and nearly continuous set of thin, bright FUV emission filaments and filament clusters extending approximately 20 degrees along the remnant's northeastern limb.
The line of FUV filaments starts at RA = 10$^{\rm h}$ 10$^{\rm m}$ and Dec = $-24\degr$ and extends southward to the bottom left of the figure at
RA = 11$^{\rm h}$ 40$^{\rm m}$ and Dec = $-32\degr$. 

 Because the morphology of Antlia's FUV filaments is similar to that seen in the proposed SNRs G354-33 and G249+24, we examined MDW H$\alpha$ images of the FUV filaments and adjacent regions above the $-32\degr$ Declination limit of the MDW survey. Overall, we found excellent positional agreement between FUV and H$\alpha$ filaments.
 
\begin{figure*}
\begin{center}
\includegraphics[angle=0,width=17.0cm]{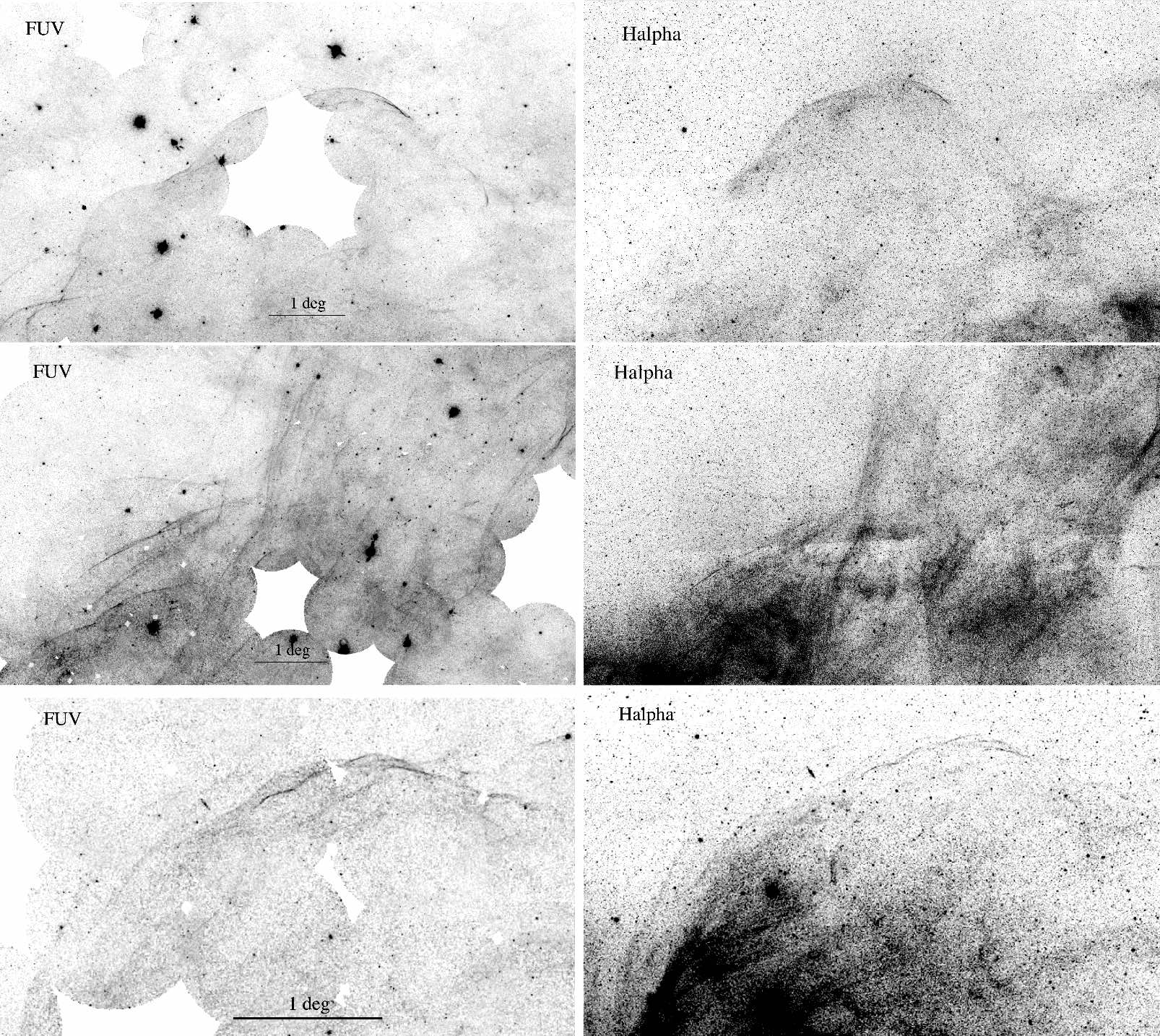}
\caption{Comparison of GALEX FUV emission vs.\ MDW H$\alpha$ image mosaic of the northern, eastern, and southeastern portions of the Antlia SNR showing a close agreement of UV and H$\alpha$ emission filaments along its eastern boundary. 
There is considerable H$\alpha$ emission but little FUV 
 emission in the nebula's interior. North is up, East to the left. \label{G3_FUV_vs_MDW}
}
\end{center}
\end{figure*}

 To illustrate this agreement
 Figure~\ref{G3_FUV_vs_MDW} shows a comparison of GALEX FUV and MDW survey H$\alpha$ images for three sections along the Antlia's northeastern boundary. While we find there is a close correlation of the outer FUV filaments with many H$\alpha$ filaments, there are also considerable differences in terms of the nebula's internal emissions. That is, far more H$\alpha$ emission is seen `behind', i.e., to the west, of the sharp, outlying filaments than seen in the FUV images. This is most striking in the lower panel images of Figure~\ref{G3_SHASSA} for the remnant's southeastern limb region. 
 While the SHASSA image shows considerable internal diffuse and filamentary emission, the brightest H$\alpha$ emissions are found along Antlia's northwestern, western and southeast limbs.
 
%%%%%%%%%%%%%%%%%%%%%%%%%%%%%%%%%%%%%%%%%%%%%%%%%%%%%%%%%%%%%%%%%%%
%%% Figure 16  G3: SHASSA image and the North Region in MDW images
%%%%%%%
%%%%%%%%%%%%%%%%%%%%%%%%%%%%%%%%%%%%%%%%%%%%%%%%%%%%%%%%%%%%%%%%%%%%%

 %%%%%%%%%%%%%%%%%%%%%%%%%%%%%%%%%%%%%%%%%%%%%%%%%%%%%%%%%%%%%%%%%%%
%%% Figure 17:  G3:  FUV and MDW images: North Region
%%%%%%%
%%%%%%%%%%%%%%%%%%%%%%%%%%%%%%%%%%%%%%%%%%%%%%%%%%%%%%%%%%%%%%%%%%%%%%
%%%%%%%%%%%%%%%%%%%%%%%%%%%%%%%%%%%%%%%%%%%%%%%%%%%%%%%%%%%%%%%%%%
%%% Figure 18:   G3:  SALT slit positions and spectra
\begin{figure*}
\begin{center}
\includegraphics[angle=0,width=16.0cm]{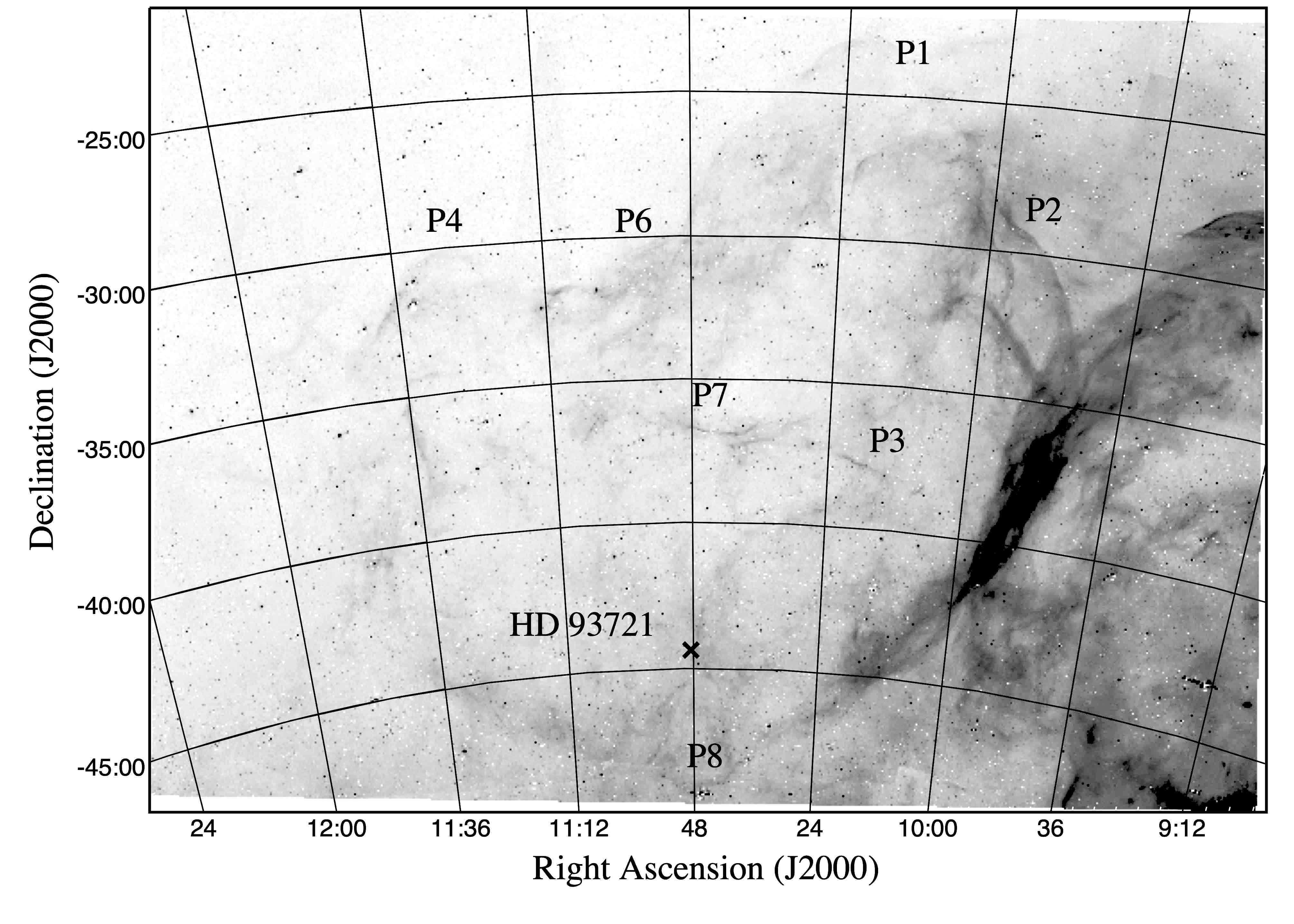}
\includegraphics[angle=0,width=0.49\textwidth]{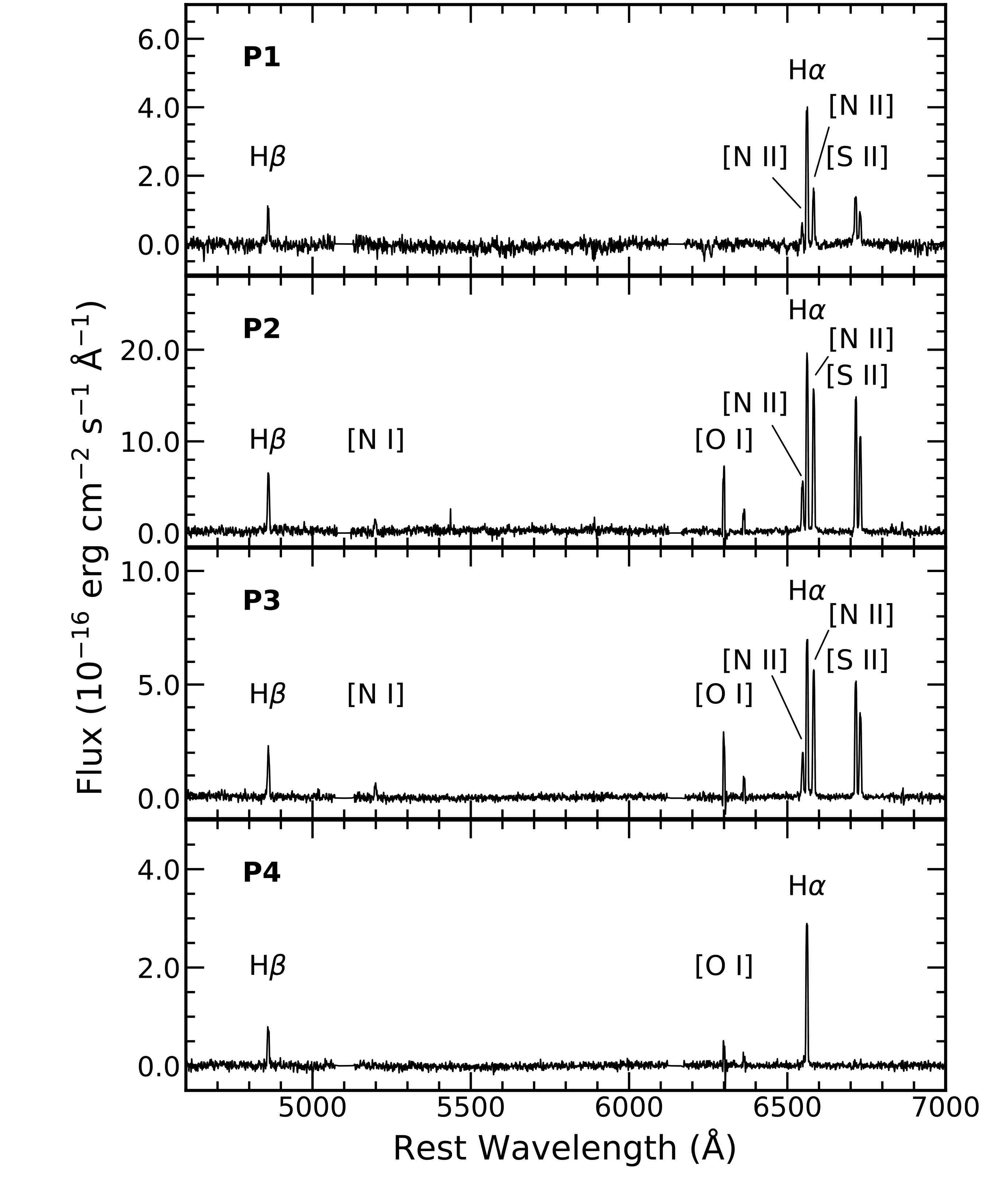}
\includegraphics[angle=0,width=0.49\textwidth]{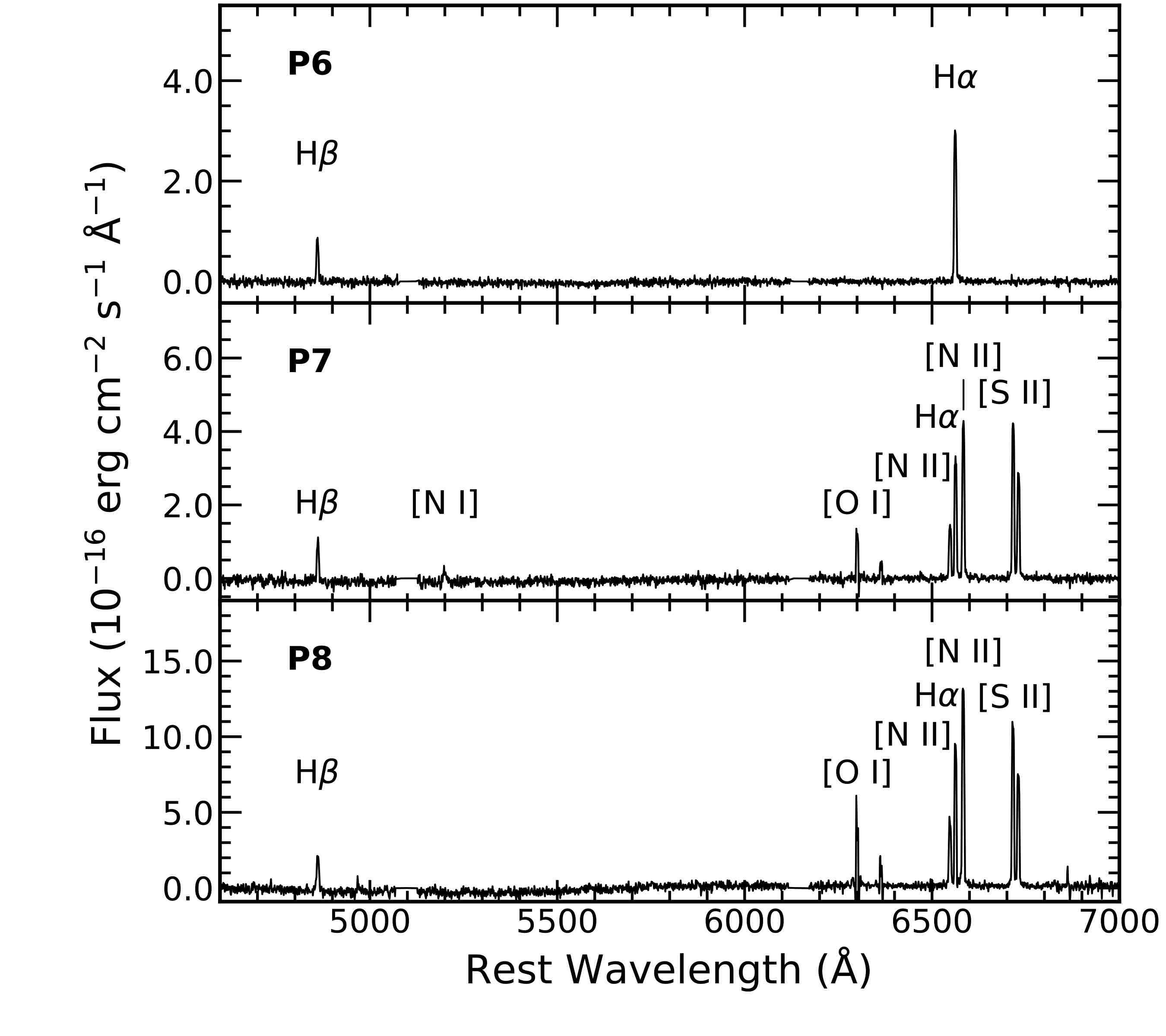}
\caption{{\it{Top:}} Continuum subtracted SHASSA H$\alpha$ image mosaic of the Antlia remnant showing the locations of seven positions where SALT spectra were obtained.
{\it {Bottom:}} Observed optical spectra at these seven positions.
\label{G3_slits_spectra}
}
\end{center}

\end{figure*}
%%%%%%%%%%%%%%%%%%%%%%%%%%%%%%%%%%%%%%%%%%%%%%%%%%%%%%%%%%%%%%%%%%%%%%

%%%%%%%%%%%%%%%%%%%%%%%%%%%%%%%%%%%%%%%%%%%%%%%%%%%%%%%%%%%%%%%%%%%%%%
%%% Antlia line fluxes
%\startlongtable
\begin{deluxetable*}{lccccccc}[htp]
\tiny
\tablecolumns{8}
\tablecaption{Observed Emission Line Fluxes for Antlia Filaments }
\tablehead{\colhead{Emission Line} & 
\multicolumn{7}{c}{\underline{~~~~~~~~~~~~~~~~~~~~~~~~~~~~~~Filament Position ~~~~~~~~~~~~~~~~~~~~~~~~~~~~~~~~~~~~~~~}} \\ 
\colhead{(\AA)} & \colhead{P1} & \colhead{P2}  & \colhead{P3} & \colhead{P4} & \colhead{P6} & \colhead{P7} & \colhead{P8} }
%\colhead{(\AA)} & \colhead{F($\lambda$)} & \colhead{F($\lambda$)} & \colhead{F($\lambda$)} &  \colhead{F($\lambda$)} & \colhead{F($\lambda$)} & \colhead{F($\lambda$)} & \colhead{F($\lambda$)}  } 
\startdata
  H$\beta$    4861                   & 100     &     100 &   100   & 100     &  100      &  100   & 100     \\
 $[$\ion{O}{3}$]$ 5007               & $< 10~~ $  &  $< 10~~ $ &  $< 10~~$ &  $< 10~~$ & $< 10~~$    & $< 10~~$ & $< 10~~ $  \\
 $[$\ion{N}{1}$]$ 5200               & \nodata &     23  &   29    &  \nodata & \nodata  &     20 & \nodata \\
 $[$\ion{O}{1}$]$ 6300                & \nodata &     93  &   98    &  \nodata &  \nodata &    101 &   122    \\
 $[$\ion{N}{2}$]$ 6548               &  57     &     89  &    97   &  \nodata &  \nodata &    137 &   181   \\
  H$\alpha$   6563                   &  430    &    300  &   339   &    379   &   368    &    304 &   365   \\
 $[$\ion{N}{2}$]$ 6583               &  167    &    247  &   275   &  \nodata &  \nodata &    403 &   542   \\
 $[$\ion{S}{2}$]$ 6716               &  146    &    239  &   247   &  \nodata &  \nodata &    406 &   442   \\
 $[$\ion{S}{2}$]$ 6731               &  91     &    167  &   184   &  \nodata &  \nodata &    283 &   307   \\
 F($[$\ion{S}{2}$]$)/F(H$\alpha$)    & 0.55    &   1.35  &  1.27   &  \nodata &  \nodata &  2.27  &  2.05    \\
 $[$\ion{S}{2}$]$ 6716/6731          &  1.60   &   1.43  &  1.34   &  \nodata &  \nodata &   1.43 &   1.44  \\
 $E(B-V)$                            &  0.35   &   0.00  &  0.12   &  0.23    &    0.20  &   0.01  &  0.19  \\
 F(H$\alpha$) erg cm$^{-2}$ s$^{-1}$ & 2.74E-15& 1.29E-14& 4.75E-15& 1.91E-15 & 2.06E-15 &2.16E-15& 6.06E-15  \\
\enddata
\label{Table_2}
\end{deluxetable*}
%%%%%%%%%%%%%%%%%%%%%%%%%%%%%%%%%%%%%%%%%%%%%%%%%%%%%%%%%%%%%%%%%%

\subsubsection{Optical Spectra}

If \citet{McCull2002} is correct in their assessment that the Antlia H$\alpha$ nebula is a SNR, as seemingly supported by the GALEX FUV and our H$\alpha$ images described above, the remnant would then be both many times larger than any confirmed SNR and be located at an unusually high Galactic latitude.  On the other hand, considering its size and  location immediately adjacent to the enormous Gum Nebula with its extended emission, outer `filamentary wisps', and numerous H~I wind blown bubbles \citep{Brandt1976d,Chanot1983,Purcell2015}, there is some uncertainty about its true origin.

Optical spectra are a powerful tool for confirming the presence of shock emission and hence can be used to investigate the SNR origin of optical nebulosities.
Consequently, we obtained long-slit, low-dispersion spectra of seven filaments and emission clumps distributed across the whole of the Antlia nebula to explore the nature of its optical emissions.

Figure~\ref{G3_slits_spectra} shows the location of the seven filaments observed in the Antlia remnant, along with the resulting spectra. Table 2 lists the relative emission line strengths uncorrected for extinction. Listed relative line strengths are believed accurate to 10\%.
Based on the observed H$\alpha$/H$\beta$ ratios of 3.0 to 4.30, we find a variable degree of extinction across the Antlia nebula, 
with $E(B - V)$  values ranging from 0.0 to 0.35
assuming an intrinsic ratio of 3.0 and an R value of 3.1. Here we have chosen an H$\alpha$/H$\beta$ value greater than the
theoretical value of 2.87 for 10$^{4}$ K due to the likelihood of significant
collisional excitation of the n = 3 level at postshock temperatures seen in SNRs. Such a range of extinction values is not unexpected in view of the nebula's large dimensions.

These optical spectra show clear evidence that the seven filaments observed distributed across the whole of the Antlia nebula exhibit line ratios indicative of shock emissions. In fact, emissions at Positions 4 and 6, located along Antlia's southeastern limb display non-radiative, pure Balmer line emission.\footnote{Note: The [\ion{O}{1}] emission seen in the spectrum for Position 4 is residual imperfect [\ion{O}{1}]  sky emission.} Such shock spectra are usually seen in situations where the shock velocities are quite high ($>$ 500 km s$^{-1}$) as seen in young Type Ia SNRs like Tycho and SN~1006. However, similar Balmer dominated spectra has also been seen in much older remnants (G107.+9.0; \citealt{Fesen2020}).  Relatively slow shocks can remain non-radiative if the ambient density is sufficiently low, like that expected in the Galactic halo.  This requires that the cooling time of the shocked gas should be long enough to prevent the forbidden lines from becoming bright.  Scaling the cooling times of \cite{Hartigan1987} with $n_0^{-1}$, the non-radiative shocks at the edge of a 100,000 year old remnant in a density of 0.1 $\rm cm^{-3}$ would be at least 170 \kms.

The other five filaments show optical spectra commonly seen in evolved SNRs like the Cygnus Loop and IC~443. In all these cases,
the I([\ion{S}{2}])/I(H$\alpha$) ratio exceeds the 0.4 criterion for shock emission. Indeed, filaments 7 and 8 display ratios above 2, which are among the highest values observed in SNRs. Interestingly, these five filaments lack appreciable [\ion{O}{3}] $\lambda\lambda$4959,5007 emission. Weak or absent  [\ion{O}{3}] emission indicates shock velocities less than about 70 km s$^{-1}$. The high H$\alpha$/H$\beta$ ratio in Position 1 of the Antlia remnant suggests a substantial contribution of collisional excitation at fairly low temperatures to the Balmer lines, and this is borne out by the low [N II] and [S II] to H$\alpha$ ratios. That suggests shock speeds around 70 km s$^{-1}$.  
Given these slow shock spectra along with pure Balmer emission spectra along the remnant's leading edges, there must be a considerable range of shock velocities present in the nebula. This is not too surprising given the large size of Antlia and a nearly 18\degr \ range in Galactic latitude. 

%%%%%%%%%%%%%%%%%%%%%%%%%%%%%%%%%%%%%%%%%%%%%%%%%%%%%%%%%%%
%%%%%%%%%%%%%%%%%%%   DISCUSSION  %%%%%%%%%%%%%%%%%%%%% 
%%%%%%%%%%%%%%%%%%%%%%%%%%%%%%%%%%%%%%%%%%%%%%%%%%%%%%%%%%%

\section{Discussion}

Based on the data presented above, 
we find that two suspected, high latitude SNRs, namely G354-33 and Antlia, are likely bona fide SNRs.  Moreover, their UV emission filaments appear best at marking these objects' forward shock front locations. 
Below we discuss supporting evidence for our conclusions, followed by estimates regarding their physical dimensions and general properties.
We then briefly comment of the usefulness of far UV imaging for SNR detection particularly in areas far off the Galactic plane.

\subsection{G354.0-33.5}

There are several observations that support the
SNR identification of this large FUV, H$\alpha$, and radio emission nebula. At its Galactic latitude of $-33.5\degr$, there are no bright and nearby early type stars near the center of this UV shell that might have generated its UV emission shell through stellar winds. There is also no known nearby recurrent nova projected inside, and the emission shell is at least two orders of magnitude larger and does not have a similar appearance to known nova shells.

Then there is the large 1420 MHz radio emission shell roughly centered on and lying entirely within the borders of the FUV emission filaments. The FUV filaments also exhibit a shock-like morphology such as commonly seen in Galactic SNRs.
Furthermore, its nearly continuous ring of thin filaments surrounds a broad, diffuse H$\alpha$ emission shell. In summary, its shock-like filamentary appearance and the positional agreement between UV, H$\alpha$ and radio emissions for such a large object located far off the Galactic plane leaves few viable options other than a SNR origin.

While we have not yet obtained optical spectra of its H$\alpha$ and UV filaments, based on 
the morphology of the H$\alpha$ emission filament along its northwestern limb and the lack of [\ion{O}{3}] emission, we suspect this remnant may resemble that of nonradiative filaments seen in a few remnants such as those that surround the northern and eastern limb of the Cygnus Loop
\cite{Fesen1982,Raymond1983,Hester1994,Sankrit2000,Medina2014}.
A nonradiative or `Balmer dominated' shock is one that heats plasma to a relatively high temperature and has encountered the plasma either so recently or is of such a low density that the gas has had insufficient time to radiatively cool and present optical line emissions. Such filaments are UV bright \citep{Shull1979,Raymond2003} a property consistent with GALEX FUV images for this object. Although the confirming test for the presence of nonradiative filaments in this remnant is spectra, the deep H$\alpha$ and
[\ion{O}{3}] images presented in Figure~\ref{G1_Chile} supports this conclusion.

Centered at a Galactic latitude around $-33.5\degr$, G354-33 would rank as the highest Galactic latitude SNR found yet. In addition, if not for the Antlia remnant, this object would also be among the largest angular dimensions among the roughly 300 SNRs in the \citet{Green2019} catalogue. Because of its $\sim11\degr \times 14\degr$ size and the fact that the largest known SNRs have physical dimensions $\sim$100 pc, we can make some crude estimates of its distance and properties.

If its diameter is taken to be 100 pc, then its distance is around 400 pc. Based on the presence of its stronger FUV than NUV emission, its shock velocity is likely greater than 100 km s$^{-1}$ but probably less than 150 km s$^{-1}$, according to Figure 4 of \citet{Bracco2020}. That shock speed is too high for this remnant to be in the snowplow phase of SNR evolution, but the presence of radiative shock waves around much of the rim indicates that it is no longer in the Sedov phase, at least in those places.  That suggests that it is in the pressure-driven shell phase, when the recently shocked gas has cooled, but there is still relatively hot, high-pressure gas in the interior. 

From the calculations of \cite{Cioffi1988}, this velocity range and a diameter of around 90 pc would be consistent with an age $\simeq$ 10$^{5}$ years and a preshock density of 0.1 $\rm cm^{-3}$ (see Fig.\ 14 of \citealt{Fesen2020}).  One might expect detectable X-ray emission during the pressure-driven shell phase, but \citet{Shelton1998} has discussed halo SNRs in which the interior gas is too cool to produce X-rays, but still rich in high ionization states such as \ion{O}{6}.

\subsection{G249.7+24.7}

GALEX FUV, SHASSA, and MDM images show it displays
a highly filamentary morphology like that commonly seen in
SNRs. Moreover, optical spectra of its filaments reveal
emission-line ratios consistent with the presence of shocks.
The discovery of its X-ray emission along with associated radio
emission and archival Parks radio data \citep{Becker2021}
leave little doubt that this object is a new SNR. 
With an angular dimension of 4.5\degr, the G249+24 remnant is
among the largest of known SNRs. Situated at Galactic latitude of over 24 degrees, G249+24 also lies farther from the Milky Way's plane than any other confirmed SNR -- that is, if it were not for G354-33. 

The detection of high-velocity \ion{Na}{1} absorptions in the
spectrum of the star HD~83636 which has a Gaia EDR3 estimated 
distance of 386 pc, sets a maximum limit on G249+24's distance
of less than 400 pc. This is a bit less than the lower distance range of 400--500 pc estimated by \citet{Becker2021}.
If high-velocity \ion{Na}{1} absorption lines are confimed by subsequent higher resolution echelle spectra, then the G249+24 remnant would then also rank among the closest SNRs known and with an especially robust distance limit.

The $\simeq$400 pc upper limit on the distance implies an upper limit on its diameter of about 30 pc.  It also provides an estimate of the ambient density.  \citet{Fesen2020} used the separation between the sharp H$\alpha$ filaments of G107.0+9.0 and the diffuse [\ion{O}{3}] emission to determine that remnant's
preshock density.  The angular separation between the H$\alpha$ and [\ion{O}{3}] seen in Figure~\ref{G2_pos1} is similar to that seen in G107.0+9.0,
but here the distance upper limit is 2.5 times smaller than the 1 kpc distance adopted for G107.0+9.0.  Therefore, the preshock density is at least 0.25 cm$^{-3}$.  

Because the emission from G249+24 is so faint, the density cannot be much higher than 0.25 cm$^{-3}$, indicating that the distance is close to the upper limit.  This is in accord with its morphology, which suggests that G247+24 is significantly above the Galactic disk and interacting with it on its southern side, indicating that it is more than 100 pc from the plane, and therefore more than 250 pc distant.  

The pure Balmer line emission at P1 indicates the that remnant is still in the Sedov phase there, while the emission and absorption spectra indicate shock speeds $\sim$60 - 100 km s$^{-1}$ in most of the rest of the remnant. The faint X-ray emission and the 0.1 keV temperature found by \citet{Becker2021} also indicate that the remnant has left the Sedov phase and entered the pressure-driven shell phase.  Assuming that the cooling occurred recently, when the shock speed fell below 400 \kms, we can combine the Sedov equations for SNR radius and shock speed to estimate and age around 15,000 years for a $10^{51}$ erg explosion.  Thus the parameters of G247+24 are similar to those of the Cygnus Loop \citep{Fesen2018} but is much fainter because it lacks the dense clouds that make parts of the Cygnus Loop bright at optical and X-ray wavelengths. 

\subsection{The Nature of the Antlia Nebula}

\citet{McCull2002} claimed this extraordinarily large emission nebula was a likely SNR based on its appearance on the deep H$\alpha$ image of the VTSS survey and on the presence of diffuse soft X-ray emission in its interior. However, this conclusion does not appear to be widely accepted, as measured by the remnant having attracted little subsequent attention. 

This situation might in part be due to a reluctance by SNR researchers to accept its huge $20\degr \times 26\degr$ angular size, more than 3-5 times larger than the largest known confirmed SNRs, plus its location so close to the even larger 36\degr diameter Gum Nebula with its complex of large emission shells and wind-blown bubbles plus the embedded Vela SNR. 
Consequently, except for a far UV study by \citet{Shinn2007} and a brief AAS abstract by \citet{Orchard2015}, the Antlia remnant has not be studied in any detail, leaving open its  properties and nature.

Our GALEX FUV mosaics show a well-defined shell in H$\alpha$ with many individual and overlapping filaments that strongly resemble ISM shocks. In addition, the locations of sharp UV filaments along the boundary of the object's H$\alpha$ emission are consistent with a SNR where such UV filaments mark the location of the remnant's shock front.

In addition, results of our seven optical spectra of Antlia's filaments clearlu indicate shock emissions throughout the nebula and hence a SNR. These spectra include two textbook cases of non-radiative Balmer dominated spectra (Positions 4 \& 6), plus several other filaments exhibiting high [\ion{S}{2}]/H$\alpha$ line ratios well above the 0.4 value distinguishing shocked from photoionized nebulae. We conclude that if the Antlia remnant was not so large, it would be an easy case for supernova remnant classification. 

%%%%%%%%%%%%%%%%%%%%%%%%%%%%%%%%%%%%%%%%%%%%%%%%%%%%%
%%% Figure 19:  VTTS and SHASSA images of G3/Antlia
%%%%%%%%%%%%%%%%%%%%%%%%%%%%%%%%%%%%%%%%%%%%%%%%%%%%%%
\begin{figure*}[hbp]
\begin{center}
\includegraphics[angle=0,width=8.5cm]{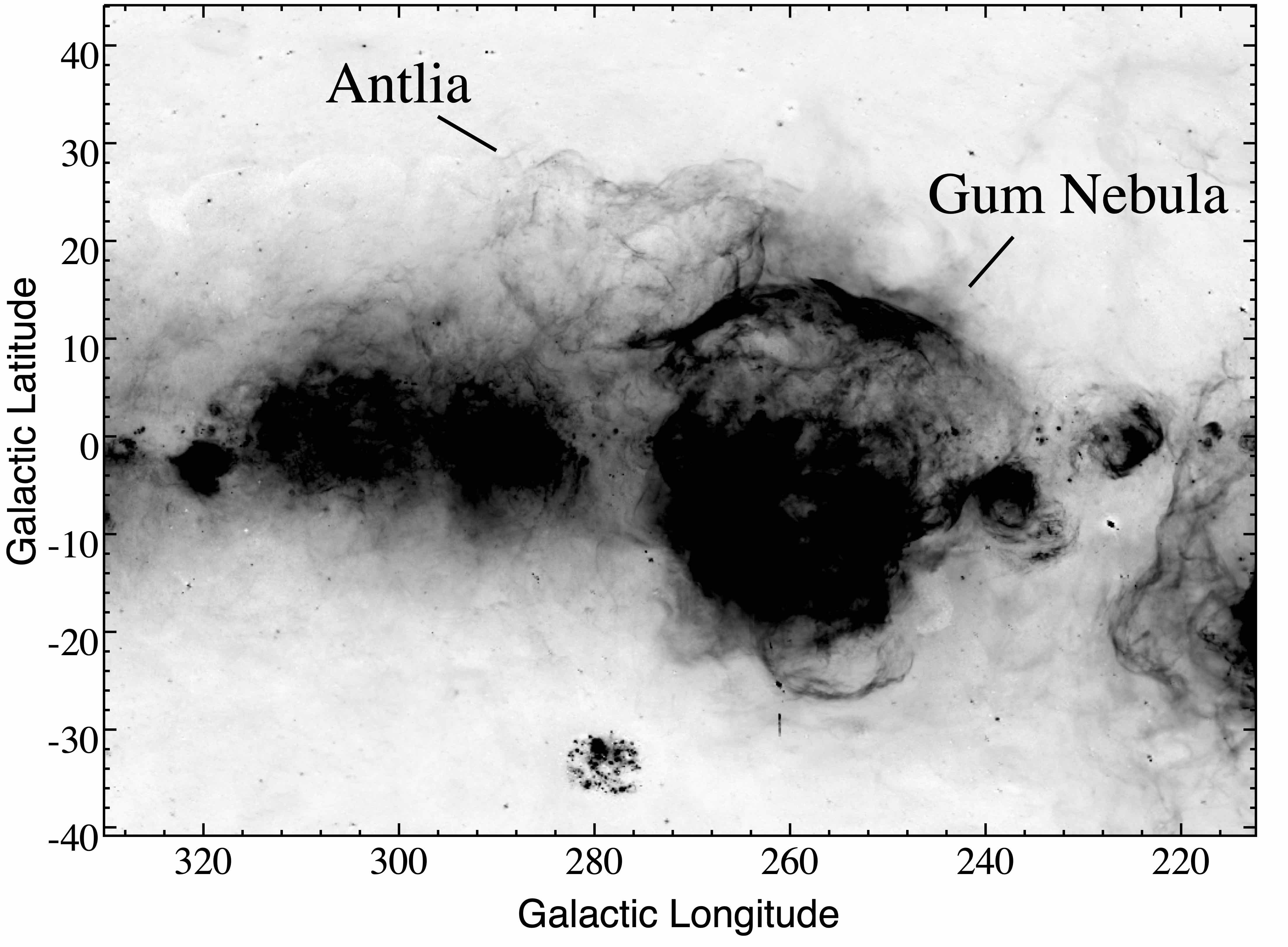}
\includegraphics[angle=0,width=8.0cm]{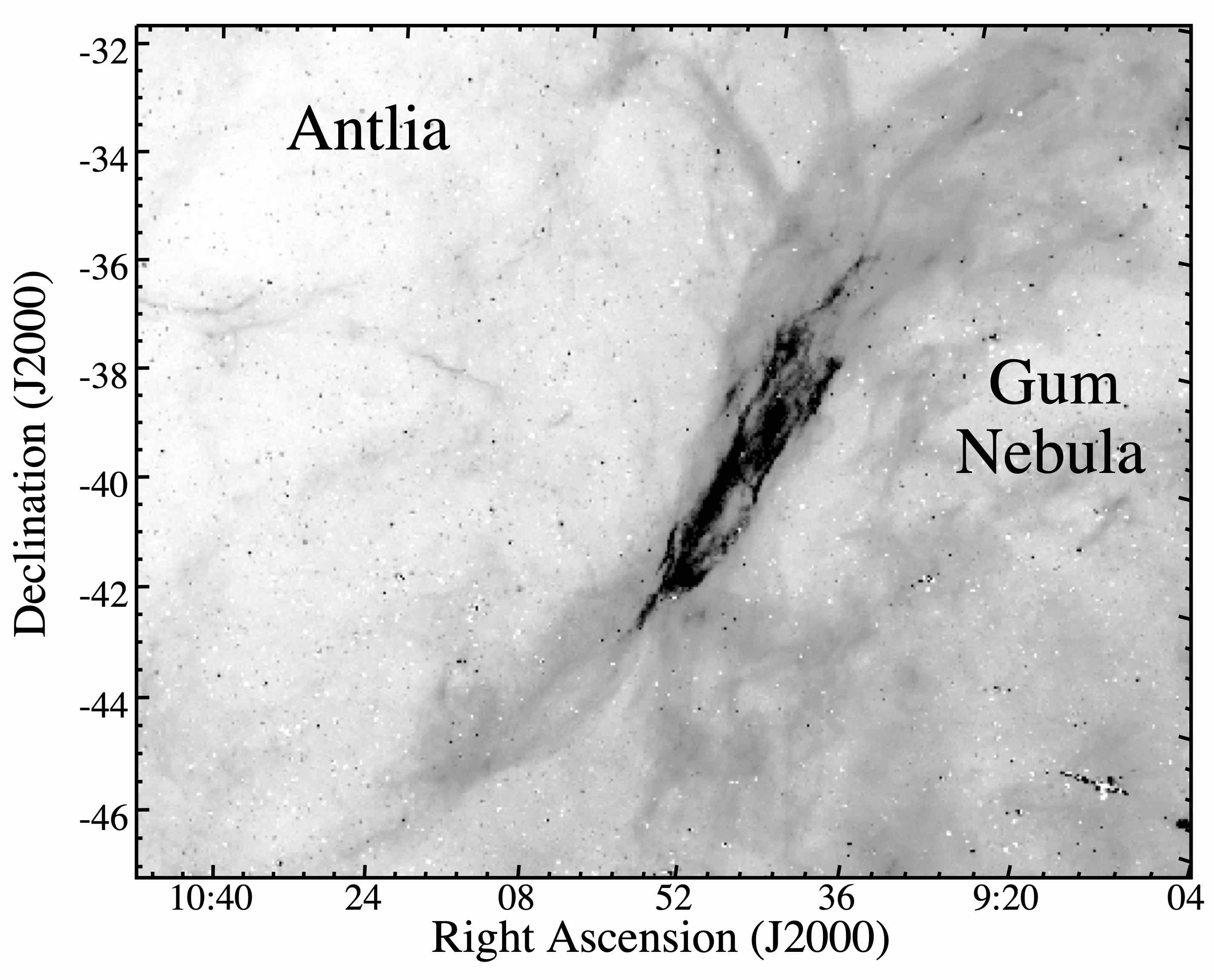}
\caption{$Left:$ VTSS H$\alpha$ image of the Galactic plane around the Gum Nebula with the location of the Antlia SNR marked. $Right:$  The continuum subtracted SHASSA H$\alpha$ image of the Antlia remnant showing the presence of bright filaments at the overlapping Antlia and Gum Nebula region suggestive of a physical interaction.
\label{Gum}
}
\end{center}
\end{figure*}
%%%%%%%%%%%%%%%%%%%%%%%%%%%%%%%%%%%%%%%%%%%%%%%%%%%%%%%%%%%
%%%%%%%%%%%%%%%%%%%%%%%%%%%%%%%%%%%%%%%%%%%%%%%%%%%%%
%%% Figure  20: VTTS and SHASSA images of G3/Antlia
%%%%%%%%%%%%%%%%%%%%%%%%%%%%%%%%%%%%%%%%%%%%%%%%%%%%%%
\begin{figure*}[ht]
\begin{center}
\includegraphics[angle=0,width=18.0cm]{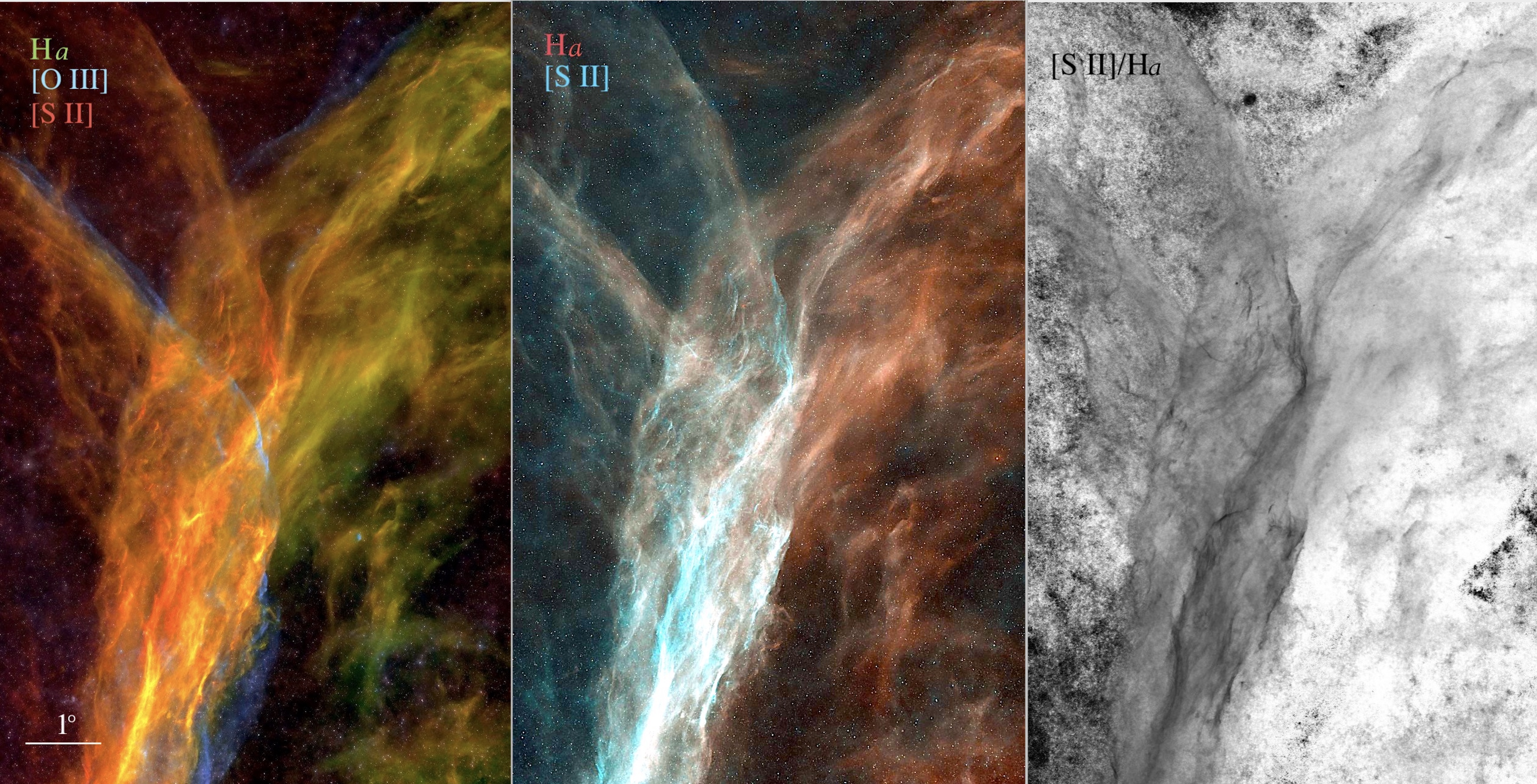}
\caption{Wide-field images of the northwestern Antlia--Gum interaction 
(approximate center: 9$^{\rm h}$ 42$^{\rm m}$, $-35\degr$ 47$'$).
$Left:$ Composite of H$\alpha$, [\ion{O}{3}], and [\ion{S}{2} images. 
$Center:$ Composite of H$\alpha$ and [\ion{S}{2} images.
$Right:$ Ratio of H$\alpha$ and [\ion{S}{2} images.
\label{Ludgate}
}
\end{center}
\end{figure*}
%%%%%%%%%%%%%%%%%%%%%%%%%%%%%%%%%%%%%%%%%%%%%%%%%%%%%%%%%%%

If the FUV and optical images and spectra presented here were not enough to confirm the Antlia nebula as a SNR, the recently released eRosita All-Sky soft X-ray 0.3--2.3 keV image reveals a large emission shell at the site of the Antlia remnant \citep{Predehl2020,Becker2021}. Both the size and morphology of this X-ray bright limb shell matches that of the remnant's H$\alpha$ as shown in Figure~\ref{G3_SHASSA}. Although a full analysis of Antlia's X-ray emission properties is needed to assess its properties, the positional and morphological similaritiy of this X-ray shell to its FUV and H$\alpha$ emissions are strong additional indicators for a SNR classification.

The physical size of the Antlia SNR is uncertain due to its unknown distance. \citet{McCull2002} estimated a wide range of possible distances, from 60 pc to 320 pc, and believed the remnant to be extremely old, at least 1 Myr, and hence in the final snowplow phase of SNR evolution. 
However, our optical spectra do not support such an old object or a distance less than 200 pc.

A 1 Myr old SNR is expected to have a a very low expansion velocity of around 10 to 20 km s$^{-1}$ \citep{Chevalier1977}.
In contrast, our optical spectra show line emissions indicating shock velocities between 50 and 150 km s$^{-1}$, making it far younger ($< 10^{5}$ yr) remnant in the pressure-driven shell phase. 
Moreover, there are other data supporting a much younger remnant with relatively high-velocity gas inside.

Before the Antlia nebula was discovered, \citet{Bajaja1989} reported  high-velocity clouds in the  remnant's direction, with
\citet{Penprase1992} reporting high-velocity \ion{Ca}{2} absorption lines in the spectrum of the B9/A0 III/IV star HD 93721 which lies in the direction of the Antlia SNR (see Fig.\ 15). This star has a estimated Gaia Early Data Release 3 (EDR3) distance of $512\pm7$ pc. \citet{Penprase1992} found  at least 10 \ion{Ca}{2} absorption components with v$_{\rm LSR}$ ranging from $-65$ km s$^{-1}$ to $+75$ km s$^{-1}$. Because a second star, HD 94724 (d = 210 pc) lying in the same general direction did not show any high-velocity components, they concluded that the high-velocity absorbing cloud's distance was between 200 and 500 pc. However, because not all stars behind a SNR display high-velocity absorption lines, a lack of high-velocity components does not provide a robust minimum distance estimate.

%%%%%%%%%%%%%%%%%%%%%%%%%%%%%%%%%%%%%%%%%%%%%%%%%%%%%%
%%% Figure  G70 FUV and Halpha comparsion
%%%%%%%%%%%%%%%%%%%%%%%%%%%%%%%%%%%%%%%%%%%%%%%%%%%%%%
\begin{figure*}
\begin{center}
\includegraphics[angle=0,width=8.0cm]{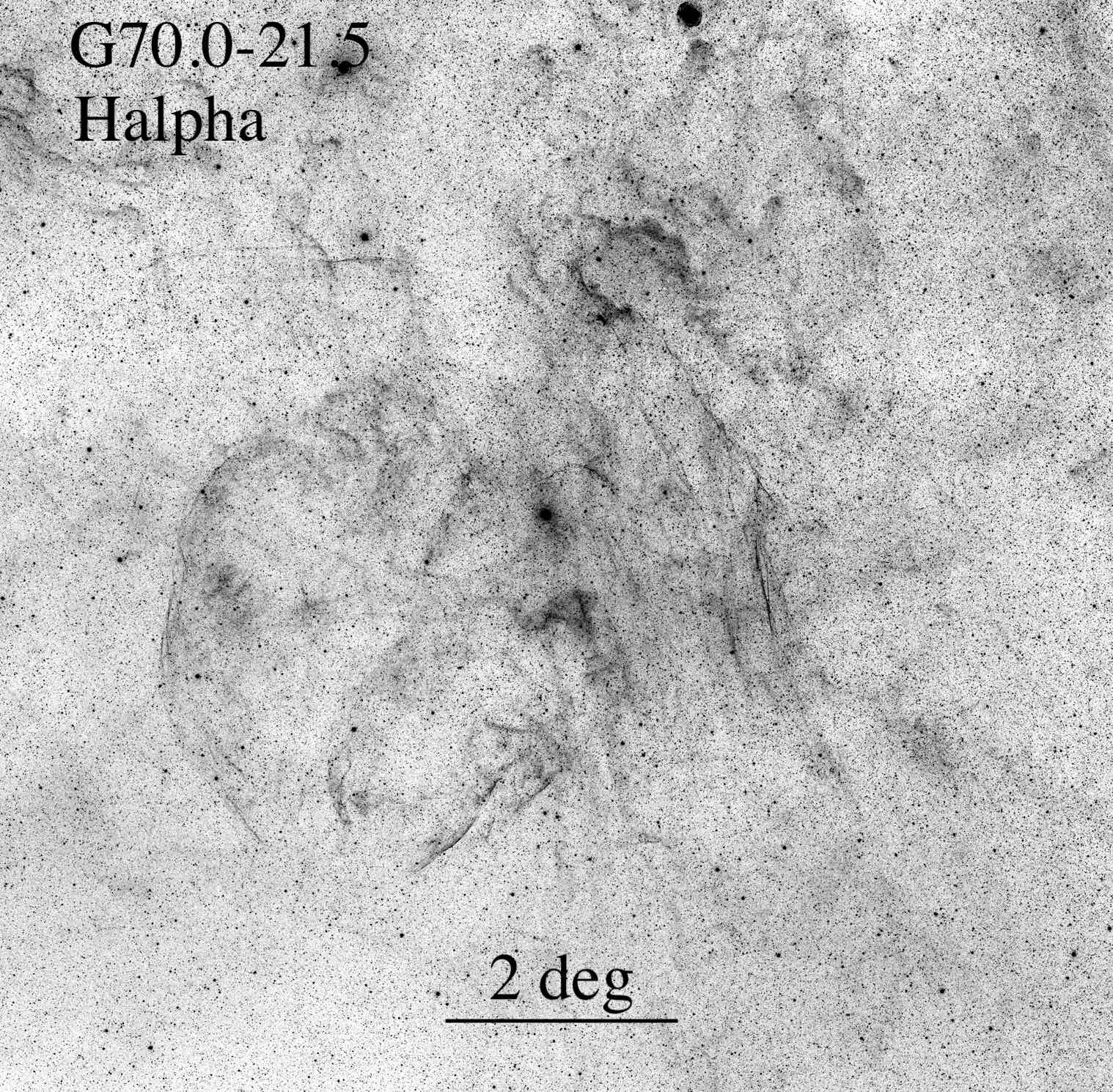}
\includegraphics[angle=0,width=8.0cm]{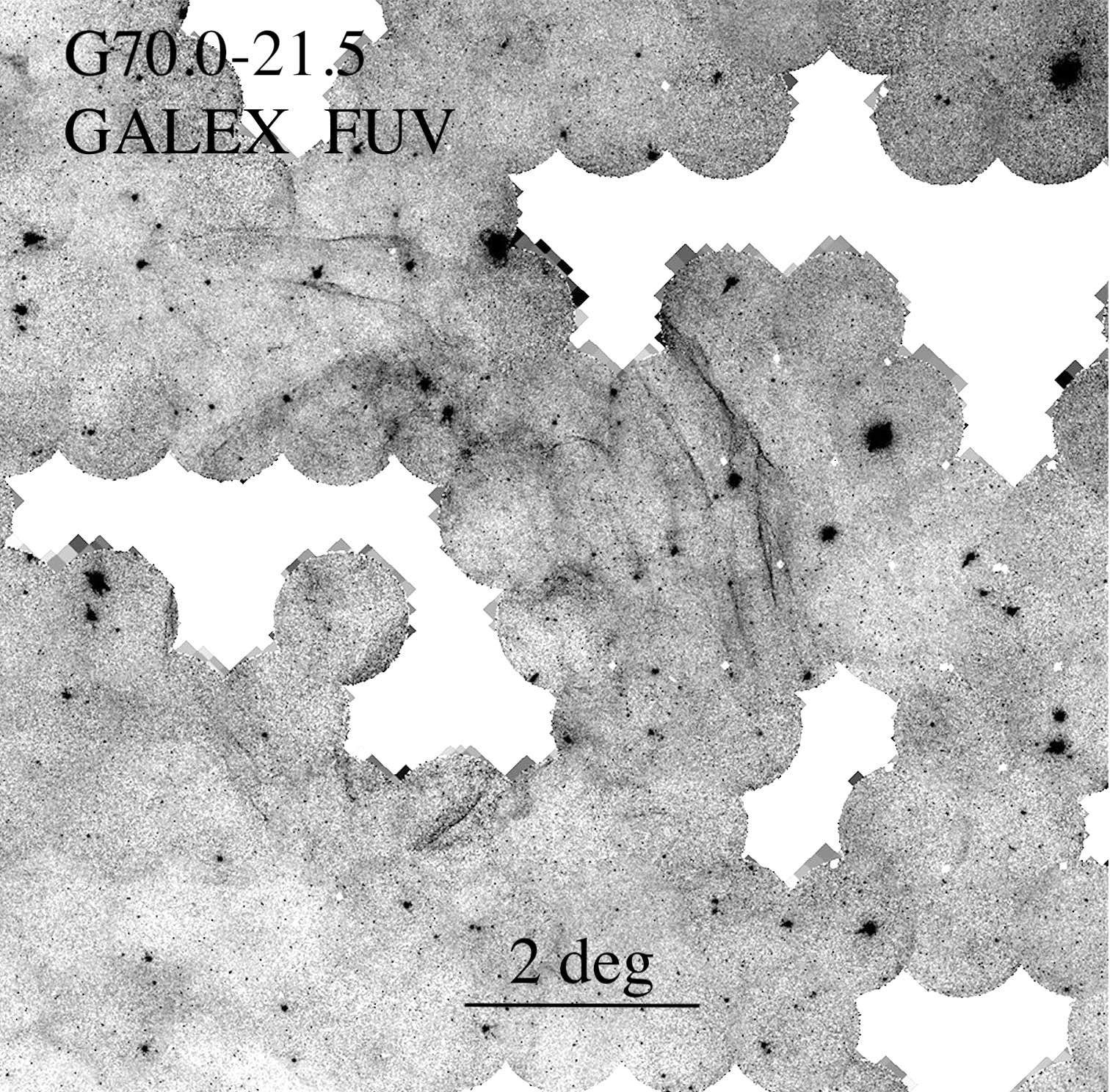}
\caption{Comparison of MDW H$\alpha$ and GALEX FUV images of the high-latitude remnant G70.0-21.5. 
North is up, East to the left.
\label{G70}
}
\end{center}
\end{figure*}
%%%%%%%%%%%%%%%%%%%%%%%%%%%%%%%%%%%%%%%%%%%%%%%%%%%%%%%%%%%

An alternative means of estimating the distance to the Antlia remnant is its apparent collision with the Gum Nebula \citep{Gum1952,Gum1955,Brandt1971,Sivan1974,Chanot1983}. 
Although the positional coincidence or abutment of the remnant's southwestern rim with the Gum Nebula's bright northeastern emission cloud has been noted in some studies of the Gum Nebula \citep{Shinn2007,Iacob2014,Purcell2015}, no proposal has been made about there being evidence for a physical interaction between the two nebulae.

Nonetheless, the SHASSA imaging of this region strongly indicates such a physical collision has occurred between the Antlia remnant and the Gum Nebula. The left panel of Figure~\ref{Gum} presents a section of the VTSS H$\alpha$ image of the Galactic plane showing the Antlia emission shell, similar to the VTSS image presented in the \citet{McCull2002} Antlia discovery paper. This shows Antlia's position relative to the Gum Nebula. The remnant's bright southern limb coincides with the northern rim of the Gum Nebula. 

The right panel of Figure~\ref{Gum} shows a low contrast version our SHASSA image of Antlia.  An unusual line of bright filaments nearly 10\degr in length can be seen in the projected overlap region of both shells (see also top panel of Fig.~\ref{G3_slits_spectra}). The simplest explanation for such bright filaments situated only in the exact Antlia/Gum overlap region is that the Antlia remnant has collided with the outer northeastern rim of the Gum Nebula.

Support for an Antlia--Gum collision is shown in Figure~\ref{Ludgate} where wide-field 
H$\alpha$ (+ \ion{N}{2}]), [\ion{O}{3}], and [\ion{S}{2}] images are presented for the northern section of the Antlia--Gum Nebula overlap region. The left panel is a color composite image revealing [\ion{O}{3}] bright emission filaments out ahead of the [\ion{S}{2}] bright emission filaments marking the location
of a local shock front. The center panel highlights the much stronger [\ion{S}{2}] emission filaments compared to the Gum Nebula's more diffuse emission. Because ISM shocks display stronger [\ion{S}{2}] compared to that
of H$\alpha$ as demonstrated in the Antlia spectra (see Table 4), the
observed strong [\ion{S}{2}] emission in the proposed collision filaments leads
support for a collision scenario. This conclusion is further indicated by the right panel which shows the observed ratio of F([\ion{S}{2}])/F(H$\alpha$ + [\ion{N}{2}]) with the stars removed. The sharp dark filaments in this panel image represent ratios between 1 and 1.3, values consistent with the measured spectral line strengths of Antlia's shock filaments (see Table 4).

If the Antlia remnant's expansion has led to a collision of it with an
outer portion of the Gum Nebula, then the Antlia's distance must be roughly the same as for the Gum Nebula. Unfortunately, the distance to the Gum Nebula is poorly known, with values ranging from $200 - 500$ pc \citep{Brandt1971,Zealey1983,Graham1986,Woer2001,Kim2005,Howarth2019}. 
Nonetheless, distances much below 200 pc like suggested by \citet{McCull2002} would appear to be ruled out, as would the suggestion by \citet{Tetzlaff2013} that the 1.2 Myr old PSR J0630-2834 was formed in the Antlia remnant at a distance $\approx$135 pc.  

If we adopt recent distances estimate to the Gum of 300 - 450 pc, which include estimates for some of the Gum Nebula's ionizing stars ($\zeta$ Pup; \citealt{Howarth2019}), then the Antlia remnant's physical diameter is $\sim120 - 180$ pc. However, given the Gum Nebula's 36\degr \ diameter, its outer parts extend over a distance of some 200 to 300 pc along the line of sight, this greatly complicates using a collision scenario to constrain the Antlia remnant's distance.
 
\subsection{The Power of FUV Emissions for Finding Interstellar Shocks}

A few large Galactic SNRs  and superbubbles have been recently studied in terms of their far UV emissions. For example, studies of far UV emissions have been reported for the Vela SNR \citep{Nish2006}, the Cygnus Loop \citep{Seon2006}, the Lupus Loop \citep{Shinn2006}, and the Orion-Eridanus Superbubble \citep{Kregenow2006}.  Many of these made use of the SPEAR imaging spectrograph \citep{Edel2006}.

However, there have been few papers reporting discoveries of large interstellar emission structures using FUV emissions. One such paper is that of \citet{Bracco2020} who reported finding a 30\degr \ long UV arc in Ursa Major using GALEX images.  That work was a follow-up to an earlier detection of a much shorter 2.5\degr \ filament by \citet{McCull2001} using deep H$\alpha$ imaging. However, only through the GALEX's FUV images was the full extent of this faint, long interstellar filament finally revealed.

The FUV emission of moderate velocity shocks is dominated by the hydrogen 2-photon continuum, resonance lines of C IV and Mg II, and intercombination lines such as C III] and Si III].  \cite{Bracco2020} computed the GALEX FUV and NUV count rates for shocks from the \cite{Sutherland2017} MAPPINGS models.  Shocks slower than about 100 \kms are dominated by the 2-photon continuum, while faster shocks have strong contributions from C IV and other lines.  The predicted ratios range from about 0.1 to 0.9.  Based on our own model calculations, a reddening $E(B-V)$ $\sim$ 0.1  does not affect the ratios very much, though it reduces both the NUV and FUV count rates by a factor of 2.   A major caveat, however, is that scattering in edge-on sheets of emitting gas, in particular SNR filaments, can strongly reduce the intensities of resonance lines such as C IV \citep{Cornett1992}, reducing the FUV/NUV ratio.

Above we have presented large GALEX FUV mosaic images of two large nebulae which appear to be true Galactic SNRs, but were either only suspected or missed in most radio studies which have concentrated their searches near the Galactic plane. The object G249+24 also appears to have been missed in wide FOV optical surveys due to its relatively weak H$\alpha$ emissions. 

Both our investigations and that of \citet{Bracco2020} indicate that broad, far UV imaging can be an especially useful means for detecting and distinguishing interstellar shocks and, in some cases, is more sensitive compared to H$\alpha$ imaging. As noted by \citet{Bracco2020}, both line emissions and two photon continuum emissions contribute to FUV emission in shocks with velocities above $\simeq 50$ km s$^{-1}$.

To illustrate the usefulness of far UV emission imaging for detecting SNRs, 
\citet{Bracco2020} noted the presence of networks of thin FUV filaments in both the Antlia SNR and the recently discovered Galactic remnant G70.0-21.5 \citep{Boumis2002,Fesen2015,Raymond2020}. In  Figure~\ref{G70} we show a comparison of MDW's  H$\alpha$ image and the GALEX FUV image of G70.0-21.5. Until the study of the SNRs discussed here, this remnant at $b = -21.5\degr$ had been the remnant with the highest Galactic latitude. This figure nicely demonstrates how that the FUV image makes the remnant easy to detect and helps to define its full dimensions despite the many missing individual GALEX images. Knowing about the existence of these GALEX FUV images could have helped \citet{Fesen2015} and \citet{Raymond2020} in their analyses of this remnant in regard to the remnant's true physical size.  In summary, therefore, far UV images appear to be an especially useful tool for identifying interstellar shocks like those found in SNRs, albeit though best suited for high Galactic latitude searches.

\section{Conclusions}

We have investigated the nature of two large and suspected supernova remnants located at unusually high Galactic latitudes through GALEX far UV emission mosaics and optical images and spectra.
This research also has uncovered one new  Galactic SNR. Our findings include:

1) The large radio emission shell, G354.0-33.5, seen in
1420 MHz and 1.4 GHz radio polarization maps is very likely a SNR. The remnant exhibits numerous sharp FUV emission filaments in a thin, unbroken shell with angular dimensions of $11\degr \times 14.0\degr$. It also exhibits a coincident H$\alpha$ emission shell.

2) A group of bright, sharp FUV emission filaments coincident with numerous but faint H$\alpha$ filaments appears to be a previously unrecognized SNR, G249.7+24.7.  Optical spectra of several filaments show  evidence for the presence of 50 - 150 km s$^{-1}$ shocks. Deep H$\alpha$ images reveal a highly filamentary morphology like that seen in evolved SNRs. 
\citet{Becker2021} independently discovered this remnant as a circular $4.4\degr$ shell of diffuse, soft X-ray emission (0.1 keV) with an estimated radio spectral index of
$-0.69 \pm0.08$.  Moderate dispersion optical spectra of two stars lying in the direction of this remnant have been found to exhibit high-velocity, red and blue shifted \ion{Na}{1} absorption lines in the range of $55 - 73$ km s$^{-1}$, with one star showing an added blue absorption feature at  $-125$ km s$^{-1}$. The closest of these two stars lies at a Gaia estimated distance of 386 pc which, if confirmed by follow-up higher dispersion data, sets a strict maximum distance to the G249.7+24.7 SNR at less than 400 pc making it one of the closest Galactic SNRs known.

3) Despite its enormous angular dimensions ($20\degr \times 26\degr$), our GALEX FUV mosaic images, plus wide-angle H$\alpha$ images and optical spectra strongly support a SNR origin for the Antlia nebula. This conclusion is in line with a reported $\pm 70$ km s$^{-1}$ \ion{Ca}{2} absorption in one background star. We estimate an age $\sim$10$^{5}$ yr, which is an order of magnitude less than the earlier estimate $\sim$1 Myr. We also find the remnant is in likely in direct physical contact along its southwestern rim with the Gum Nebula.

4) Our investigation of suspected SNR located at unusually high Galactic latitudes ($> 15\degr$) highlights the value of UV images to detect interstellar shocks.

Follow-up work on these three objects could include optical spectra of the FUV and H$\alpha$ filaments of the G354-33 remnant, wide-field [\ion{O}{3}] $\lambda$5007 imaging of G249+24 exploring its [\ion{O}{3}] emission,
and high-dispersion echelle spectra of the two apparent background stars toward G249+24 to investigate our reported high-velocity \ion{Na}{1} absorption lines.
Higher resolution H$\alpha$ images and optical spectra of the series of the long and very bright filaments seen in southwestern limb of the Antlia remnant could also provide an additional test of our conclusion regarding Antlia's collision with the Gum Nebula. 

Although we have found that all three nebula in our study are SNRs, only the G249.7+24.7 remnant has a distance or upper limit estimate, leaving us with only approximate physical parameters and evolutionary status. This problem can be addressed especially well for the other two objects, G354.0-33.5 and Antlia, given their enormous angular dimensions and hence offering numerous potential background stars to explore high-velocity ISM absorptions. Given the era of accurate Gaia parallaxes, obtaining high-dispersion spectra looking for high-velocity 
\ion{Na}{1} $\lambda\lambda$5890,5896 and \ion{Ca}{2} $\lambda$3934 absorptions in background stars like that done for several SNRs could be quite fruitful
(e.g., Vela: \citealt{Jenkins1976J,Danks1995,Cha2000}; IC~443: \citealt{Welsh2003}; S147: \citealt{Sallmen2004}; Cygnus Loop: \citealt{Fesen2018}; W28: \citealt{Ritchely2020}).
Such observations might also help gauge the range of shock velocities across the very large and expansive remnants of G354.0-33.5 and Antlia.

Our findings that both the G354.0-33.5 and Antlia nebula are real SNRs was a bit of a surprise. Out of the 300 or so confirmed SNRs, there are only about a dozen SNRs larger than two degrees and less than half that number located more than ten degrees off the Galactic plane \citep{Green2015,Green2019}. Now with
the realization that some Galactic remnants can reach angular sizes several times larger than even Vela, it encourages the investigation of other large emission shells suspected as possible SNRs or previously viewed as unlikely SNR candidates. It also raises the value of SNR searches in the $\mid b \mid$ $> 10\degr$ range, generally viewed before as unproductive.

Finally, we note that although far UV imaging appears to be a sensitive new tool for uncovering the presence of interstellar shocks, it is most useful in searching  uncomplicated regions, like the objects discussed here located in the Galactic halo. However, the number of similar but unrecognized high latitude Galactic remnants is probably pretty small. Nonetheless, given the dozens of unconfirmed but seemingly likely or suspected SNRs in the literature (see list in \citealt{Green2019}'s catalogue), many new SNR discoveries may be aided by the use of UV imaging. 

\bigskip

We thank Eric Galayda and the entire MDM staff for making the optical observations possible, and the SALT Observatory and Resident Astronomer staff for obtaining the excellent RSS spectra despite disruptions due to COVID-19 restrictions. We also thank R.\ Benjamin for helpful discussions. This work made use of the Simbad database, NASA's Skyview online data archives, and the Max Planck Institute for Radio Astronomy Survey Sampler. This work is part of R.A.F's Archangel III Research Program at Dartmouth. D.M.\ acknowledges support from the National Science Foundation from grants PHY-1914448 and AST-2037297.

%\facilities{MDM Observatory (OSMOS), SALT (RSS)}

%\software{PYRAF \citep{pyrafcite}, Astropy v4.0 \citep{AstropyCiteA,AstropyCiteB}, ds9 \citep{ds9cite}, L.A.\ Cosmic \citep{vanDokkum2001}, OSMOS Pipeline (thorosmos: \url{https://github.com/jrthorstensen/thorosmos})}

\bibliography{ref2.bib}
\end{document}